\documentclass[a4paper,11pt]{article}
\pdfoutput=1 
\usepackage{jcappub}
\usepackage{graphicx}
\pdfoutput=1 
\usepackage{color}
\usepackage{makecell}
\usepackage{afterpage}
\usepackage[dvipsnames]{xcolor}
\usepackage{multirow}
\usepackage{cleveref}
\usepackage[normalem]{ulem}		
\usepackage{xspace}
\usepackage{subcaption}
\usepackage{booktabs}
\usepackage{csquotes}
\usepackage{placeins}
\usepackage{centernot}
\usepackage{mathtools}
\usepackage{stmaryrd}
\usepackage{float}
\usepackage{listings}
\definecolor{cadmiumgreen}{rgb}{0.0, 0.42, 0.24}
\definecolor{deepblue}{rgb}{0,0,0.5}
\definecolor{deepred}{rgb}{0.6,0,0}
\definecolor{deepgreen}{rgb}{0,0.5,0}

\definecolor{SGC}{HTML}{9B4F96}

\newcolumntype{L}[1]{>{\raggedright\arraybackslash}p{#1}}
\newcolumntype{C}[1]{>{\centering\arraybackslash}p{#1}}

\newcommand{\class}{{\sc class}\xspace}
\newcommand{\classnet}{{\sc classNet}\xspace}
\newcommand{\classfull}{{\sc classFull}\xspace}
\newcommand{\cobaya}{{\sc cobaya}\xspace}
\newcommand{\paperone}{{\sc CosmicNet i}\xspace}

\title{CosmicNet II: Emulating extended cosmologies with efficient and accurate neural networks}

\author[1]{Sven G\"unther,}
\author[1]{Julien Lesgourgues,}
\author[1]{Georgios Samaras,}
\author[2]{Nils Sch\"oneberg,}
\author[1]{Florian Stadtmann,}
\author[1]{Christian Fidler,}
\author[3]{and Jes\'us Torrado}

\affiliation[1]{Institute for Theoretical Particle Physics and Cosmology (TTK), \\ RWTH Aachen University, D-52056 Aachen, Germany.}
\affiliation[2]{Dept. F\'isica Qu\`antica i Astrof\'isica, Institut de Ci\`encies del Cosmos (ICCUB), Facultat de F\'isica, Universitat de Barcelona (IEEC-UB), Mart\'i i Franqu\'es, 1, E08028 Barcelona, Spain}
\affiliation[3]{Service de Physique Th\'eorique, Universit\'e Libre de Bruxelles, Boulevard du Triomphe CP225, B-10503 Brussels, Belgium}

\emailAdd{fidler@physik.rwth-aachen.de}
\emailAdd{sven.guenther@rwth-aachen.de}
\emailAdd{lesgourg@physik.rwth-aachen.de}
\emailAdd{samaras@physik.rwth-aachen.de}
\emailAdd{schoeneberg@physik.rwth-aachen.de}
\emailAdd{stadtmann@physik.rwth-aachen.de}
\emailAdd{torrado@physik.rwth-aachen.de}

\abstract{
In modern analysis pipelines, Einstein-Boltzmann Solvers (EBSs) are an invaluable tool for obtaining CMB and matter power spectra.
To significantly accelerate the computation of these observables, the CosmicNet strategy is to replace the usual bottleneck of an EBS, which is the integration of a system of differential equations for linear cosmological perturbations, by trained neural networks. This strategy offers several advantages compared to the 
direct emulation of the final observables, including very small networks that are easy to train in high-dimensional parameter spaces, and which do not depend by construction on primordial spectrum parameters nor observation-related quantities such as selection functions. In this second CosmicNet paper, we present a more efficient set of networks that are already trained for extended cosmologies beyond $\Lambda$CDM, with massive neutrinos, extra relativistic degrees of freedom, spatial curvature, and dynamical dark energy. We publicly release a new branch of the {\sc class} code, called {\sc classnet}, which automatically uses networks within a region of trusted accuracy. We demonstrate the accuracy and performance of {\sc classnet} by presenting several parameter inference runs from Planck, BAO and supernovae data, performed with {\sc classnet} and the {\sc cobaya} inference package.
We have eliminated the perturbation module as a bottleneck of the EBS, with a speedup that is even more remarkable in extended cosmologies, where the usual approach would have been more expensive while the network's performance remains the same. We obtain a speedup factor of order 150 for the emulated perturbation module of \class. For the whole code, this translates into an overall speedup factor of order 3 when computing CMB harmonic spectra (now dominated by the highly parallelizable and further optimizable line-of-sight integration), and of order 50 when computing matter power spectra (less than 0.1 seconds even in extended cosmologies).
}

\begin{document}

\hfill{\small TTK-22-25}

\maketitle

\newpage
\section{Introduction}
With the advent of precision cosmology, Bayesian model comparison and parameter inference from current and upcoming data are playing an increasingly important role, while becoming ever more numerically expensive. This Bayesian inference is most often based upon a variety of two-point statistics, such as the CMB angular power spectra or the matter power spectrum. To compute these correlation functions at the linear level (and possibly also at the non-linear one using various algorithms and recipes), one employs Einstein-Boltzmann solvers (EBS) such as \textsc{camb} \cite{camb_code} or \class \cite{class_overview,class_approximations}. After each new data release and for each new proposed theoretical ingredient, the testing of different model and data combinations requires huge grids of such inference runs, each of which combines tens of thousands of individual executions of the underlying EBS. While each execution itself only takes a a few seconds, the accumulated runtime usually adds up to many thousands of core hours.

The runtime of a single execution is usually dominated by two steps: the solution of the coupled set of differential equations including the Boltzmann hierarchy, which gives the source functions for cosmological observables at first order in linear perturbation theory, and the integration of these source functions along the line-of-sight. While the second bottleneck can be feasibly circumvented with new integration strategies 
(for example \cite{Campagne:2017xps,Schoneberg:2018fis}) and more efficient parallelization, the former remains problematic due to its sequential nature. Since the source functions have a very smooth dependence on the underlying cosmological parameters as well as the wavenumber and time under consideration, it is tempting to replace the first bottleneck by a quick emulator.

While for computationally expensive N-body runs the use of emulation has already been broadly adopted \cite{Heitmann:2008eq,Heitmann:2009cu,Lawrence:2009uk,Agarwal:2012ew,Agarwal:2013aea,Heitmann:2013bra,Lawrence:2017ost,Euclid:2018mlb,Ho:2021tem,Euclid:2020rfv,Arico:2021izc,DeRose:2018xdj,McClintock:2018uyf,Zhai:2018plk,Zennaro:2021bwy,Kobayashi:2020zsw}, in the context of EBS emulation has only recently become more common. Almost two decades ago first proposals of power spectrum emulation have been put forth in the context of \texttt{DASh} \cite{dsh_code} \texttt{CMBwarp} \cite{cmbwarp_code} or \texttt{PICO} \cite{pico_code} based on polynomial representations, but these haven't been able to keep up with the increasing accuracy requirements imposed by the newer WMAP \cite{WMAP:2012nax} and Planck \cite{Planck:2013pxb,planck2015,planck2018} missions due to the large number of required polynomial coefficients. Instead, the first use of neural networks to emulate the cosmological power spectra was performed by \texttt{CosmoNet} \cite{Auld:2006pm,Auld:2007qz} at accuracy sufficient for inference with WMAP data. Finally, in \paperone \cite{Albers:2019rzt} a first fully-integrated neural network emulator of the source functions capable of predicting recent Planck data \cite{planck2018} has been presented. Soon after, various groups have released neural network emulators aimed at predicting directly the final observable spectra: 
galaxy clustering and cosmic shear harmonic power spectrum for {\sc CosmicMemory} \cite{Manrique-Yus:2019hqc},
3D matter power spectrum within {\sc bacco} \cite{Arico:2021izc}, 
CMB harmonic power spectra $C_\ell$ and 3D matter power spectrum for {\sc CosmoPower} \cite{SpurioMancini:2021ppk}, 
CMB harmonic power spectra for {\sc connect} \cite{Nygaard:2022wri}.  Other neural network emulators have been developed for more specific observables such as the 21cm power spectrum, see e.g. \cite{Bevins:2021eah}. Some groups have also invested in other emulator techniques to predict the final observables, see for example \cite{Mootoovaloo:2021rot,Ho:2021tem}.

Thus, the strategy of {\sc CosmicNet} is very different from that of all other EBS emulators on the market: it is designed for singularly emulating one bottleneck within the EBS execution -- the prediction of the source functions -- while other emulators are directly targeting the final observables. We believe that none of these two strategies is generally better: depending on the user needs,  {\sc CosmicNet}, {\sc CosmoPower}, or {\sc connect} could be be better suited. On the one hand, once the networks have been trained for the purpose of a given parameter estimation (that is with the right set of model parameters / observables / accuracy) emulators like 
{\sc CosmoPower} or {\sc connect} will always be faster, since they emulate the whole sequence of EBS  tasks. The advantages of the {\sc CosmicNet} strategy are more on the side of robustness, range of application, and (re)training speed. As a matter of fact:
\begin{itemize}
\item The source functions emulated in the {\sc CosmicNet} strategy are independent of several assumptions on the cosmological model (e.g. on the primordial power spectrum,  which could be arbitrarily complicated) and on the observables (for instance, the matter density or lensing source functions predicted by the network can be convolved with whatever selection function of a given survey). This means that there are several situations in which the  {\sc CosmicNet} networks can be readily used while other networks would need a specific retraining.
\item When considering some new physics -- beyond the extended cosmology already considered in this work -- that affects the evolution of cosmological perturbations (e.g. non-minimal assumptions concerning neutrinos, dark matter, dark energy or gravity), the  {\sc CosmicNet} networks do need retraining, like those of other emulators. However, the task of emulating the source functions is simpler than that of predicting e.g. harmonic power spectra, thanks to the smoothness of these functions and the way the line-of-sight integration already accounts for shifts in the geometry (caused e.g. by curvature or different dark energy models). As a consequence, the {\sc CosmicNet} strategy involves smaller and simpler networks than other methods, which are faster to retrain -- especially when the number of free cosmological parameters increases.
\item When the goal is to evaluate only the matter power spectrum, no line-of-sight integral is required and EBS codes only have one bottleneck -- precisely the one that {\sc CosmicNet} is able to remove. Then, already with the {\sc CosmicNet} strategy, the EBS code is so fast that there is no point in accelerating it further, since the evaluation of the data likelihood would be typically slower. Instead, when the goal is to evaluate harmonic power spectra $C_\ell$\,, the {\sc CosmicNet} strategy does not provide such a speed-up as other emulators. Nevertheless, with {\sc CosmicNet}, the acceleration might be sufficient to reduce the EBS evaluation time below the one of the likelihood evaluation -- making further acceleration pointless. Moreover, our hope is to remove the second bottleneck in the future, with a more systematic use of new algorithms (similar to \cite{Schoneberg:2018fis}) and high parallelisation schemes on future CPUs.
\end{itemize}
 In this paper, we will often come back to this comparison, and we will discuss the assets of the {\sc CosmicNet} strategy in more details.
In summary, we acknowledge that the use of codes such as  {\sc CosmoPower} or {\sc connect} is optimal for several tasks, but we believe that the {\sc CosmicNet} strategy offers sufficient advantages for being pursued in parallel.

The first version of {\sc CosmicNet} discussed in \cite{Albers:2019rzt} and the current released version of {\sc CosmoPower} \cite{SpurioMancini:2021ppk} only targeted the vanilla $\Lambda$CDM model.
\enlargethispage*{2\baselineskip}
Within the context of this work we extend the approach to cover non-standard models including the presence of dark radiation, curvature, non-zero neutrino masses, and a time-dependent equation of state for the dark energy. This grants the emulator a much larger flexibility in covering many of the most common inference runs, which typically address one or two-parameter extensions of the vanilla model. Further, the performance of these extended networks in these particular extensions of $\Lambda$CDM hints at their stability and usability in more general non-standard scenarios. Finally, compared to our previous \paperone implementation, we propose a reduction in network sizes (allowed by smarter network design), a more efficient interconnection with the remainder of the \class code, and a more advanced selection of the training and prediction domains. Altogether, these improvements allow for a much more efficient training and evaluation of the networks, leading to the final elimination of the numerical solution of the Boltzmann equations as a bottleneck of EBS codes.

Our concrete implementation of the {\sc CosmicNet} strategy and {\sc CosmicNet} neural networks is implemented in a new public branch of the \class code called \classnet available at \url{https://github.com/lesgourg/class_public}. We will refer to the original code without neural networks as \classfull. Note however that the {\sc CosmicNet} purposes are very general and could be implemented in other EBSs.

We present the new pragmatic architecture of our neural networks in  \cref{sec:architecture}. We discuss the training strategy employed for these networks in \cref{sec:training_cn}, and show the corresponding performance in terms of accuracy and speed in \cref{sec:results}. We finally conclude our discussion in \cref{sec:conclusion}.
\section{A new pragmatic network architecture}\label{sec:architecture}

As in \paperone, we recognize that the angular power spectra are derived from the line-of-sight approach, and can thus be written as convolutions of the underlying source functions $S(k,\tau)$ with the various radial projection functions.\footnote{In flat space, these are spherical Bessel functions and their derivatives, while in curved spacetime more general hypergeometric functions need to be used.}
The precise splitting of the source functions among the available radial projection functions is not unique, as different options can be equivalent through partial integration. Within this work, we will base ourselves on the splitting traditionally used in \class \cite{Lesgourgues:2013bra}, which includes four source functions.
The perturbations occuring in these source functions are typically found from integrating a system of coupled ordinary differential equations \cite{gauges}. The objective of this work is to provide a fast and effective way of skipping this computational step. Explicitly, we may write (see \paperone or \cite{Lesgourgues:2013bra})
\begin{align}
	S_{T_0} &= g \cdot (F_0 + \phi) + e^{-\kappa} 2\phi' + (g \theta_b/k^2)'~, \label{eq:def_T0}\\
	S_{T_1} &= e^{-\kappa} \, k \, (\psi-\phi)~,\\
	S_{T_2} &= g (G_0 + G_2 + F_2)/8~, \\
	S_{P} &= \sqrt{6} S_{T_2}~.
\end{align}
Here $k$ is the comoving wavenumber, $g$ is the visibility function, $\kappa$ is the optical depth, $\phi$ and $\psi$ are the Bardeen potentials, $F_0$ and $F_2$ are the photon temperature multipoles (in the notation of \cite{gauges}), $G_0$ and $G_2$ are the photon polarization multipoles, and $\theta_b$ is the baryon velocity divergence. All the perturbations that appear in equations (\ref{eq:def_T0}-\ref{eq:def_weyl}) are formally {\it transfer functions}, that is, perturbations normalised to the initial condition ${\cal R}=1$, where ${\cal R}$ is the usual curvature perturbation.
\enlargethispage*{3\baselineskip}
We note that the polarization source function can simply be expressed as $\sqrt{6}$ times the $T_2$ source function, given the equations derived in the optimal hierarchy scenario \cite{Tram:2013ima}, which have been proven to be quite accurate despite being based on an approximation \cite{Pitrou:2020lhu}. In order to also enable the computation of matter power spectra, galaxy power spectra and lensing corrections of the CMB, compared to \paperone we have added the source functions of the matter (m) and cold dark matter+baryon (cb) overdensities as well as the lensing potential. These are simply given, respectively, by
\begin{align}
  S_{\delta_\mathrm{m}} &= \delta_\mathrm{m} + 3 H \theta_\mathrm{m}/k^2~, \\
  S_{\delta_\mathrm{cb}} &= \delta_\mathrm{cb} + 3 H \theta_\mathrm{cb}/k^2~, \\
  S_{\phi+\psi} &= \phi+\psi~. \label{eq:def_weyl}
\end{align}
Note that the source functions for m and cb overdensities include velocity corrections that make them gauge independent. These corrections are most relevant on super-Hubble scales. Like the others, these two source functions can be convolved along the line of sight with radial functions in order to compute angular power spectra (see e.g. \cite{DiDio:2013bqa}). They can also simply be squared and subsequently multiplied by the primordial curvature spectrum in order to form the three-dimensional linear Fourier power spectrum $P_x(k,z)$ with $x=\mathrm{m},\mathrm{cb}$.

We could choose a similar approach as in \paperone and use one network per source function, that is, six networks (three for temperature source functions, two for density source functions, and one for the lensing source function). However, building on the experience with \paperone, we recognize that the given components can be further decomposed into more fundamental building blocks which are comparatively straightforward (and thus fast) to predict. In particular, splitting the CMB source function components into parts sourced either during recombination or reionization proved extremely efficient. For this purpose, we define a function $g_\mathrm{reco}(z)$, which is simply equal to the visibility function $g(z) = -\kappa'(z) e^{-\kappa(z)}$ assuming that there is no reionization. Then, we derive the re-ionization contribution\footnote{In practice, we compute $g_\mathrm{reco}(z)$ and $g_\mathrm{reio}(z)$ simultaneously, during a single evaluation of the thermodynamics module of \class. Indeed, reionization is implemented as an additive contribution to the ionization fraction ($x_e$) and therefore also to the scattering rate $\kappa' \propto x_e$. Thus one can simply record the value of $x_e$ before adding reionization contributions and use this value to compute $g_\mathrm{reco}(z)$. The additional contribution to the ionization fraction from reionization is simply added as $x_e = x^\mathrm{reco}_e + x^\mathrm{reio}_e$, with $x^\mathrm{reio}_e(z)$ defined by the user (by default, it is a hyperbolic tanh function dependent on a power of the redshift, but the \class input features several other options).} as $g_\mathrm{reio}(z) \equiv g(z) - g_\mathrm{reco}(z)$. Performing the splitting into reco/reio components for $T_0$ and $T_2$\,, and additionally disentangling the contribution from the integrated Sachs-Wolfe effect (ISW) in $T_0$ gives us a total of nine components that have to be predicted. However, it turns out that the prediction of the components related to large-scale clustering ($\phi+\psi, \delta_\mathrm{m}, \delta_\mathrm{cb}$) are sufficiently similar to be predicted by the same network.\footnote{Note that the function $\phi'$ needed for $S_{T_0, \mathrm{ISW}}$ should not be aggregated to the same network. Indeed, our network for $\phi+\psi, \delta_\mathrm{m}, \delta_\mathrm{cb}$ is optimised for accurately estimating these functions during structure formation, at redshifts relevant for the computation of power spectra and CMB lensing effects. $\phi'$ needs to be accurately predicted additionally at early times, around and after photon decoupling, in order to compute the early ISW contribution to the temperature spectrum. It is more efficient to assign this task to a dedicated network.}

\enlargethispage*{2\baselineskip}
To summarize, in this work we define nine components that are predicted using seven networks as follows:
\begin{align}
  S_{T_0} &= S_{T_0, \mathrm{ISW}} + S_{T_0, \mathrm{reco}} + S_{T_0, \mathrm{reio}}\\
  &= e^{-\kappa} \underbrace{2\phi'}_{\mathrm{[N1]}} + \underbrace{g_\mathrm{reco} \cdot (F_0 + \phi) + (g_\mathrm{reco} \theta_b/k^2)'}_{\mathrm{[N2]}}
  + \underbrace{g_\mathrm{reio} \cdot (F_0 + \phi) + (g_\mathrm{reio} \theta_b/k^2)'}_{\mathrm{[N3]}}~,\\
  S_{T_1} &= e^{-\kappa} \, k \, \cdot \underbrace{(\psi-\phi)}_{\mathrm{[N4]}}~,\\
  S_{T_2} &= S_{T_2,\mathrm{reco}} + S_{T_2, \mathrm{reio}} =
  g_\mathrm{reco} \underbrace{(G_0 + G_2 + F_2)/8}_{\mathrm{[N5]}}
  + g_\mathrm{reio} \underbrace{(G_0 + G_2 + F_2)/8}_{\mathrm{[N6]}} ~,\label{eq:source_ST2_definition}\\
	S_x &= (1+ \underbrace{\epsilon_x}_{\mathrm{[N7]}_{x\kern-3pt}} ) \cdot A_x \qquad \mathrm{with} \,\, \quad x \in \{ {\phi+\psi}, \, {\delta_\mathrm{m}}, \, {\delta_\mathrm{cb}}\}~.
\end{align}
Here we explicitly show for each part of each source function the corresponding neural network $\mathrm{[N}i\mathrm{]}$ with $i \in \{1..7\}$ that is used to predict it, as well as the analytical approximations $A_x$ for the source functions that further simplify the task for the network (see \cref{sec:approx}). These splittings are motivated by the nature of the underlying perturbations to be predicted, which are explicitly listed for each network below.

We note that the network predictions are always for a given grid in wavenumbers $k$ (of size $N_k=\mathcal{O}(700)$) and are separately obtained for each time ($\tau$) -- that is, $\tau$ is one of the input parameters to the networks, while the output returns a vector of source functions for all values of $k$ in the grid. We further point out that the input to all networks always includes at least the cosmological parameters defining a given cosmology (except those relating to the primordial power spectrum), which in our case are explicitly given as\footnote{We adopt here some common notations for the cosmological parameters. The correspondance with usual CLASS names is given by $\Omega_\mathrm{b} h^2=${\tt omega\_b}, $\Omega_\mathrm{m} h^2=${\tt omega\_m}, $H_0=${\tt H0}, $\kappa_\mathrm{reio}=${\tt tau\_reio}, $\Omega_k=${\tt Omega\_k}, $w_0=${\tt w0\_fld}, $w_a=${\tt wa\_fld}, $\Omega_\nu h^2=${\tt omega\_ncdm}, $\Delta N_\mathrm{eff}=\,${\tt N\_eff-3.044}. The conversion to {\tt N\_ur} depends on the assumed number of neutrinos. For our default choice of three neutrinos, $\Delta N_\mathrm{eff}=\,${\tt N\_ur-0.00641}.}
\begin{equation}
	\mathrm{Cosmo~inputs} = \{\Omega_\mathrm{b} h^2, \Omega_\mathrm{m} h^2, H_0, \kappa_\mathrm{reio}, (\Delta N_\mathrm{eff}, \Omega_k,  \Omega_\nu h^2,w_0, w_a)\}~.\label{eq:cosmo_inputs}
\end{equation}
Here, the $\Lambda$CDM parameters are the physical baryon density $\Omega_\mathrm{b} h^2$, the physical matter density $\Omega_\mathrm{m} h^2$ (including baryons, cold dark matter and possibly massive neutrinos), the Hubble parameter $H_0$ and the optical depth of reionization $\kappa_\mathrm{reio}$. The additional parameters in parenthesis are related to simple extensions of the $\Lambda$CDM model that are commonly considered in the literature: the effective neutrino number corresponding to additional free-streaming dark radiation $\Delta N_\mathrm{eff}$, a curvature density fraction $\Omega_k$, some dark energy equation of state in the CLP expansion characterized by $w(z) = w_0 + w_a (1-a)$ \cite{Chevallier:2000qy,Linder:2002et}, and finally the physical density of massive neutrinos $\Omega_\nu h^2 \approx \sum m_\nu/(93.14\mathrm{eV})$ related to the sum of neutrino masses (see \cite{planck2018} for a more detailed description of these extensions). Compared to \paperone, we added the last five parameters to demonstrate the flexibility of the underlying network architecture in predicting any non-$\Lambda$CDM cosmology. This list of input parameters can in principle by quite simply expanded by any additional desired parameter.

Our networks are designed to output the source functions for a discrete set of wavenumbers $k$. This set needs to be chosen with care, since it should be adequate for all cosmologies, while \class usually computes such an array for each cosmological model. Indeed, \class{} includes a sophisticated algorithm that spaces the $k$ values in order to densely sample regions in which the source functions are important and quickly-varying, and less densely sample regions in which they are subdominant or slowly-varying. Thus, when executing the \classfull code, each cosmology leads to a different $k_\mathrm{min}$, $k_\mathrm{max}$ and a different set of intermediate values. Our strategy is to let \classnet adopt a fixed optimal $k$ grid by comparing the grid of all models in the training set and building a unique conservative grid out of these.\footnote{More precisely, \classnet identifies three $k$-grids in the training set: the one with the smallest $k_\mathrm{min}$\,, the one with the largest $k_\mathrm{max}$\,, and the one with the largest number of values. It then defines a fixed $k$-grid containing all the values of the latter grid, completed by a relatively dense logarithmic sampling of values on each edge, down to $k_\mathrm{min}$ and up to $k_\mathrm{max}$\,.}

We will now go through each network and summarize their input, output and architecture. The input always consists of at least the above \enquote{Cosmo inputs} plus conformal time~$\tau$. When relevant, we rescale $\tau$ by $\tau_\mathrm{reio}$ (or $\tau_\mathrm{reco}$), because some source functions oscillate with a phase that is set by $\tau_\mathrm{reio}$ (or $\tau_\mathrm{reco}$): using the rescaled input variable $\tau/\tau_\mathrm{reio}$ (or $\tau/\tau_\mathrm{reco}$) reduces the dependence of such source functions on cosmology. On top of this, since neural networks can efficiently handle redundent input, we may pass additional quantities. 
Our strategy for the definition of the input results from trying several options and selecting those which give the best compromise between training time (for a given targeted accuracy) and evaluation time:
\\[1\baselineskip]
\begin{itemize}
\item First, we may pass some functions of time -- derived at negligible computational cost by the background or thermodynamics modules of \class{} --  that match some expected behavior of the source functions, and thus make their prediction easier. Together with  $\tau$, $\tau/\tau_\mathrm{reio}$, or $\tau/\tau_\mathrm{reco}$ these functions constitute the \enquote{Tau input} of each network.
\item Second, in order to ease the task of each network, we may pass explicit analytical approximations to the targeted source functions. In \paperone, we relied massively on this approach for the CMB networks. The drawback is that this strategy leads to large $k$-dependent input vectors, and thus to wide networks that are slow to evaluate. In this work, we found that omitting such approximations leads to a better compromise for all CMB source functions. The new networks are much shallower and faster to evaluate. For the particular network used to predict ($\delta_\mathrm{b}$, $\delta_\mathrm{cdm}$, $\phi+\psi$) and derive large scale structure observables, we do have at our disposal some efficient analytical approximations. However, we do not pass them as input to the network: instead, we ask the network to predict only the ratio between the true and the approximate source function, as detailed in Eq.~(\ref{eq:source_ST2_definition}).
\end{itemize}
Note that, as is common for any deep neural network, we employ a non-linear activation function, in our case the leaky ReLU:
\begin{equation}
\textrm{LReLU}(x)=\begin{cases}
x & x\ge 0 \\
\beta x & x < 0
\end{cases}\label{eq:leakyrelu}
\end{equation}
with $\beta = 0.25$ as has been shown in \paperone to provide optimal performance for our purpose.
In the following we provide an explicit list of network layouts and input quantitites.

\begin{figure}
	\centering
	\includegraphics[width=6cm]{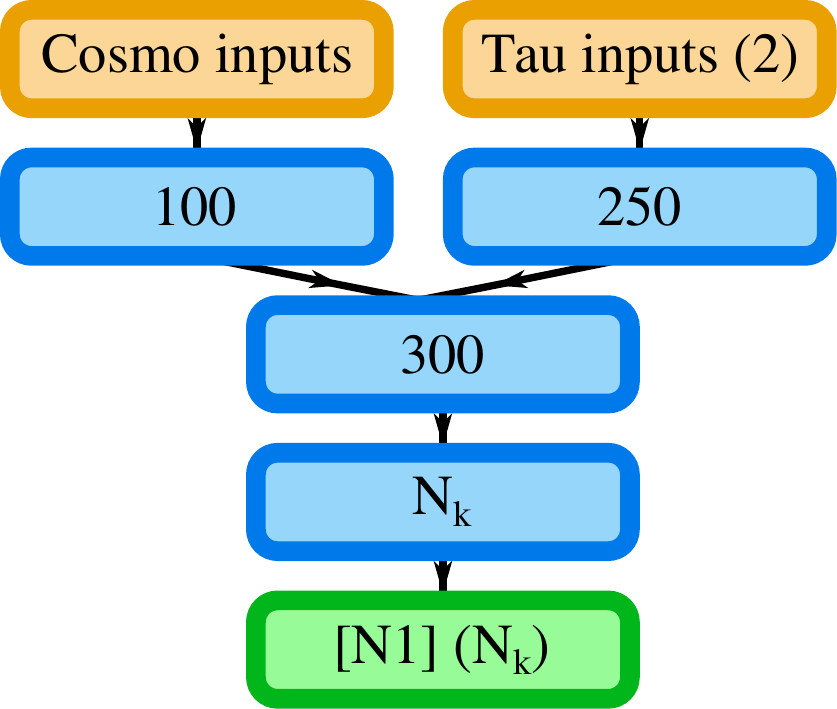}
	\caption{\label{fig:layout_ST0_ISW} Layout of the [N1] network predicting the ISW contribution to the $S_{T_0}$ source function. The input layers are depicted as dark orange, the hidden layers are shown as blue, and the output layer is shown in green. The numbers in brackets depict the number of neurons, and $N_k$ is the number of wavenumbers $k$ to be predicted.}
\end{figure}

\paragraph{Network [N1] ($T_0$ ISW):}
For the network [N1], the source function that should be predicted corresponds to most of the ISW effect (network [N4] contains a sub-dominant ISW contribution. Indeed, while [N1] captures the contribution of the ISW when both potentials would be equal, [N4] captures the difference of the potentials sourced by anisotropic stress). The network target and input read
\enlargethispage*{2\baselineskip}
\begin{equation}
	T_0,\mathrm{ISW} = 2 \phi'~, \qquad \qquad \mathrm{Tau~inputs} = \{\tau, D(\tau)\}~,\label{eq:t0isw_def}
\end{equation}
where $D(\tau)$ is the scale-independent growth function of matter density fluctuations for the given cosmology. This source function has comparatively a simple behavior and a small contribution to the final temperature spectrum. Thus, in practice, we found that a very small and fast network could be adopted without significant loss of accuracy. The layout is displayed in \cref{fig:layout_ST0_ISW}. We choose a deep feed-forward neural network which only takes as input the parameters of \cref{eq:cosmo_inputs} and the two $\tau$-dependent numbers of \cref{eq:t0isw_def}. These inputs are separately fed into two fully connected layers of sizes 100 and 250, and then combined into a single 300 neuron layer. One more layer is imposed between that layer and the output layer. The former layers both have the same width, given by the dimension $N_k$ of the wavenumber grid. For this network, all the values of $\tau$ at which \class needs to store the source functions (for future use by other modules) are considered for the loss, which is defined as a mean square against the true $2\phi'$.

\paragraph{Network [N2] ($T_0$ reco):}
The network [N2] is designed to capture all non-ISW effects for the $T_0$ source function that arise from recombination. Explicitly, this means that it attempts to model the intrinsic, Sachs-Wolfe, and Doppler contributions with the following input:
\begin{equation}
\resizebox{0.9\textwidth}{!}{$
	T_0~\mathrm{reco} = g_\mathrm{reco} (F_0 + \phi) + \partial_\tau (g_\mathrm{reco} \theta_b)/k^2~,~~ \mathrm{Tau~inputs} = \{ \tau/\tau_\mathrm{reco} , g_\mathrm{reco}(\tau), g_\mathrm{reco}'(\tau), e^{-\kappa(\tau)}\}~.$
}\label{eq:t0reco_def}
\end{equation}
Interestingly, we found that further splitting the function into sub-components did not increase computational efficiency, as each of the components is approximately as difficult to predict as their composition. We further note that the contribution from this source function peaks around recombination due to the $g_\mathrm{reco}(\tau)$ function and can be ignored for $\tau>4 \cdot \tau_{reco}$ (this is around $\sim$5\% of $\tau_{0}$ in a standard $\Lambda$CDM cosmology), saving a good amount of computational resources. This is particularly important since the underlying components of this source function are highly oscillatory around recombination, transitioning from tightly coupled baryonic acoustic oscillations to the decoupled regime. As a first order approximation, one could expect the function to simply behave as $\sin(k r_s) e^{-k^2/k_D^2}$ as expected at leading order in perturbation theory. However, in practice the transition into the decoupled regime strongly impacts the shape of the oscillations, leading to frequency shifts and non-trivial damping. As such, the modeling of this source function requires the most advanced networks employed within this work.

\begin{figure}
	\centering
	\includegraphics[width=\textwidth]{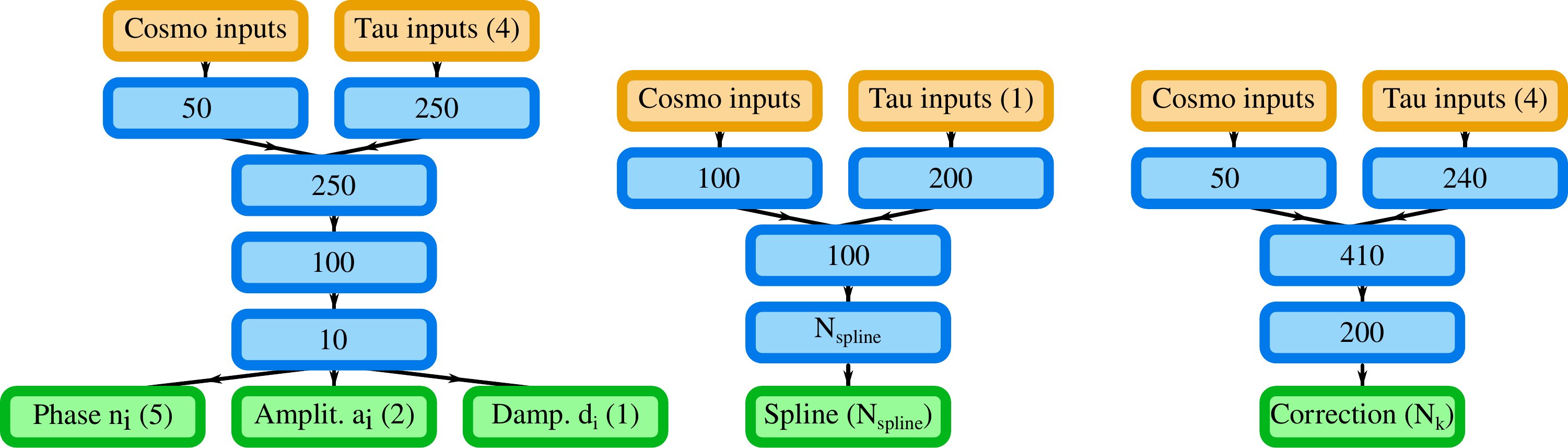}
	\caption{\label{fig:layout_ST0reco} Layout of the [N2] network predicting the $T_0$ reco contribution, similar to \cref{fig:layout_ST0_ISW}. \textbf{Left:} Approximation network, \textbf{Middle:} Spline network, \textbf{Right:} Correction network (see the text for a definition of the role of these networks).}
\end{figure}

A layout of the network [N2] is displayed in \cref{fig:layout_ST0reco}. Essentially, the output arises from the combined prediction of three networks. A first network attempts to predict a best-fitting oscillatory function with a given phase, amplitude, and damping (see details below). A second network takes care of predicting the overall offset of the zero point of accoustic oscillations, using a very smooth spline function that is simply added on top. Finally, a third correction network takes care of the remaining deviations of the assumed functional shape from the true result which cannot be captured by the previous two networks.

The first network tries to predict the parameters $n_i$ (phases), $a_i$ (amplitudes), and $d_i$ (damping) of the following analytical approximation to the source function:
\begin{equation}
\resizebox{0.9\textwidth}{!}{$
	T_0~\mathrm{reco}  \simeq \left[a_1 \cos(n_1 + k r_s (1+n_2) + k^2 n_3) + a_2 \mathrm{sinc}(k r_s (1+n_4) + k^2 n_5)\right] \cdot \exp(-k^2/\widetilde{k}_D^2),$}
\end{equation}
where $\mathrm{sinc}(x) = \sin(x)/x$\,, and the damping is computed as $\widetilde{k}_D^{-2} = k_D^{-2} (1+d_1)$. Here $r_s$ is the sound horizon of the given cosmology, whereas $k_D$ is an approximation for the damping scale. Both of these quantities are computed within the \class thermodynamic module.\footnote{These quantities are defined by \class{} as $r_s=\int_{0}^{\tau_\mathrm{rec}} \! d\tau \,c_s$ and
$k_D^{-2} =\left[ \int_{0}^{\tau_\mathrm{rec}} \! d \tau \, \frac{1}{6 \kappa'} \frac{R^2+16/15(1+R)}{(1+R)^2} \right]$, where $\tau_\mathrm{rec}$ is the location of the peak of the visibility function $g_\mathrm{rec}(\tau)$, $c_s^2 = \frac{1}{3(1+R)}$ is the sound speed of the photon-baryon fluid, and $R = 4\rho_b/3\rho_\gamma$ is the baryon loading.}

This approximation is basically an extension of the results predicted at leading order in perturbation theory, which would correspond to $n_i=d_1=0$ (see e.g. equation (5.57) in \cite{Hu:1995em}). We additionally allow for a running of the frequency with wavenumber ($n_i\neq0$) and a correction to the damping scale ($d_1\neq0$).

In principle the cosmological parameters and all relevant time-dependent information listed in \cref{eq:t0reco_def} could be passed to all networks. In practice, we found that for the spline network only information about $\tau/\tau_\mathrm{reco}$ is relevant. The other networks do profit from the additional information contained in $g_\mathrm{reco}(\tau)$, $g_\mathrm{reco}'(\tau)$, $e^{-\kappa(\tau)}$.

The second network takes care of predicting the offset of the acoustic oscillations caused mainly by gravitational forces on baryons. Analytically, at leading order, we expect this offest to be given by $-R \, \phi$, but we do not impose this assumption. We define $N_\mathrm{spline}$ nodes equally spaced in $k^{1/3}$ between $k_\mathrm{min}$ and $0.6$~Mpc$^{-1}$, such that small $k$'s are more efficiently sampled. The role of the network is then to predict $N_\mathrm{spline}$ interpolation coefficients. We find that with $N_\mathrm{spline}=12$ the network nicely capture the full offset in the relevant range.

The third network simply predicts a correction factor at each point in the $k$-grid. Thus its final layer has a width of $N_k$. Note that the second and third network are technically degenerate. However, we do not train them simultaneously. During the first four training epochs, we only train the first (approximation) and second (spline) networks, while the third network is de-activated. In the remaining epochs, we keep the spline network approximately fixed through a small learning rate (see \cref{ssec:hyperparameters}), and we keep training only the first (approximation) and third (correction) networks.

\begin{figure}
	\centering
	\includegraphics[width=6cm]{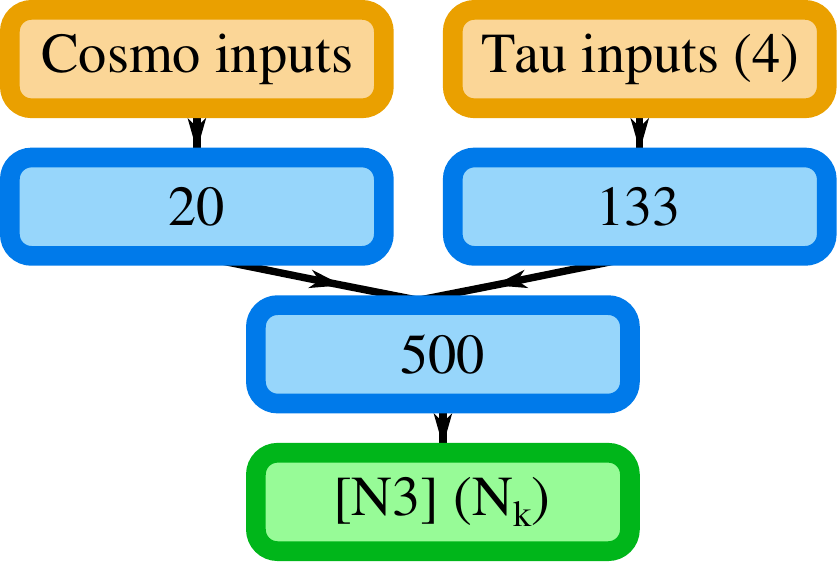}
	\caption{\label{fig:layout_ST0reio} Same as \cref{fig:layout_ST0_ISW}, but for the network [N3]}
\end{figure}

\enlargethispage*{2\baselineskip}
\paragraph{Network [N3] ($T_0$ reio):}
The network [N3] is designed to capture all non-ISW effects for the $T_0$ source function that arise from reionization in the late universe. Explicitly, this means that it attempts to model

\begin{equation}
\resizebox{0.9\textwidth}{!}{$
	T_0~\mathrm{reio} = g_\mathrm{reio} (F_0 + \phi) + \partial_\tau (g_\mathrm{reio} \theta_b)/k^2~,~~ \mathrm{Tau~inputs} = \{\tau/\tau_\mathrm{reio}, g_\mathrm{reio}(\tau), g_\mathrm{reio}'(\tau), e^{-\kappa(\tau)}\}~.$
}\label{eq:t0reio_def}
\end{equation}
It should be noted that at late times during reionization the photon overdensity contribution $F_0$ is sub-dominant compared to the gravitational potential~$\phi$ as well as the baryon~velocity~$\theta_b/k^2$. The latter are almost completely driven by the growth of cold dark matter fluctuations during structure formation. Thus the $k$-dependent shape of the source function is relatively simple and a small and fast network structure suffices (see figure \ref{fig:layout_ST0reio}). Moreover, due to the vanishing of this source function at early times, the network is only computed for $\tau \geq 0.6 \cdot \tau_\mathrm{reio}$, saving another significant amount of computational work.

\paragraph{Network [N4] ($T_1$):}

\begin{figure}
	\centering
	\includegraphics[width=6cm]{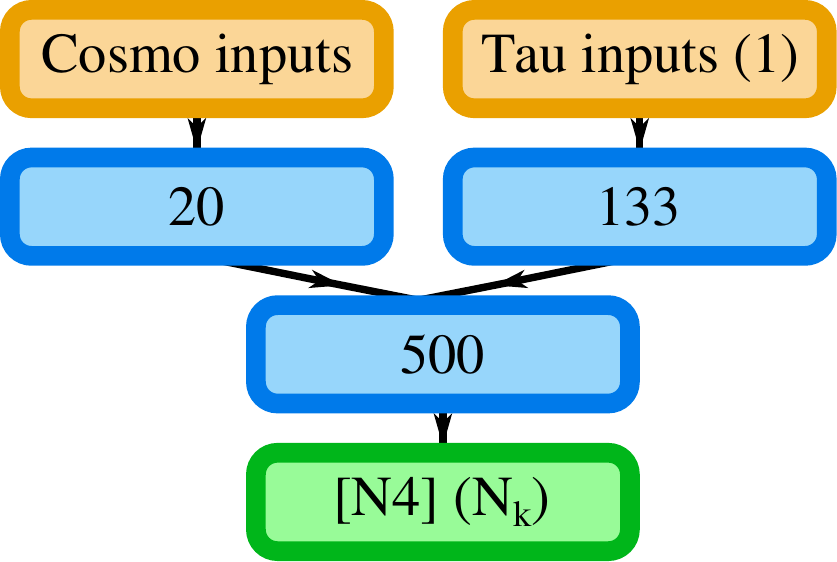}
	\caption{\label{fig:layout_ST1} Same as \cref{fig:layout_ST0_ISW}, but for the network [N4].}
\end{figure}

The network [N4] is designed to capture the $T_1$ source function, which in Newtonian gauge is simply
\begin{equation}
	T_1 = \psi-\phi~,~~ \mathrm{Tau~inputs} = \{\tau\}~.
\end{equation}
In absence of decoupled ultra-relativistic species, the two metric fluctuations would be exactly equal and this function would vanish. However, decoupled neutrinos and photons have a small shear stress that lead to $\phi \neq \psi$ and alters sub-dominantly the overall ISW effect. On the one hand, the shear oscillation pattern as a function of $\tau$ and $k$ is rather complicated. On the other hand, high precision on $T_1$ is not necessary in order to achieve a good accuracy in the final predicted observables. Thus we can keep this network as small and fast as the previous one (see figure \ref{fig:layout_ST1}).

\paragraph{Network [N5] ($T_2$ reco):}

The network [N5] is responsible for computing the contributions to the $T_2$ source function arising from recombination. Explicitly, this is
\begin{equation}
	T_2~\mathrm{reco} = (G_0+G_2+F_2)/8~, \qquad \mathrm{Tau~inputs} = \{\tau/\tau_\mathrm{reco}\}
\end{equation}
with the usual $F_\ell$ and $G_\ell$ from \cite{gauges}. Note that we do not include $g_\mathrm{reco}$ in the definition, unlike for $T_0~\mathrm{reco}$ (see \cref{eq:source_ST2_definition}). We actually noticed that the sharp cutoff introduced by $g_\mathrm{reco}$ made the source function harder to compute. A similar factorization is of course not possible for the $T_0$ source function due to the term proportional to $g'(z) \theta_b/k^2$.

\begin{figure}
	\centering
	\includegraphics[width=6cm]{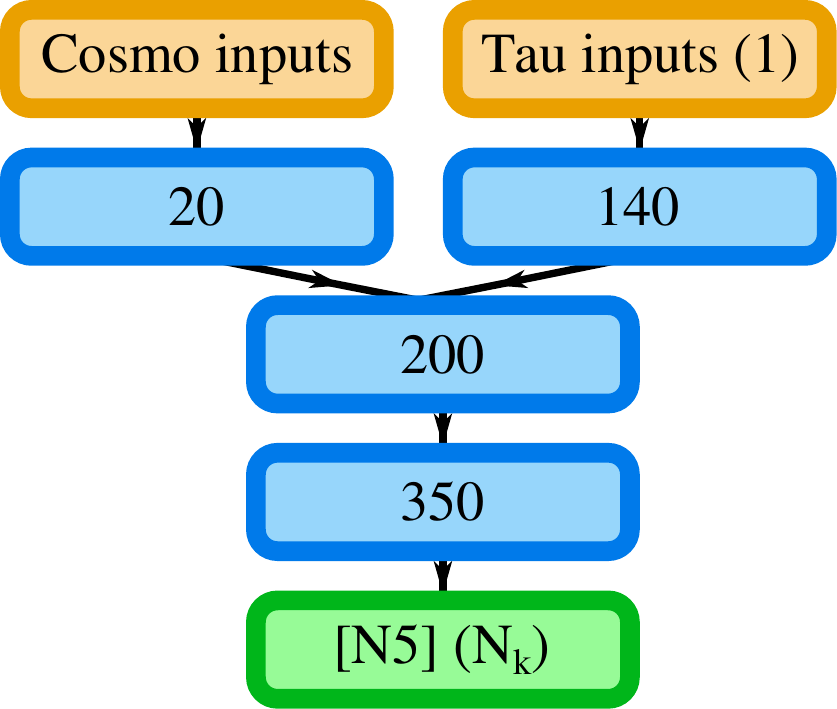}
	\caption{\label{fig:layout_ST2_reco} Same as \cref{fig:layout_ST0_ISW}, but for the network [N5].}
\end{figure}

The modeling of the polarization multipoles $G_\ell$ in principle requires a complicated network structure similar to the [N2] network above. However, in this case even a relatively simple network architecture managed to vastly outperform more complicated architectures based on analytical approximations (similar to [N2]).
The reason is that the polarization multipoles are sourced only indirectly through the higher photon temperature multipoles (in particular $F_2 = 2\sigma_\gamma$), leading to a simpler and more regular oscillatory structure of the perturbations. Furthermore, in this case the time pre-factor can be factorized and needs not be modeled independently. Note that while for the temperature autocorrelations the contribution from the $T_2$ source function is notoriously subdominant, it is the only (and hence dominant) component for the E-polarization autocorrelation spectra.
In the end, we found that a medium size and fully connected feed-forward network provides the best performance while allowing for sufficient accuracy. The layout of this network is displayed in \cref{fig:layout_ST2_reco}. By only computing the network at $\tau \leq 4 \cdot \tau_\mathrm{reco}$\,, one can save again a considerable amount of computational resources.

\paragraph{Network [N6] ($T_2$ reio):}

The network [N6] predicts contributions to the $T_2$ source function arising from reionization. Explicitly, we write
\begin{equation}
	T_2~\mathrm{reio} = (G_0+G_2+F_2)/8~, \qquad \mathrm{Tau~inputs} = \{\tau/\tau_\mathrm{reio}\}
\end{equation}
Note that this is completely the same as for the [N5] network. Both networks are only differentiated by the times $\tau$ at wich they are predicted (and thus the times for which the loss is computed). This splitting is relevant because the function $G_0+G_2+F_2$ behaves very differently around recombination and reionization.

\begin{figure}
	\centering
	\includegraphics[width=6cm]{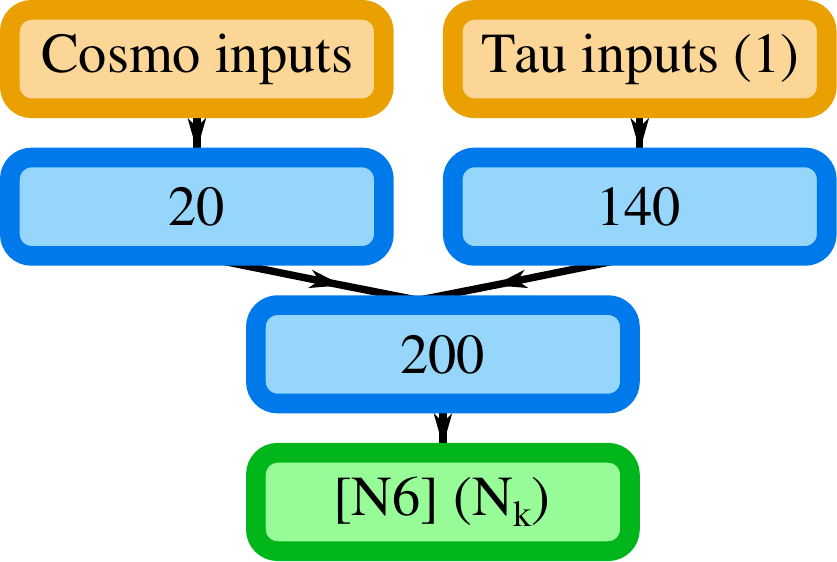}
	\caption{\label{fig:layout_ST2_reio} Same as \cref{fig:layout_ST0_ISW}, but for the network [N6].}
\end{figure}

Unlike the baryon velocity or the metric potential, the photon shear and polarization multipoles remain highly oscillatory even until late times. However, like for the [N5] network, it turns out that even an extremely small network is able to vastly outperform more complicated configurations (see figure \cref{fig:layout_ST2_reio}.). In this case, the network is only evaluated at $\tau \geq 0.6 \cdot \tau_\mathrm{reio}$ in order to save computational resources.

\paragraph{Networks [N7]:}

The network [N7] is designed to predict both weak lensing observables (such as the lensing potential angular power spectrum used for CMB lensing) and the matter power spectrum. This feat is possible due to the Poisson equation relating metric fluctuations and matter overdensities, which allows for a single network to predict multiple functions. Additionally, unlike in the previous networks, here we explicitly use analytical approximations of the expected result. The network is only meant to predict the relative difference between the true source functions and their analytical approximation. This considerably eases the task of the network, whose output is a smooth function varying within a small range, unlike the source functions themselves that span several orders of magnitude.

The targeted source functions are the lensing potential $\phi+\psi$ and the matter overdensities $\delta_\mathrm{m}$ and $\delta_\mathrm{cb}$.
\begin{figure}
	\centering
	\includegraphics[width=9cm]{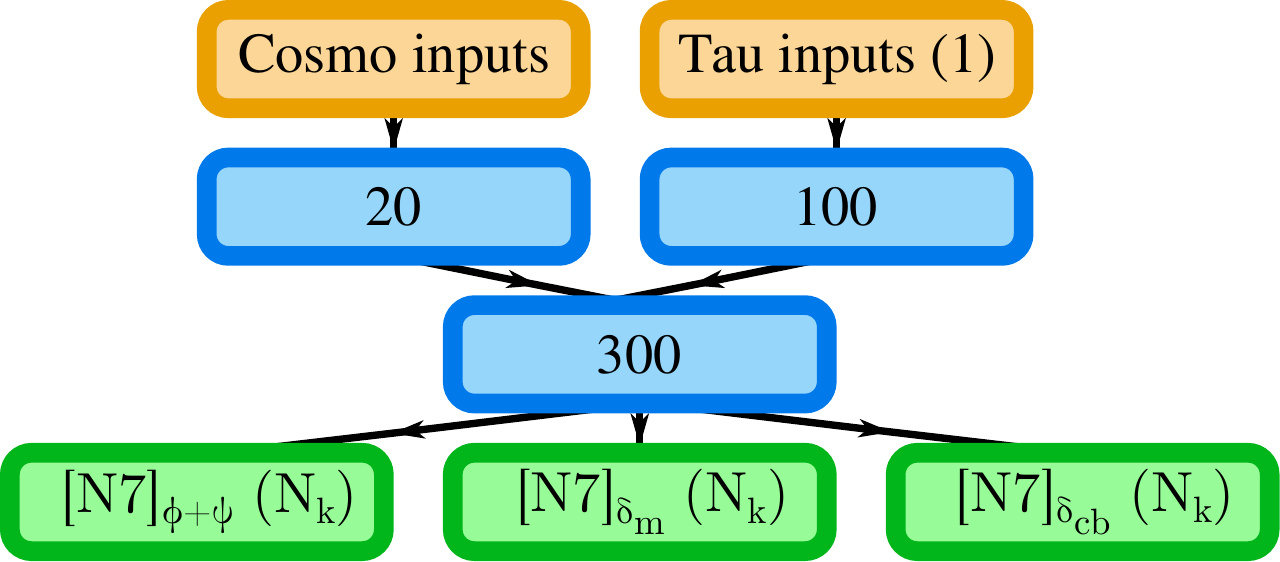}
	\caption{\label{fig:layout_phi} Same as \cref{fig:layout_ST0_ISW}, but for the network [N7].}
\end{figure}
In this case, there is only one time-dependent input $\tau$ in addition to the cosmological parameters. As apparent from \cref{fig:layout_phi} the network that predicts all three source functions can be made surprisingly simple due to the relatively predictable behavior of the source functions.

Since the targeted source functions vary by many order of magnitudes throughout time and wavenumber, instead of evaluating the loss based on absolute differences between the predicted and true source functions, we use a mean-squared error on the relative deviation between them.

For the analytic approximations, we use the Hu \& Eisenstein transfer functions $T_\mathrm{m}$ and $T_\mathrm{cb}$ of reference~\cite{hu_eisenstein}. Note that these transfer functions are normalized differently than the \class ones. We thus need to re-normalize them in a way that we detail in Appendix~\ref{sec:approx}. Additionally, the Hu~\&~Eisenstein approximation applies to the $\Lambda$CDM model extended to spatial curvature and massive neutrinos. We also describe in Appendix~\ref{sec:approx} how to extend their range of validity to models with arbitrary relativistic degrees of freedom and dynamical dark energy.

\paragraph{Notable differences with respect to \paperone}

We point out that the network architecture presented in this work has dramatically changed compared to \paperone. This is a result of various different advancements in understanding the problem.

First, we recognized that the convolutional neural networks used in \paperone are relatively slow compared to fully connected deep networks. The use of convolutional neural networks was motivated by correlations between local features in the data. This correlation exists, but deep fully connected networks simply outpaced the convolutional networks for this particular task.

Furthermore, we tried to reduce all networks to simple (and thus fast) designs, abandoning the cumbersome pre-computation of Bessel functions and sines or cosines that was adopted in \paperone. This allowed for a great speedup of the network execution and training time necessary to cover the extended parameter space of this work.\footnote{The $T_0~\mathrm{(reco)}$ network is the only one that could not be radically simplified. As we shall see in section \ref{ssec:performance_breakup}, it has a large contribution to the overall execution time.} For this reason, we traded off between the complexity in implementation (corresponding almost directly to execution speed) and the accuracy of the approximations passed to the networks in order to reach a maximally accurate and fast network architecture. We believe that the current network design is very efficient and worth publishing as such. Still, we remain confident that future iterations on the network architecture might lead to further simplification and speedup.

\section{Training strategy}
\label{sec:training_cn}

In this section we explain how the networks that will be released together with this work have been trained.
We also provide a documentation/tutorial of the code in \cref{sec:implement}, where the user can find notes on how to train their own networks.

\subsection{Domain/Parameter space}
\label{ssec:domain_cn}
The training region should be sufficiently large to contain combinations of cosmological parameters relevant for most parameter estimations from current data. However, a larger covered parameter region will also require larger networks (with more weights and longer training/execution times) to achieve a given desired accuracy. Therefore, the training region must be adjusted in such way to provide a good balance between small networks that are rarely used during parameter estimation, and frequently used but rather large networks. This means that we must first define a set of likelihoods that we want to sample efficiently with \classnet, and then infer an optimal training domain.  During an MCMC exploration of the parameter space, when a model falls outside of the trained domain, the code defaults to the solution of the full ODEs, that is, \classfull is used instead of \classnet. As long as this occurs rarely, the incurred penalty in execution time is minimal. Thus, our goal is to design the trained domain in such a way that --
in an MCMC run based on the least constraining likelihoods in our set of \enquote{relevant likelihoods} -- most chain points belong to the trained domain, but not necessarily all points.

\subsubsection*{Including extended cosmological models}
In \paperone, we only trained our networks within the framework of the minimal $\Lambda$CDM model, that is, with only four relevant cosmological parameters $\{\Omega_\mathrm{b} h^2, \Omega_\mathrm{m} h^2, H_0, \kappa_\mathrm{reio}\}$. We considered the Planck 2018 \cite{planck2018,planck2018_1,planck2018_2} likelihood as the standard (least-constraining) likelihood defining the physically-allowed region, and used it to delineate the boundaries of our domain.´

In this work, we enlarge the parameter space to account for extended cosmologies with additional free parameters $\{\Delta N_\mathrm{eff}, \Omega_k,  \Omega_\nu h^2,w_0, w_a\}$. We apply bounds to exclude unphysical regions from the training and validation domains. In the current release of our networks, these bounds are
\begin{align}
w_0+w_a&\leq-\frac{1}{3}~, \label{eq:parbounds_w0} \\
\Delta N_\mathrm{eff}&\geq 0~, \label{eq:parbounds_N}\\
\Omega_\nu h^2&> 1.70698158 \cdot 10^{-5}~, \label{eq:parbounds_h}\\
\kappa_\mathrm{reio}&>0.004~. \label{eq:parbounds_tau}
\end{align}

Note that in principle one could relax the first bound to $w_0+w_a \leq 0$ only (to avoid unphysically large contribution of the dark fluid perturbations). Note also that we only consider models with enhanced radiation density, $\Delta N_\mathrm{eff}\geq 0$, and not scenarios with some entropy release or low-temperature reheating potentially leading to $\Delta N_\mathrm{eff}\leq 0$.
The lower bound on $\Omega_\nu h^2$ just comes from taking the limit $m_\nu \longrightarrow 0$ for neutrinos with a standard decoupling history and temperature evolution. Finally, the lower bound on $\kappa_\mathrm{reio}$ comes from the value of the optical depth $\kappa(z)$ at the redshift at which $\kappa_\mathrm{reio}$ is usually computed, $z=40$, in a scenario with no reionization at all and with cosmological parameters close to the Planck best fit. The precise values of these limits are not particularly important, since the code automatically switches to the \classfull mode outside of the validation domain.

If we only used Planck 2018 as our least constraining likelihood, the training domain would include values of $\{\Omega_k, w_0, w_a\}$ that are very strongly excluded by current BAO and SNIa data, due to the existence of \enquote{geometrical degeneracies} at the level of CMB observables. We prefer to make the networks more efficient at the cost of reducing their scope. For this purpose, we include BAO and supernovae data in our least constraining data set. If our scheme is efficient in the Planck+BAO+SNIa case,  it will be even more efficient for any joint fit with additional or more recent data sets, as long as those are not in strong tension with Planck+BAO+SNIa. More precisely, we infer our training domain from the following experimental data:
\begin{itemize}
\item Cosmic Microwave Background (CMB) data from the 2018 release of the Planck survey, including low-multipole TT + EE and high-multipole TT + TE + EE polarised data \cite{planck2018_1}, together with the Planck 2018 CMB lensing power spectrum reconstruction \cite{planck2018_2}.
\item Baryonic Acoustic Oscillations (BAO), in particular those used as external data in the Planck 2018 release, combining data from the 6df Galaxy Survey \cite{bao_sixdf}, SDSS DR7 Main Galaxy Sample \cite{bao_dr7} and the BAO power spectrum from SDSS DR12 \cite{bao_dr12}.
\item The Pantheon Supernova (SN) type Ia sample data \cite{pantheon} (including data from the Pan-STARRS1 (PS1) Medium Deep Survey).
\end{itemize}
We stress again that the role of the BAO+SNIa data set is mainly to restrict the training domain to plausible values of $\{\Omega_k, w_0, w_a\}$. We will see later that our networks are still efficient and accurate when fitting Planck data alone with the minimal $\Lambda$CDM model. We also remind the reader that, in runs featuring either a less constraining data combination or some data in tension with Planck+BAO+SNIa, \class will automatically switch to the \classfull mode for any model outside of the training region, while for models within the training region the benefit of using the networks will remain.

\subsubsection*{A new pragmatic training domain}
In \paperone, the trained domain consisted in a hypercube based on the Planck 2018 \cite{planck2018,planck2018_1,planck2018_2} best-fit value $\widehat{x}_i$ and standard deviations $\sigma_i$ for each parameter $x_i$. The cube was centered at $\widehat{x}_i$ with width $10\sigma_i$\,. However, due to correlations between parameters, only a small fraction of the entire cube was usually explored during parameter estimation. Thus, the network of \paperone were relatively deep and long to train, but most of the trained region was never used.

Here we propose a more efficient scheme based on an ellipsoidal (rather than cubic) domain. We focus on a Gaussian approximation to the Planck+BAO+SNIa likelihood around its best-fit model.\footnote{The cosmological parameters at the best-fit model are given by $H_0=68.9$~km/s/Mpc, $\kappa_\mathrm{reio} = 0.0494$, $\Omega_\mathrm{b} h^2=0.0224$, $\Omega_\mathrm{m} h^2 = 0.144$, $\triangle N_\mathrm{eff}=0.066$, $\Omega_k=-9.66\cdot10^{-5}$, $\Omega_\nu h^2=2.75\cdot 10^{-4}$, $w_0=-0.944$, $w_a=-0.281$, $A_s=2.079\cdot10^{-9}$, $n_s=0.971$. \label{footnote:bestfit}} Explicitly, we can define a difference in log likelihood (or equivalently in $\chi^2$ values) as
\begin{align}
\Delta \chi^2 = \left(x-\widehat{x}\right)^TC^{-1} \left(x-\widehat{x}\right)~,\label{eq:chi2_parameter_region}
\end{align}
where $x$ is the vector of parameters, $\widehat{x}$ the best-fit of Planck+BAO+SNIa, and $C$ is its covariance matrix (see below).

Given the well known properties of multi-variate Gaussian distributions, it is possible to show that a value of $\Delta \chi^2$ greater than some threshold $T$ has a certain probability $p$ to occur during sampling. By adjusting this threshold $T$, we can define a region that the sampling algorithm is very unlikely to leave. To derive a reasonable threshold, we adjust the threshold to get a \enquote{theoretical}\footnote{The way we determine a \enquote{theoretical} threshold given a probability of fallback is based on the multivariate Gaussian approximation of \cref{eq:chi2_parameter_region}. If all parameters were truly Gaussian distributed, then the $\Delta \chi^2$ would be distributed according to the $\chi^2$ distribution. Then, given the cumulative probability distribution $F_D(x)$ of the chi square distribution with $D$ degrees of freedom, one simply has $T(p) = F_D^{-1}(p)$ for a parameter space of $D$ dimensions. When referring to the hyperellipsoid in terms of $s \sigma$, we refer to the corresponding probability $p = \mathrm{erf}(s/\sqrt{2})$ (e.g. for $2 \sigma$ the $p \approx 95.4\%$) and the corresponding parameter space inside the threshold $T(p)$.} fallback probability of $5.7 \cdot 10^{-7}$ (this corresponds to $5\sigma$). In practice the likelihood is not perfectly Gaussian (violating \cref{eq:chi2_parameter_region}) and we impose additional bounds on the parameters (see \cref{eq:parbounds_N,eq:parbounds_h,eq:parbounds_tau,eq:parbounds_w0}), such that in practice the fraction of points for which \classfull needs to be used in a fit to Planck+BAO+SNIa data is closer to $7.0 \cdot 10^{-3}$ for the model with all parameters of \cref{eq:cosmo_inputs}.

Applying this threshold to equation \cref{eq:chi2_parameter_region} defines our region of validity for \classnet. Outside of this hyperellipsoid, our code automatically switches to the \classfull mode. Note that while our networks only depends on 9 cosmological parameters (as in \cref{eq:cosmo_inputs}), the full model that we fit to Planck+BAO+SNIa in order to compute a covariance matrix includes two additional parameters $\{A_s\,, n_s\}$ for the primordial power spectrum. The covariance matrix that we use in equation \cref{eq:chi2_parameter_region}  is the 9-dimensional sub-matrix that corresponds to a marginalisation over $\{A_s\,, n_s\}$. Given that we want to avoid discontinuities and inaccurate predictions at the edges of this hyperellipsoid, we conservatively extend the training region to  6$\sigma$ (a threshold of $\Delta\chi^2=59.13$ with 9 dof), while we keep the region of validation, testing and execution of the final network at the aforementioned 5$\sigma$ level ($\Delta \chi^2 = 46.12$ with 9 dof).

\newpage
\subsection{Training Data}
\label{ssec:data_nn}
To obtain training and validation data sampled evenly in the training and validation regions, we use a Latin Hypercube Sampling (LHS) of the 9-dimensional hyperellipsoid outlined above in \cref{ssec:domain_cn}. To achieve this, we calculate eigenvectors and eigenvalues of the covariance matrix $C$ and sample an axis-aligned hyperellipsoid whose lengths are given by the eigenvalues. This is done by an adapted LHS algorithm which discards sampled points outside the ellipsoid for every sampled dimension. After sampling we use the eigenvectors of $C$ to transform this set back from the eigenspace to the parameter space.\footnote{With this method, it is not immediately guaranteed that we will achieve a given desired number $M_\mathrm{target}$ of training, validation, or testing points, especially since the hyperellipsoid is also cut by the parametric constraints outlined in \cref{ssec:domain_cn}. However, a simple approach can enable us to get roughly the desired number of points. To estimate the volumetric fraction of points within the hypercube that also lie inside the cut hyperellipsoid, we draw a first sampling of the hypercube with $N$ points, of which only $M$ points are accepted. We then repeat the process with $N'=(N/M) M_\mathrm{target}$, which yields a number of accepted points of the order of $M_\mathrm{target}$. After a few iterations of this type, we obtain a LHS of the hyperellipsoid with approximately $M_\mathrm{target}$ points. These iterations over an LHS algorithm take a negligible amount of time.}

With this method, we sampled 9\,979 points for training in the training region, 980 points for validation in the validation region, and 994 points for testing in the validation region. The training and validation samples are rather small compared to other emulators \cite{cosmopower}, making the training faster. However, we will see in the next sections that this is sufficient for reaching a good accuracy, thanks to our overall strategy and network design.

On a technical side, we want to note here that the targeted $6\sigma$ training-hyperellipsoid requires to start from a hypercube slightly \textit{wider} than $12\sigma_i$ along each parameter axis. This surprising fact can be derived by using \cref{eq:chi2_parameter_region} and checking explicitly what is the maximal value of $x_i$ that a given parameter can reach.\footnote{Noting that the covariance matrix $C$ is positive definite and symmetric allows us to write it as $C = Q D D Q^{-1}$ with the orthogonal matrix $Q$ ($Q^T = Q^{-1}$) and the diagonal matrix $D$. Then we can write the transformation $x = \widehat{x} + Q D z$, which allows us to write \cref{eq:chi2_parameter_region} simply as $z^T z = (x-\widehat{x})^T \mathcal{C}^{-1} (x-\widehat{x}) = \Delta \chi^2 \leq T$. By using the theory of Lagrange multipliers and constrained optimization, we can find that the minimum/maximum value reached by some $x_i$ is simply $\nabla \left[\pm x_i - \lambda (z^T \cdot z - T)\right]$. This results in solutions of the form $z_k = \pm \sqrt{T} Q_{1k} D_k/\sqrt{\sum_i \, Q_{1i}^2 D_i}$, which after using $C_{ii} = \sigma_i^2 = \sum_j Q_{ij} D_j$ gives $z_k = \pm \sqrt{T} Q_{ik} D_k/\sigma_i$. Undoing the transformation, we find finally $x_i =\widehat{x}_i + Q_{ij} D_j z_j  = \widehat{x}_i \pm \sqrt{T} \sigma_i$, leading to a total extent of $2\sqrt{T}$.}
A probability of $6\sigma$ corresponds to $\Delta\chi^2\leq T=59.13$ in nine dimensions. The maximal extent of a parameter in one direction is simply $\sqrt{T} \sigma_i$ (see footnote), and thus we have $x_i-\widehat{x}_i\approx7.689 \sigma_i$, which corresponds to a cube with width $15.3\sigma_i$ (and not 12$\sigma_i$ as one could have naively expected).

We then use \class to calculate the training and validation data, i.e. the input/output for each training and validation point. However, for each model, we do not need to go through all the \class modules, since we only need to emulate the source functions. Thus, we can stop \class after the {\tt perturbations} module and save the source functions and all other relevant parameters (like the array of conformal times or the growth factor). This enables a significant reduction in training time.

\subsection{Training Hyperparameters}
\label{ssec:hyperparameters}

For any deep learning application, the precise layout and training history of the networks is subject to a few choices encoded in parameters commonly called ``hyperparameters''. These should be optimized to increase the speed and accuracy of the network. For many hyperparameters in our networks (such as the size and number of each hidden layers), we have performed exhaustive grid or random search optimization. For some other parameters, we simply relied on few simple tests or intuition. 

All our networks are trained for 40 epochs. As stated in section~\ref{sec:architecture}, the loss function is defined as a mean square difference between the predicted and targeted source functions $S_X(k,\tau)$ at each discrete point in the $(k,\tau)$ grid.\footnote{Except for [N7] where a relative mean square loss is used.} The loss is only computed above the minimal physical wavenumber $k_\mathrm{min}$ of each cosmological model, which depends on spatial curvature (see Appendix~\ref{app:omegak_approx} for details). The training strategy includes a choice of hyperparameters defining the learning rate.

\subsubsection*{Learning rate}

For the adjustment of the learning rate during training, we use the Adam optimizer \cite{adam}.
We also use batch learning, which means that, during each epoch, the networks are trained on a subset of the training set called a batch of data. Ideally, each batch would contain a combination of different cosmologies and of different conformal times $\tau_i$. However, for the purpose of this work, it was much faster to include in each batch a single cosmology (but all time slices $\tau_i$). However, the source functions vary strongly with cosmology (in particular, due to acoustic oscillation patterns). Thus this batch learning strategy may result in an oscillatory behaviour of the training/validation loss with each batch, and potentially a slow convergence of the networks. To enforce a limited training time, we decrease the learning rate exponentially as a function of the number of epochs $e$, even before the automatic decrement set by the Adam optimiser. We checked that this exponential decay enhances the performance of our training, although at the expense of freezing out the networks earlier, and not being able to reach arbitrarily high accuracy. In practice,  we decrease the learning rate of each network as a function of epoch $e$ as $\exp(-e/8)$, excepted for two cases.\footnote{In the network [N2] ($T_0$ reco), after epoch $e=5$, we keep the spline network fixed, in order to be able to start the training of the correction network; then, we decrease the learning rate of the correction network as $\exp(-(e-5)/8)$. In the nework [N7], the multiple simultaneous outputs favor a slightly faster decaying learning rate, and we adopt $\exp(-e/5)$ instead of the default $\exp(-e/8)$. We leave a more advanced batch selection (allowing to abandon the forced decrement in the learning rate) to future work.}

\subsubsection*{Repeated training}

Given the above discussion on the learning rate, it is expected that different trainings with different randomly initialized initial weights will tend towards different local minima or stationary points of the loss. 
As a workaround, for the two networks [N2] and [N5] with the largest contribution to the loss (corresponding to $T_0~\mathrm{reco\_no\_isw}$ ${T_2}~\mathrm{reco}$), we restarted the learning process ten times with the same hyperparameters. For the selection of the final network, we use a variety of tracers of performance, such as the validation loss, cuts through posteriors of the likelihood, and full MCMC contours. The reason for this more complex approach is that the validation loss alone does not well capture all kinds of systematic shifts in the source spectra, such as those causing a $n_s$\,-like tilt in the final power spectra. Incorporating a sensitivity to such shifts into the loss is left for future work. Our method for investigating cuts through the posterior, presented in \cref{ssec:trouble}, allows us to quickly notice if the hyperparameters of \classnet cause such systematic shifts. We additionally cross-check with a single MCMC chain for the full 11 dimensional model once a candidate network (with low validation loss and without significant deviations in the posterior) has been found. We recommend that, when the networks of \classnet are retrained, such a posterior cut (or full Bayesian sampling) is also performed in order to ensure that they reach at least the same accuracy as our original released networks.

\newpage
\section{Results\label{sec:results}}

In order to fairly evaluate the performance and accuracy of \classnet, we have to fix a given set of cosmological and precision parameters for both \classnet and \classfull. As such, we assume in all calculations three massive degenerate neutrinos with {\tt N\_ncdm=1} and {\tt deg\_ncdm=3}. The total mass of the neutrinos is specified through the input parameter $\Omega_\nu h^2$, that represents the total neutrino density today. Variations in the effective neutrino number $N_\mathrm{eff}$ are implemented through another input parameter accounting for extra ultra-relativistic degrees of freedom (beyond the $\Lambda$CDM standard prediction $N_\mathrm{eff}$=3.044). We also compute the source functions up to the wavenumber\footnote{We achieved this by always setting the input parameter {\tt \textquotesingle Pk\_max\_1/Mpc\textquotesingle} to {\tt 100} in \classnet. This means that not only the ($\delta_\mathrm{m}$, $\delta_\mathrm{cb}$, $\phi+\psi$) source functions but also the CMB source functions are computed up to $k_\mathrm{max}=100\,\mathrm{Mpc}^{-1}$. This may sound as a waste of resources, since the calculation of the CMB $C_\ell$ spectra only depends on $k < 0.6 \mathrm{Mpc}^{-1}$ (with default precision). However, at large $k$, CMB source functions are negligible and thus easy to predict: the inclusion of large $k$ values does not slow down the training of our networks in any significant way.} $k_\mathrm{max}=100\,\mathrm{Mpc}^{-1}$. In the case of requesting all outputs, we pass to \class the parameters {\tt \textquotesingle output\textquotesingle:\textquotesingle tCl, pCl, lCl, mPk\textquotesingle} and {\tt \textquotesingle lensing\textquotesingle:\textquotesingle yes\textquotesingle}, while if we consider only the matter or CMB+baryon power spectra we set instead {\tt \textquotesingle output\textquotesingle:\textquotesingle mPk\textquotesingle}.

In \cref{sec:performance} we first investigate the performance of the \classnet code, and continue to discuss the accuracy in \cref{ssec:source_functions} for the underlying source functions, in \cref{ssec:power_spectra} for the observable power spectra, and in \cref{sec:parameter_estimation} for the full parameter inference pipeline.

\subsection{Performance\label{sec:performance}}

\subsubsection{Speedup of the perturbation module\label{sec:speed_pert}}

To judge the efficiency of the NN implementation, one should do a timing of only the replaced \class{} modules (in this case the \texttt{perturbations} module). Of course, the speedup of this module does not directly translate to a speedup of the complete \class calculation. Indeed, when harmonic spectra ($C_\ell$s) are requested, the \texttt{transfer} module of \class is still a bottleneck, putting a lower bound on the evaluation time of \classnet. We recall that speeding up the \texttt{transfer} module separately (by means of new parallelization schemes and new strategies for carrying out the time-consuming line-of-sight integrals) is still in progress: thus it is interesting to focus on the performances of the \texttt{perturbations} module only. However, we will also discuss the speedup of the complete \class calculation further below. 

The \classfull \texttt{perturbations} module and the networks implemented in \texttt{PyTorch} for \classnet both support multi-core parallelism. Thus we computed the speedup factor for different numbers of threads $N_\mathrm{threads}$\,. While the training of the networks took place on a GPU, we chose to evaluate them on the CPU to generate our benchmarks, since this allows for a fairer comparison to \classfull (which does not support GPU parallelization) and we cannot reasonably assume that the machines on which the MCMC will be run must have a GPU. This also implies that our speedup factors are conservative, as the GPU performance of the deep networks is considerably faster than the CPU performance. We perform these benchmarks on
an Intel(R) Xeon(R) Gold 6140 CPU @ 2.30GHz CPU with 18 physical cores; while this CPU does support hyperthreading, we chose to only display the benchmarks for physical core numbers (and indeed, increasing the number of threads beyond 18 does not yield significantly different results). Note that a given sampler will usually run multiple instances of \class in a MPI-parallelized way, which implies that a linear scaling to a larger number of cores is not necessarily an important issue.

The benchmark consists of running \classfull{} and \classnet{} 50 times for different points of the testing set which was sampled from the $5 \sigma$ hyperellipsoid outlined in \cref{ssec:data_nn} to produce an estimate of the average time\footnote{Each benchmark run is preceded by a few \enquote{warm-up} evaluations whose results are discarded. This avoids any initial outliers due to start-up/initialization procedures, shared library loading, cache misses, etc.} needed. This sample incorporates the 9 relevant parameters of the $\Lambda$CDM+$M_\nu$+$N_\mathrm{eff}$+$\Omega_k$+$(w_0,w_a)$ model and is run using the default precision parameters of \classfull.

\begin{figure}
    \centering
    \includegraphics[height=4.4cm]{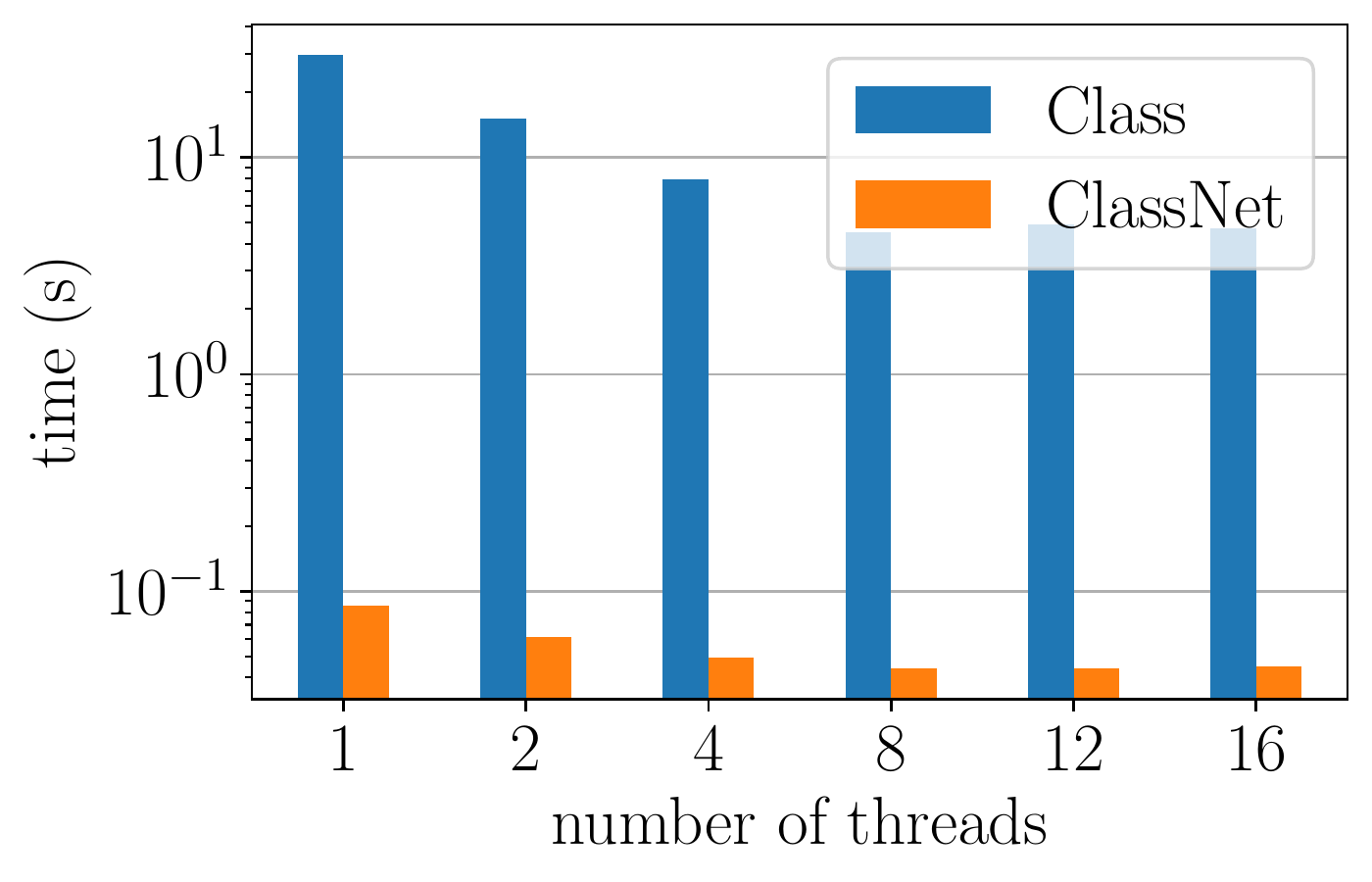}
    \includegraphics[height=4.4cm]{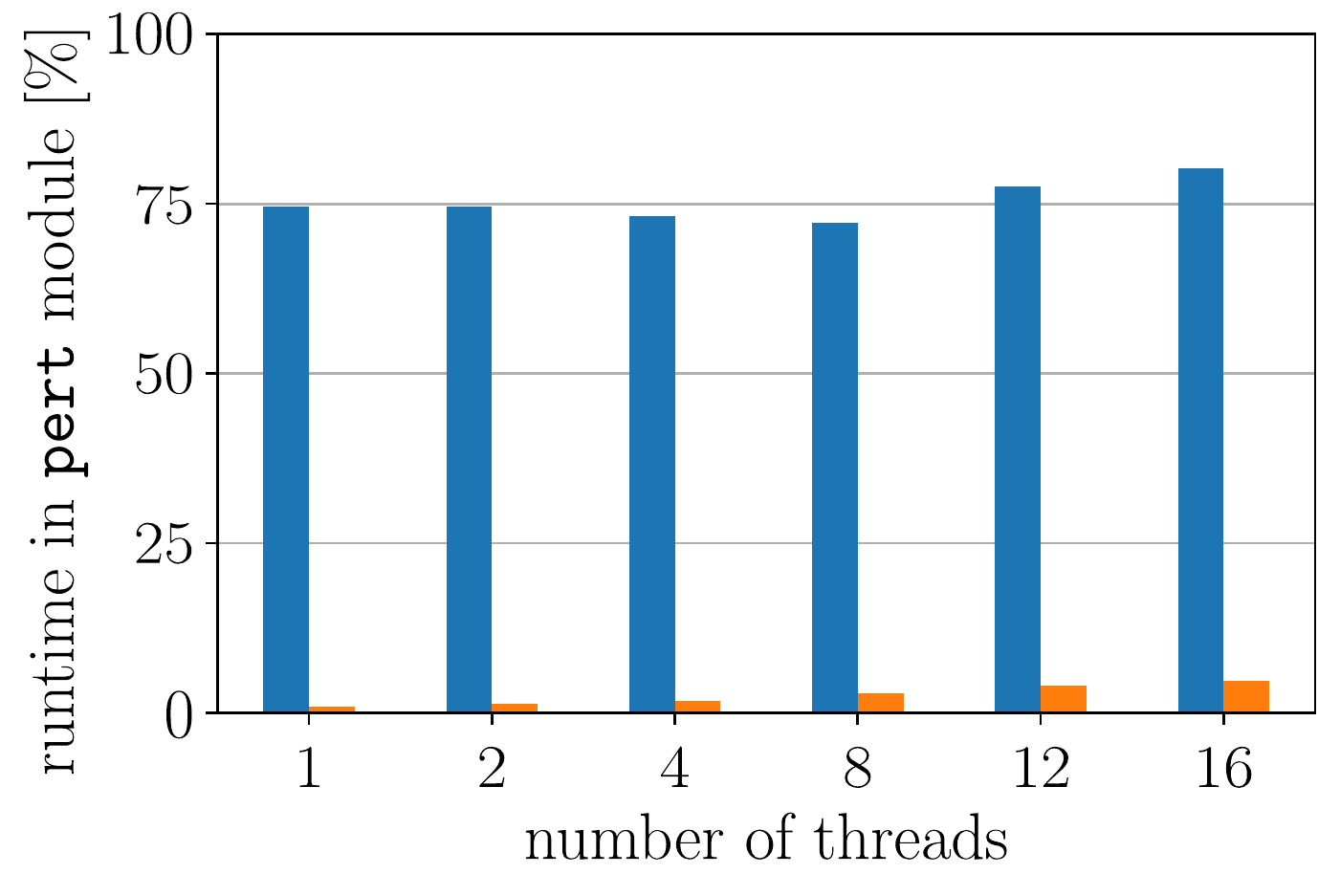}
    \caption{Execution times when \class is used to compute all CMB spectra plus the matter and baryon + CDM power spectra. \textbf{Left:} Average time spent in the \texttt{perturbations} module with and without neural networks as a function of the number of used CPU cores. Note the logarithmic scale. \textbf{Right:} Fraction of the total \class runtime spent in the \texttt{perturbations} module.}
    \label{fig:benchmark-abs-time}
\end{figure}

We define the speedup to be the relative reduction in wall-clock run time, i.e.
\begin{equation}
\textrm{speedup} = \frac{T^\texttt{perturbations}_{\mbox{\classfull}}}{T^\texttt{perturbations}_{\mbox{\classnet}}} - 1
\end{equation}
such that a speedup of $0$ corresponds to the same level of performance, and a speedup factor of $1$ means that the runtime is halved. The absolute times for the \texttt{perturbations} module are displayed in the left panel of \cref{fig:benchmark-abs-time}. For \classfull{}, the evaluation time approximately scales like $N_\mathrm{threads}^{-1}$ up to $N_\mathrm{threads}\sim 8$ and plateaus beyond, due to the impossibility to parallelise the integration of the ODEs for the largest Fourier wavenumber $k_\mathrm{max}$\,. For \classnet{}, the scaling as a function of $N_\mathrm{threads}$ is worse, due to the  overhead related to the pre- and post-processing of the NN input/output (see \cref{fig:benchmark-time-per-step}).

However, the speedup is always considerable, ranging from a factor of $\sim 300$ for one thread to $\sim 80$ for $N_\mathrm{threads}\geq 8$. This improvement by several orders of magnitude is sufficient to remove the \texttt{perturbations} module from the list of bottlenecks in the EBS solver (as shown in the right panel of \cref{fig:benchmark-abs-time} and further discussed below in section~\ref{sec:class_speedup}). Thus, the main goal in terms of speedup is achieved and further optimization would not make a great difference in terms of overall execution speed. If the other bottlenecks of \class are eventually tackled, one might want to revisit the core scaling of \classnet, but we leave this investigation to future work. 

\newpage
\subsubsection{Speedup breakdown}\label{ssec:performance_breakup}

\begin{figure}
    \centering
    \includegraphics[height=8.4cm]{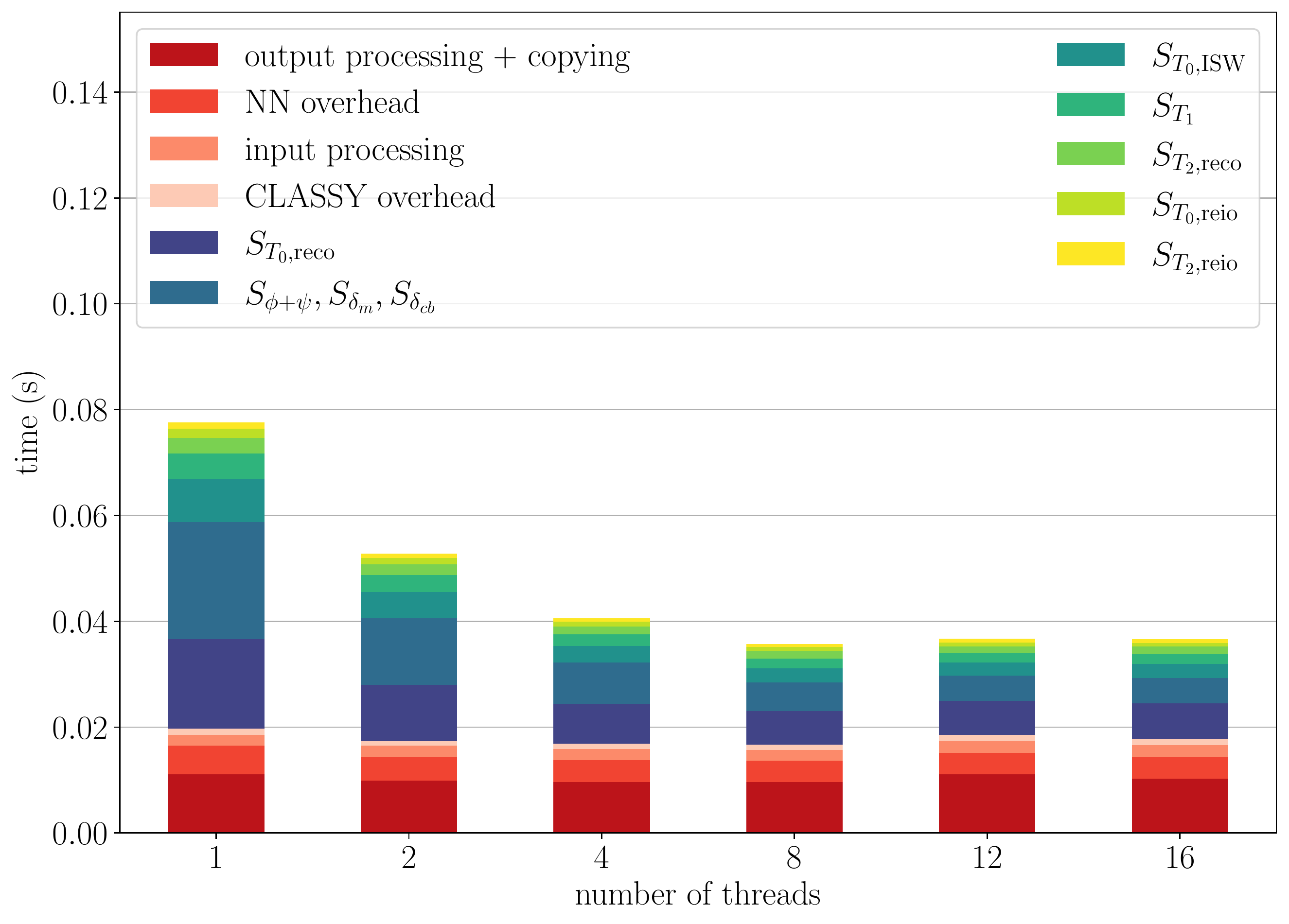}
    \caption{Time required for the individual network evaluations and pre- and post-processing. The blue/green/yellow bars correspond to the actual network evaluation time, while the reddish bars at the bottom correspond to additional overhead (which could possibly be further eliminated).}
    \label{fig:benchmark-time-per-step}
\end{figure}

\Cref{fig:benchmark-time-per-step} provides a more detailed overview of the individual contributions to the evaluation time of the \texttt{perturbations} module when emulated by \classnet. Blue, green, and yellow bars show the actual evaluation time of individual neural networks, while reddish bars account for additional overhead. We also show how the runtime depends on the number of CPU threads employed in the computation. In each of the two ``network evaluation'' and ``overhead'' categories, different contributions are ordered from the shortest (top) to the longest (bottom) when running on 8 threads.

For the overhead contributions, we differentiate between the additional runtime connected to the processing of the input and output of the neural networks, and overhead runtime caused by a variety of effects such as memory allocation (NN overhead) or the call of auxiliary functions (\textsc{classy} overhead).
Since none of these steps are parallelized, they contribute with an approximately fixed time to the total  runtime. For $N_\mathrm{threads}\geq8$, the overhead begins to dominate the runtime, leading to no significant improvement when increasing the number of threads.
We note that we do not display the additional runtime cost of loading the networks, because when \classnet is executed repeatedly within a parameter inference code such as \cobaya or {\sc MontePython}, the loading of the network only needs to be performed once per chain. Then, between different \class calls, the networks are kept in memory. Loading the networks initially only takes about $\sim30$ms.

We now review the definition and the performance of each of the steps detailed in \cref{fig:benchmark-time-per-step}, following the order in which they are executed by the code.

\pagebreak[20]
The python wrapper {\sc classy} is the primary point of contact between the \class C-code and the \texttt{PyTorch}-based networks of \classnet. Before the actual execution of the network begins, a few auxilary computations need to be performed,\footnote{ First, a function checks whether the current cosmology falls in the range of validity of \classnet using the hyperellipsoid method of \cref{eq:chi2_parameter_region}. Then, the python classes are instantiated and updated with the current cosmological parameters, and memory is allocated for each source function. Finally, the $k$-array relevant to the current cosmology is copied, and the networks are activated.} constituting the \enquote{CLASSY overhead} of \cref{fig:benchmark-time-per-step}. This part contributes to the overall runtime by a negligible amount.

Before running, the networks require some input from the \class  \texttt{background} and \texttt{thermodynamics} modules (such as conformal times of recombination/reionization, visibility function, etc.). These are obtained and copied during in the \enquote{input processing} step, which also contributes negligibly to the overall runtime.

At this stage, the networks can be evaluated.
The slowest network is [N2] ($T_0$ reco), since it has the most complex architecture -- as can be checked from \cref{fig:layout_ST0reco}. A similar execution time is taken up by [N7] ($\phi+\psi,\delta_m,\delta_{cb}$) for which the evaluation of the Hu-Eisenstein approximation takes the majority of the time.
As one would expect, the simplest networks [N3] ($T_0~\mathrm{reio}$), [N4] ($T_1$), [N5] ($T_2~\mathrm{reco}$), and  [N6] ($T_2~\mathrm{reio}$) -- which consist only of a few dense layers -- are the fastest and contribute only very little to the total runtime.
While the network [N1] ($T_0~\mathrm{ISW}$) has the same architecture, it also has a much larger number of weights, explaining its slightly longer evaluation time.

Next, the network output needs to be reshaped and extrapolated towards the low-$k$ boundary (see for example \cref{app:omegak_approx}). This step, mentioned in \cref{fig:benchmark-time-per-step} as \enquote{NN overhead}, has a reasonably small contribution to overall runtime.

Lastly, the predictions obtained from the networks must be transformed from 32bit float (used by \texttt{PyTorch}) to 64bit double precision (employed within \class) and then be copied back into the memory locations allocated within the \class C-code. This last step of \enquote{output processing+copying} takes a non-negligible amount of time, and could probably be optimized in the future. Since the overall runtime of the \texttt{perturbations} module in \classnet is already less than 100ms, we did not further pursue this direction.

\subsubsection{Complete CLASS speedup \label{sec:class_speedup}}

We can estimate the speedup of the whole \class code in the current version {\tt v3.2.0}, keeping in mind that possible improvements in the \texttt{transfer} module may lead to considerably more optimistic numbers in the future. Additionally we run the timing benchmark without considering non-linear corrections to the power matter spectrum such as \texttt{HALOFIT} or \texttt{HMCode}. Since these codes are implemented in a non parallelized way, they can create a further bottleneck when evaluating \classnet. We should also stress that the speedup depends a lot on which cosmology is used (e.g. because the \texttt{transfer} module called by both \classfull and \classnet becomes slower for growing curvature $|\Omega_k|$) and which observables are requested (e.g. because the \texttt{transfer} module also takes longer when the user asks for the calculation of number count $C_\ell$'s or cosmic shear $C_\ell$'s). To find the mean speedup factor we compute the execution time averaged over the uniformly drawn sample described in \cref{sec:speed_pert}.

As one can see in the right panel of \cref{fig:benchmark-abs-time}, the achievable \class speedup is limited by the fraction of the runtime spend in all but the \texttt{perturbations} module. When using \classfull, the \texttt{perturbations} module is dominating the average runtime with $70-80\%$ for all number of threads, while the remaining $20-30\%$ provide an upper bound on possible speedup. When using \classnet, the fraction of time spent in \texttt{perturbations} becomes negligible (1\%~for~1~thread up to 5\%~for~16~threads). Thus the {\tt perturbations} module is removed from the list of bottlenecks and the main target of this work is amply met. The resulting complete \class speedup is found to be $2.5-2.9$ for 1-8 threads and a $3.3-3.9$ for 12 or 16 threads.

Note that even a speedup by a simple factor of two or three represents a considerable amount of saved CPU$\times$hours on the scale of an entire project based on a large grid of Bayesian inference runs. 

Additionally, this speedup becomes considerably more significant when omitting the evaluation of the CMB spectra as we then skip the evaluation of the \texttt{transfer} module. When \class is set to calculate only the matter/baryon+CDM spectra up to $k_\mathrm{max}=100\,$Mpc$^{-1}$ (which is a common requirement when evaluating the likelihood of modern cosmic shear surveys), we find a total speedup of $\sim 200$ for 1 thread, $\sim 100$ for 2 threads, down to $35-50$ for 4-16 threads. On a standard computer, the total execution time is always below $0.1$ second, even on a single thread. 
It should be noted, however, that in our benchmark the execution of \classfull can take as much as $\mathcal{O}(10~\mathrm{s})$, due to the imposed high maximum wavenumber of $k_\mathrm{max} \sim 100/\mathrm{Mpc}$ and due to the computational cost of massive neutrinos and spatial curvature in the full 11-dimensional model. The speedup factor would be smaller if $k_\mathrm{max}$ was reduced. However, the small \classnet runtime remains impressive in any case.

When \classnet only computes the matter power spectrum, the remaining bottleneck that prevents acceleration by one additional order of magnitude is the \texttt{thermodynamics} module. However, some further optimisation of that module would not be too relevant at this stage. Indeed, the execution time of a typical likelihood codes for the next generation of surveys (such as e.g. Euclid) is currently of the order of a few seconds, that is, slower than \classnet by at least one order of magnitude.

\subsection{Source Functions}
\label{ssec:source_functions}

We can now proceed to investigate the precision of the prediction of \classnet compared to \classfull at the level of the source function. 

While in principle we could use the loss itself (i.e. the absolute or relative mean square difference between the $S_X$ averaged over $(k,\tau)$) to compare the predictions, here we will first simply present a quick and qualitative discussion of the differences (absolute or relative) of the source functions in the entire $(k,\tau)$ space, reserving a more quantitative description to \cref{ssec:power_spectra,sec:parameter_estimation}.
The comparisons presented in this section refer to the center of the sampled domain.\footnote{The center of the domain is located at the Planck+BAO+SNIa best-fit model of the run with all 11 parameters of \cref{eq:cosmo_inputs} with parameter values given in footnote~\ref{footnote:bestfit}.}
Note however that the training set samples from the ellipsoidal domain evenly, so we do not expect this particular choice to affect the qualitative features described below.

\begin{figure}[H]
	\centering
	\begin{subfigure}{0.99\linewidth}
		\includegraphics[width=\linewidth]{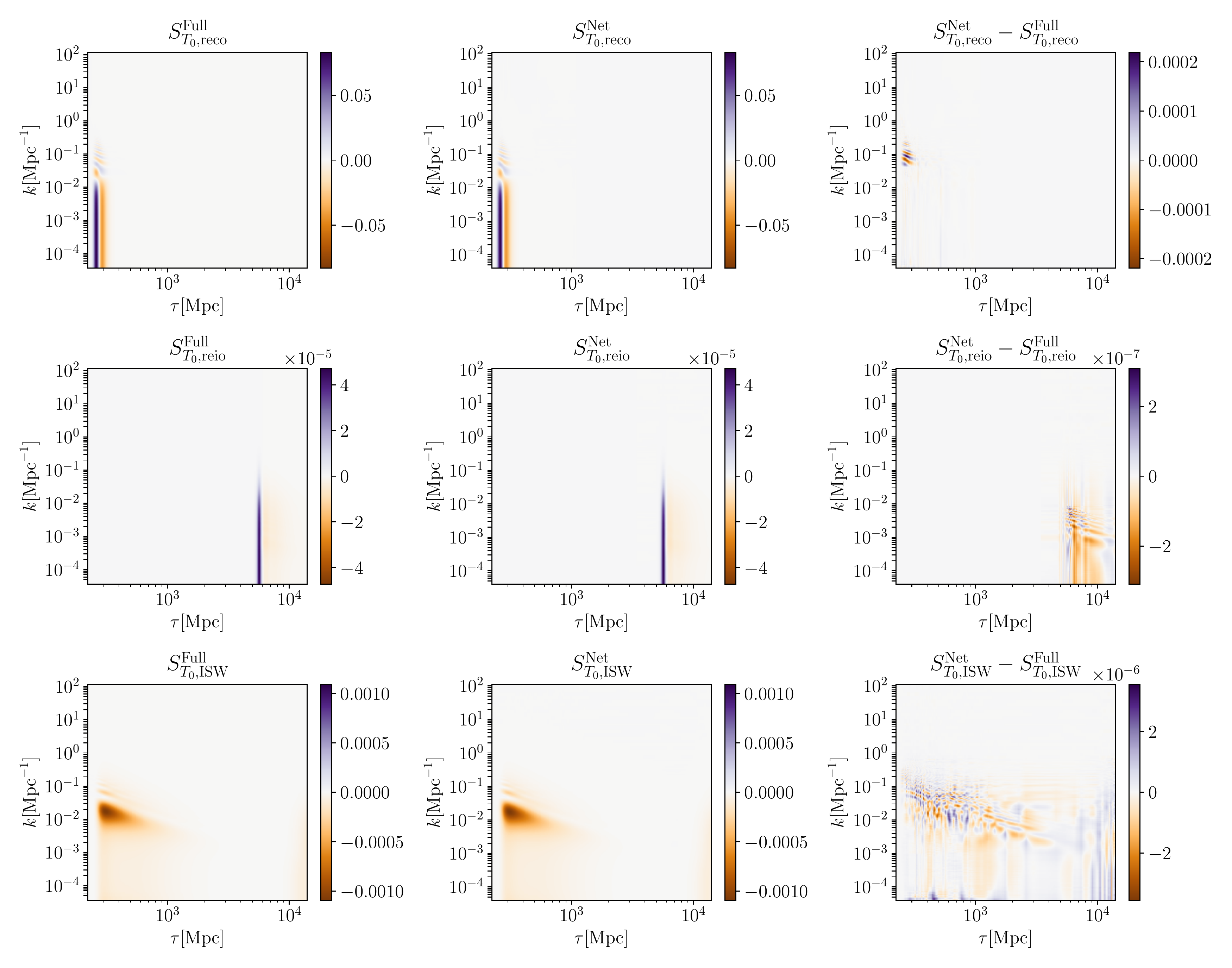}
	\end{subfigure}
	\caption{Contributions to the source functions $S_{T_0}$ from recombination (top), reionization (middle) and the ISW effect (bottom). \textbf{Left:} full calculation with \classfull, \textbf{Middle:} \classnet prediction, \textbf{Right:} absolute difference.}
	\label{fig:source_functions_t0}
\end{figure}

We display in \cref{fig:source_functions_t0} the three components of the first temperature source function $S_{T_0}$ of \class, which are defined in \cref{eq:t0reco_def,eq:t0reio_def,eq:t0isw_def} and represent the contributions from recombination, reionization and the (early and late) ISW effect. The left column contains the full numerical solution $S^\mathrm{Full}_X(k,\tau)$ from \classfull. The middle column shows the prediction $S^\mathrm{Net}_X(k,\tau)$ of \classnet. As expected, the prediction and numerical calculation are almost indistinguishable by eye. To visualize their differences, we show $S^\mathrm{Net}_X-S^\mathrm{Full}_X$ in the right column, using a different color scale (since the differences are typically two to three orders of magnitude smaller than the source functions).\footnote{Using the relative difference would be meaningless for these source functions, since they are zero or close to zero in many regions, especially those where their precise value becomes unimportant.}

All three contributions are oscillatory. As expected, the recombination contribution is non-zero only at early times, the reionization contribution at late times, and the ISW contribution at both early and late times. Interestingly, the $S_{T_0,\mathrm{reco}}$ contribution is predicted far better for wavenumbers $k<4\times10^{-2}\mathrm{Mpc}^{-1}$, while the biggest difference occurs around $k\approx 10^{-1}\mathrm{Mpc}^{-1}$. This is related to two facts:
\begin{enumerate}
	\item The sampling in $k$ that is employed within \class is sparser at $k>4\times10^{-2}\mathrm{Mpc}^{-1}$ than below, giving a lower weight to this region during the training (as the mean squared error for the network at a given time step $\tau$ is computed by summing over the given sampling points in $k$). The CMB source functions until $\ell \sim 3000$ are mostly sensitive up to $k$ until $4 \times 10^{-2} \mathrm{Mpc}^{-1}$, so higher accuracy at higher $k$ is not currently required.
	\item The underlying design of the [N2] network  contains generalized cosine and sine approximations to $S_{T_0,\mathrm{reco}}$, which cannot fully capture the shape at high $k$ where the oscillations become more sporadic. Thus, we speculate that exploring new designs for the [N2] network could dramatically improve the \classnet{} accuracy, but this is left for future work.
\end{enumerate}

The $S_{T_0,\mathrm{reio}}$ [N3] and $S_{T_0,\mathrm{ISW}}$ [N1] residuals consist of more diffuse and random noisy patterns. In any case, all three networks are able to reproduce the oscillations present in underlying source functions with sub-percent or even sub-permille accuracy. Comparing the sizes of the residuals in \cref{fig:source_functions_t0}, we observe that they are in any case subdominant compared to those of $S_{T_0,\mathrm{reco}}$\,. We will see in \cref{ssec:power_spectra} and \ref{sec:parameter_estimation} that the accuracy achieved for all three networks related to $S_{T_0}$ is sufficient to predict the $C_\ell^\mathrm{TT}$ spectrum with high precision and to estimate cosmological parameter values using current cosmological data sets.

\begin{figure}[t]
	\centering
	\begin{subfigure}{0.99\linewidth}
		\includegraphics[width=\linewidth]{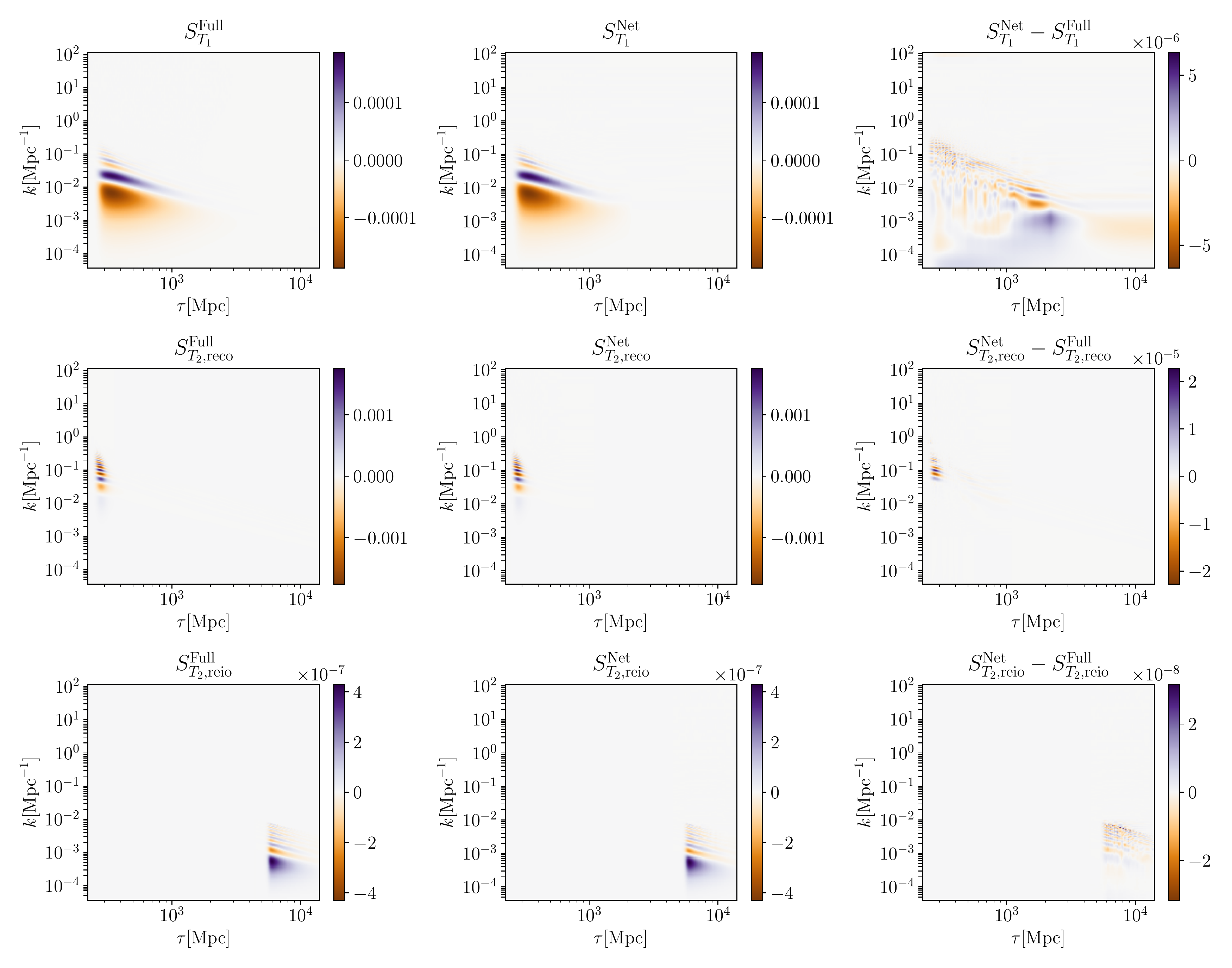}
	\end{subfigure}
	\caption{Source function $S_{T_1}$ and contributions to $S_{T_2}$ from recombination (middle) and reionization (bottom). \textbf{Left:} numerical calculation with \classfull, \textbf{Middle:} \classnet prediction, \textbf{Right:} Absolute difference.}
	\label{fig:source_functions_t1t2}
\end{figure}

\enlargethispage*{1\baselineskip}
\Cref{fig:source_functions_t1t2} shows the source function $S_{T_1}$ and the two contributions to $S_{T_2}$. The small [N4] network is able to predict the oscillations present in $S_{T_1}$ with sub-percent accuracy. The [N5] network achieves a similar precision on $S_{T_2}^\mathrm{reco\_no\_isw}$, at the expense of a deeper network architecture justified by the highly oscillatory shape of this source function. The small [N6] network predicts $S_{T_2}^\mathrm{reio\_no\_isw}$ with a precision of a few percents only, but this is sufficient for our purpose. Indeed the residuals in $S_{T_2}^\mathrm{reio\_no\_isw}$ have a totally negligible impact on $C_\ell^\mathrm{TT}$ and mainly affect the reionization peak in $C_\ell^\mathrm{EE}$ (see for example \cref{fig:power_spectra_ee_lines}). Increasing the accuracy of this network would allow for a better emulation of this peak, but due to the high cosmic variance at low $\ell$, this is unimportant for parameter estimation from current data, as will be shown in section~\ref{sec:parameter_estimation}.

\begin{figure}[H]
	\centering
	\begin{subfigure}{0.99\linewidth}
		\includegraphics[width=\linewidth]{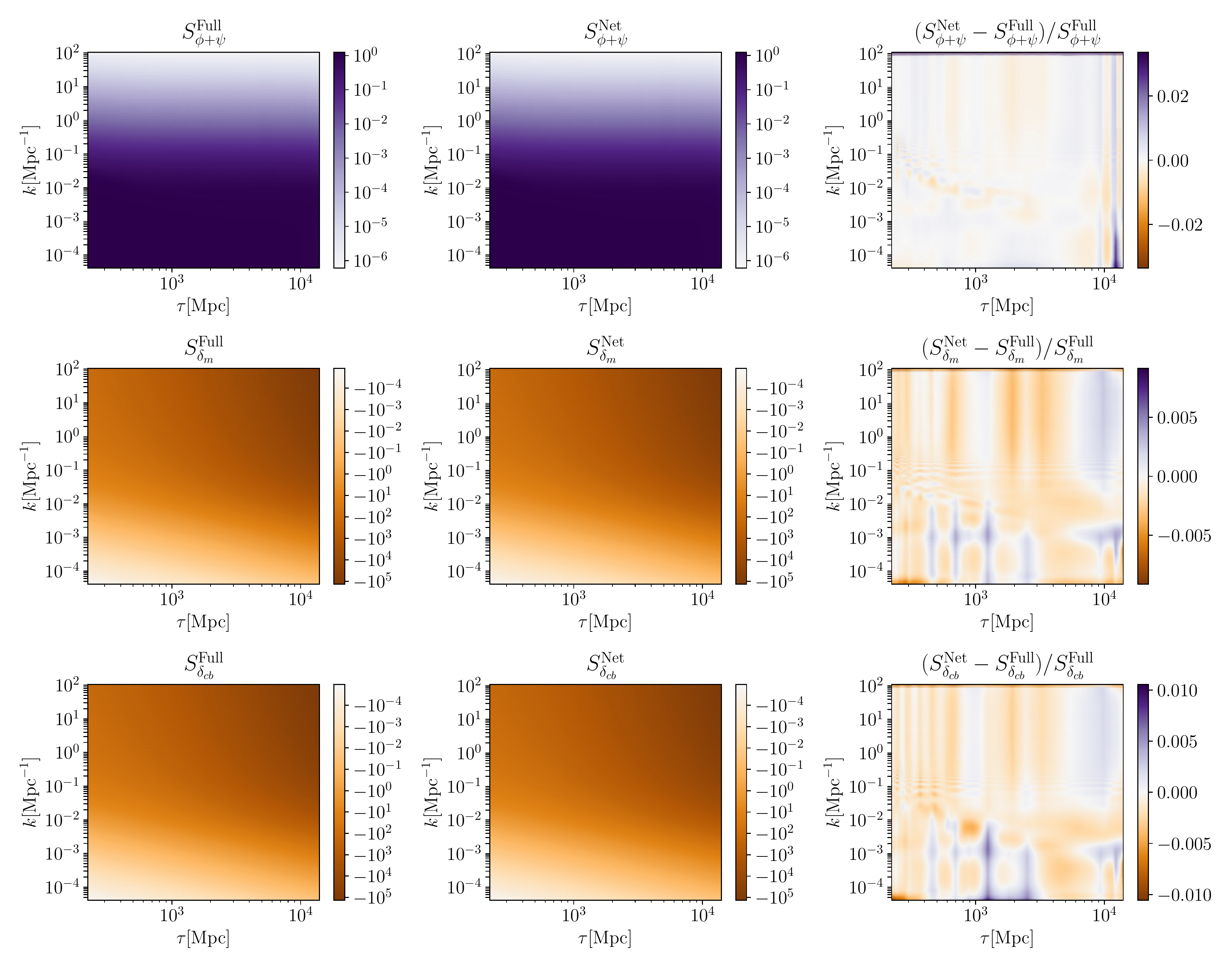}
	\end{subfigure}
	\caption{Source functions $S_{\phi+\psi}$, $S_{\delta_\mathrm{m}}$ and $S_{\delta_\mathrm{cb}}$. The left and middle plots have a logarithmic scale to include the many orders of magnitude over which these three source functions vary. $S_{\delta_m}$ and $S_{\delta_{cb}}$ are negative. \textbf{Left:} numerical calculation with \classfull, \textbf{Middle:} \classnet prediction, \textbf{Right:} Absolute difference.}
	\label{fig:source_functions_pp}
\end{figure}

In \cref{fig:source_functions_pp} we display the source functions $S_{\phi+\psi}$, $S_{\delta_\mathrm{m}}$, and $S_{\delta_\mathrm{cb}}$ predicted by the network [N7]. Since these source functions have a large dynamical range, the network [N7] is trained with a relative mean squared loss and we display the function itself using a logarithmic color scale.

In the residual of $S_{\phi+\psi}$, a 2\%-level feature at $k<10^{-3}$ and $\tau>10^4$~Mpc shows that the [N7] network fails to make an accurate prediction of the lensing potential $\phi+\psi$ for very large scales and recent times. However, only smaller scales (larger wavenumber) contribute significantly to the CMB lensing spectrum (and potentially to the cosmic shear $C_\ell$'s), such that this feature is almost irrelevant. The very careful reader may also observe a feature close to $k_\mathrm{max}$ for these three source functions, due to interpolation issues at the boundary of our grid, but this region is also irrelevant for predictions on observable scales. Apart from these two innocuous features, we find that the [N7] network predictions are accurate at least at the per-mille level.

We will see in the next section that the precision achieved by our networks is always sufficient for getting robust and accurate observable spectra, at least at the precision level required by the analysis of current data.

\subsection{Power Spectra}
\label{ssec:power_spectra}

We now present various tests of the accuracy achieved by \classnet at the level of observable spectra -- CMB harmonic power spectra $C_\ell^{XY}$ with $X,Y~\in$~\{T, E, B, $\phi$\} and matter Fourier power spectra $P_\mathrm{m}(k,z)$, $P_\mathrm{cb}(k,z)$. For the CMB spectra, we directly consider observable quantities such as the lensed CMB spectra, corrected for lensing effects by the {\tt lensing} module of \class. These spectra depend on the \classnet predictions for the temperature, polarization and lensing potential source functions.

All the tests presented in this section are based on the validation set presented in section~\ref{ssec:domain_cn}, that is, about $10^3$ cosmological models that provide a uniform sampling of a 9-dimensional ellipsoid corresponding to 5$\sigma$ deviations from the Planck+BAO+SNIa best-fit model. We stick to the 9-dimensional parameter basis of equation~(\ref{eq:cosmo_inputs}) and to \class default precision settings.

First, for each model, we compare the difference between the prediction of \classfull and \classnet for each multipole $C_\ell^{XY}$ with $X,Y~\in$~\{T, E, B, $\phi$ \} to the amplitude of cosmic variance plus Planck instrumental noise,
\begin{align}
\Delta_\ell = \frac{|C_\ell^\mathrm{Net}-C_\ell^\mathrm{Full}|}{\sigma_{C_\ell}} \qquad \qquad \text{where} \qquad  \sigma_{C_\ell} = \sqrt{\frac{2}{2 \ell +1}}\left(C_\ell^\mathrm{CV}+N_\ell\right)\label{eq:def_Dell}
\end{align}
where $C^\mathrm{CV}_\ell$ represents the contribution from cosmic variance\footnote{We recall that the contribution from cosmic variance for the $C^{XY}_\ell$ spectrum is given by $C_l^{XX}$ for $X=Y$ and $\sqrt{\frac{1}{2}(C^{XY}_\ell)^2+ \frac{1}{2}C_\ell^{XX}C_\ell^{YY}}$ in the general case.}
and $N_\ell$ reflects approximately the Planck noise level.\footnote{We use an estimate of the Planck 2015 noise level that corresponds to the noise file of the {\tt planck\_fake\_realistic} likelihood in the {\tt MontePython} {\tt v$\geq$3.0} package \cite{Brinckmann:2018cvx}.}
Note that $\sigma_{C_\ell}$ is dominated by cosmic variance until $\ell\sim 2200$ for the TT spectrum. Cosmic variance has a significant contribution up to $\ell \sim 1000$ for the EE spectrum and $\ell\sim 300$ for the lensing potential spectrum, while above the experimental error dominates.\footnote{We define this threshold as where the noise becomes ten times as large as the cosmic variance.} The noise is assumed to vanish for the TE spectrum.

The ratio $\Delta_\ell$ states how measurable the deviation of observables between \classnet and \classfull is for a given $\ell$. For instance, $\Delta_\ell\sim0.1$ means that the error induced by the neural networks is one order of magnitude below theoretical and instrumental errors for this $\ell$. The next section will confirm that the low values of $\Delta_\ell$ presented in this section allow for unbiased parameter estimation from current data.

\begin{figure}[t]
	\centering
	\begin{subfigure}{0.55\linewidth}
		\includegraphics[width=\linewidth]{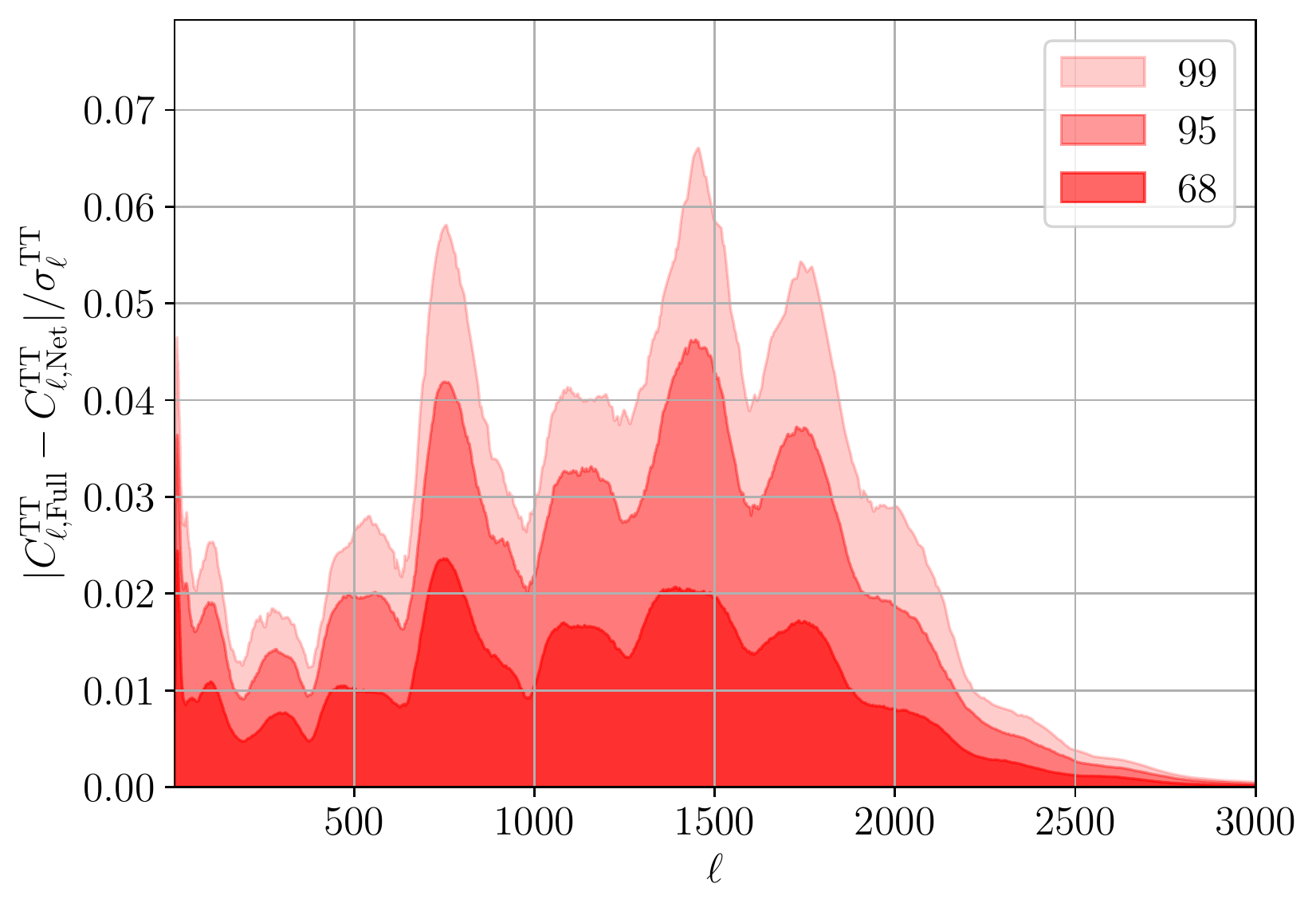}
	\end{subfigure}
	\vspace*{-1em}
	\caption{\label{fig:power_spectra_tt}
		TT angular power spectra residuals divided by the sum of cosmic variance and Planck instrumental noise, as defined in \cref{eq:def_Dell}. Regions of increasing intensity represent the 68\%, 95\% and 99\% best predicted cases within the $\sim 10^3$ cosmological models of our validation set.}

	\begin{subfigure}{0.85\textwidth}
		\includegraphics[width=0.5\linewidth]{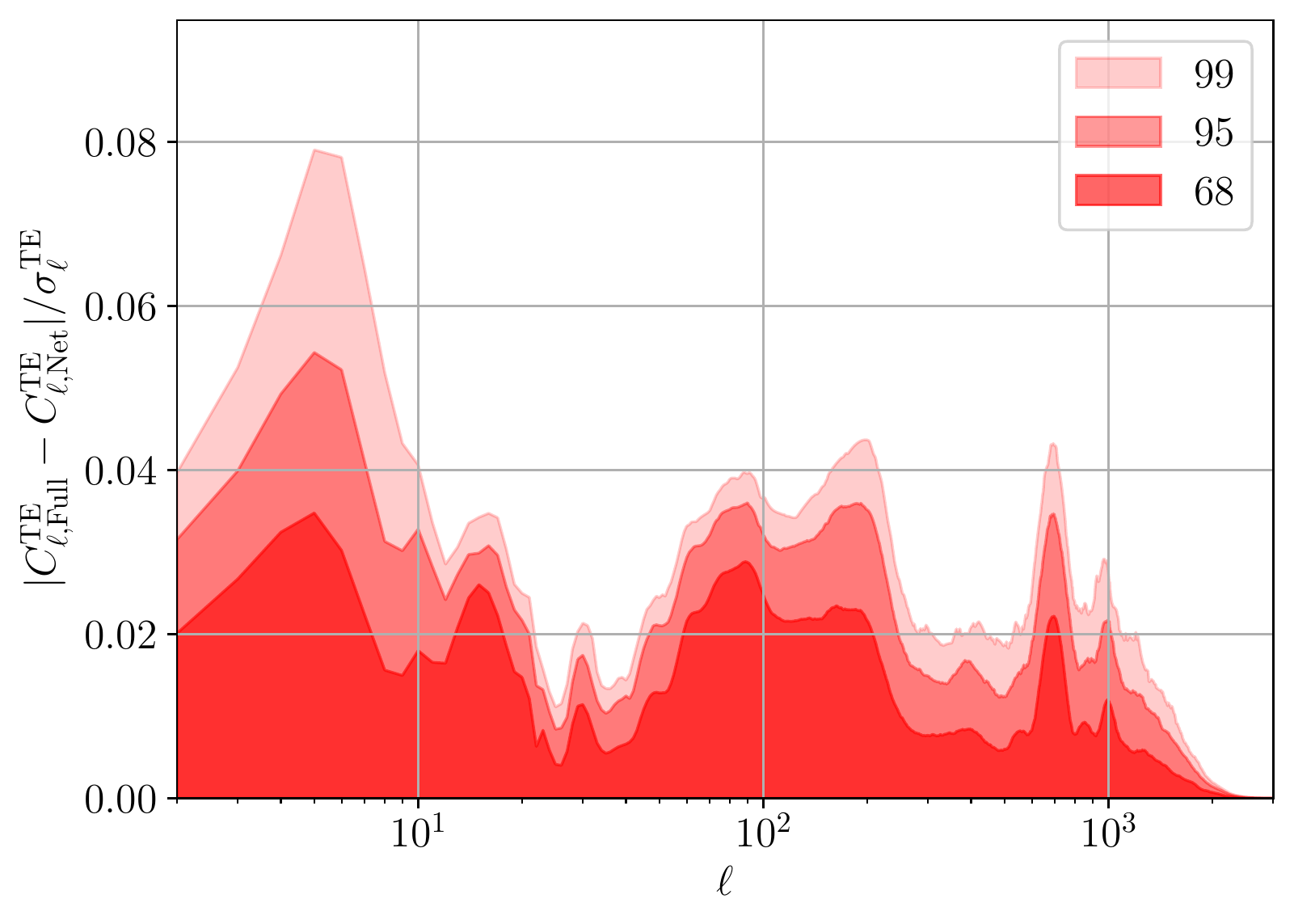}
		\includegraphics[width=0.5\linewidth]{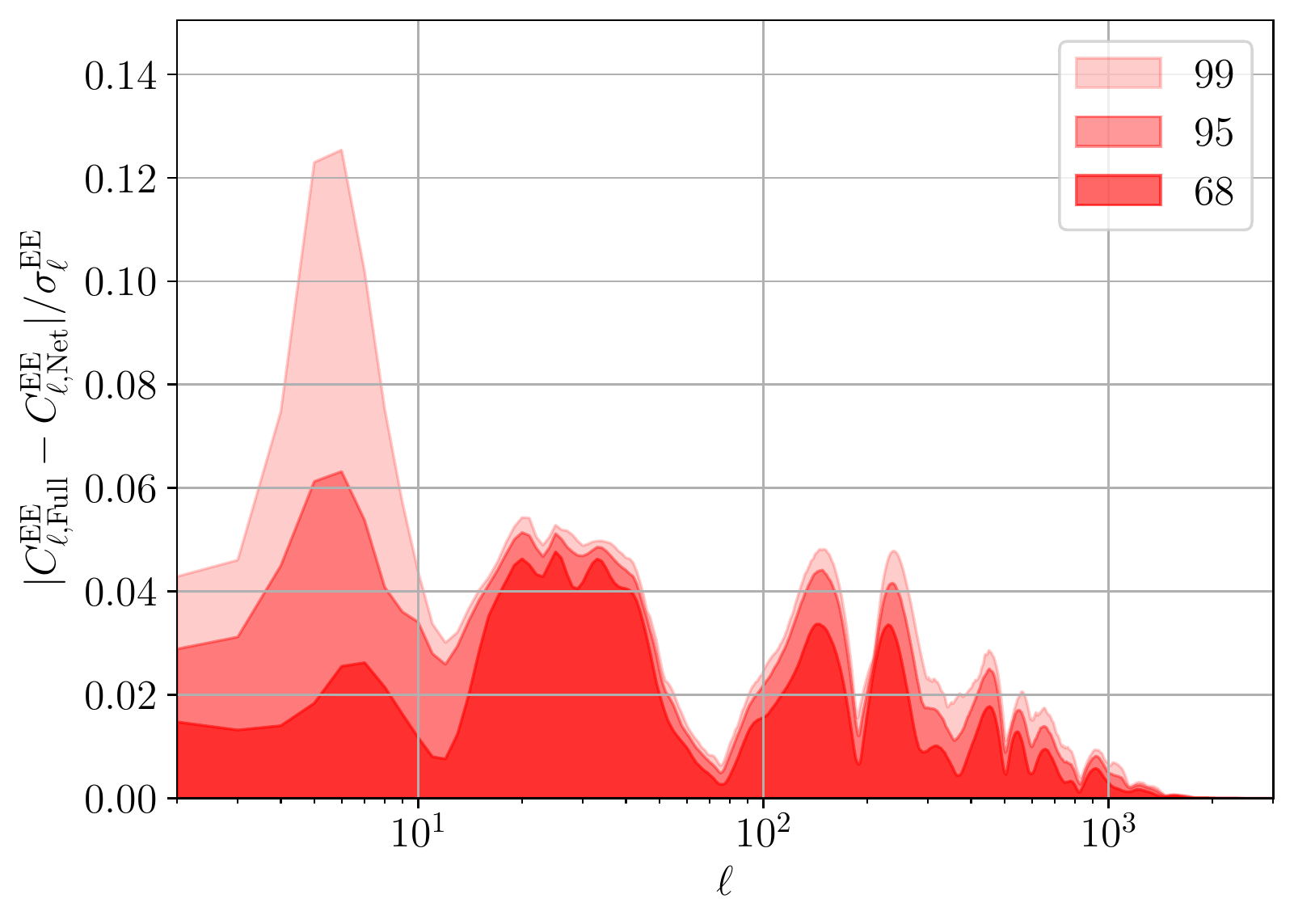}
	\end{subfigure}
	\vspace*{-1em}
	\caption{\label{fig:power_spectra_teee}Same as \cref{fig:power_spectra_tt} but for TE (left) and EE (right). Shown logarithmically in $\ell$.}
	\centering
	\begin{subfigure}{0.85\textwidth}
		\includegraphics[width=0.5\linewidth]{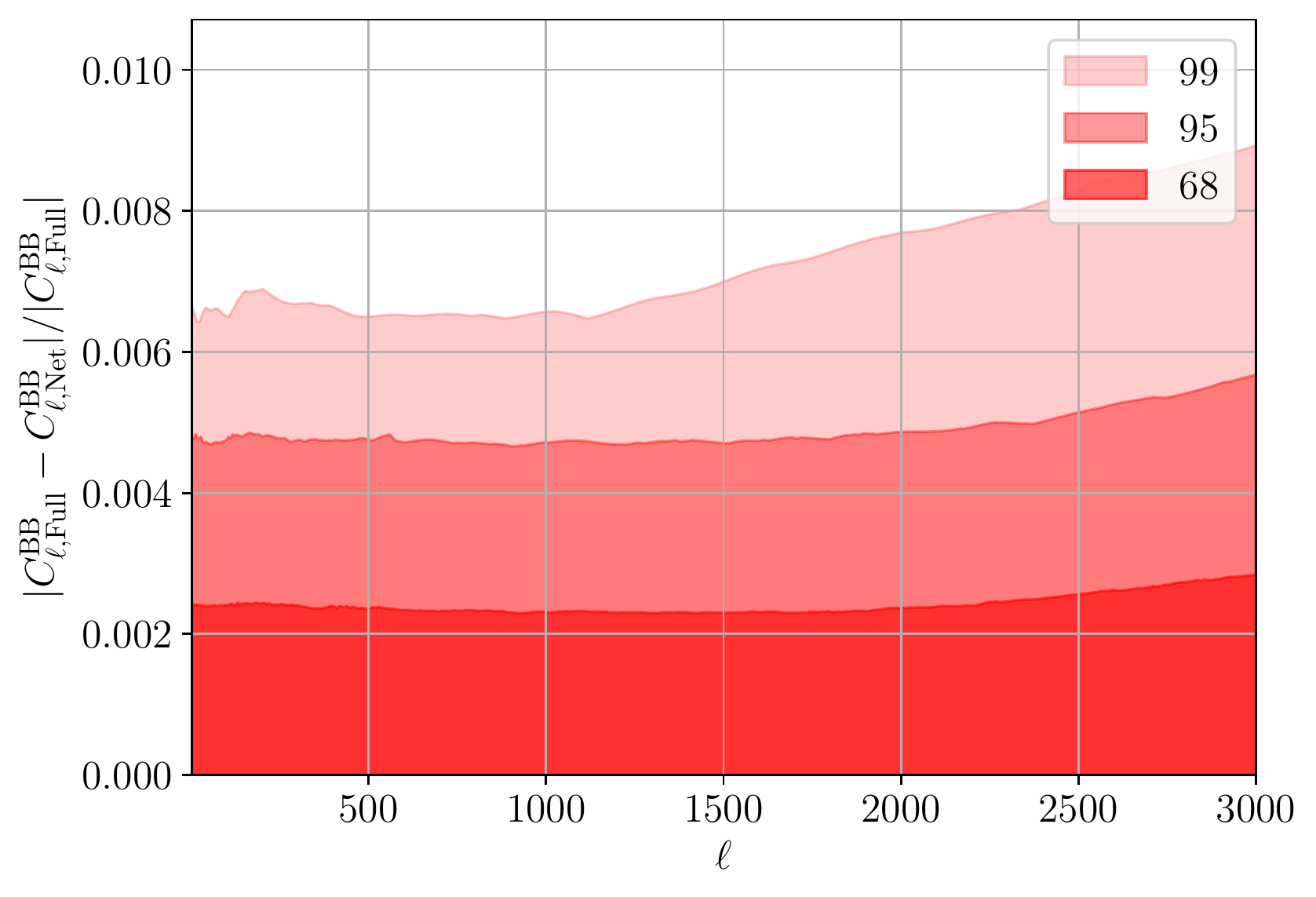}
		\includegraphics[width=0.5\linewidth]{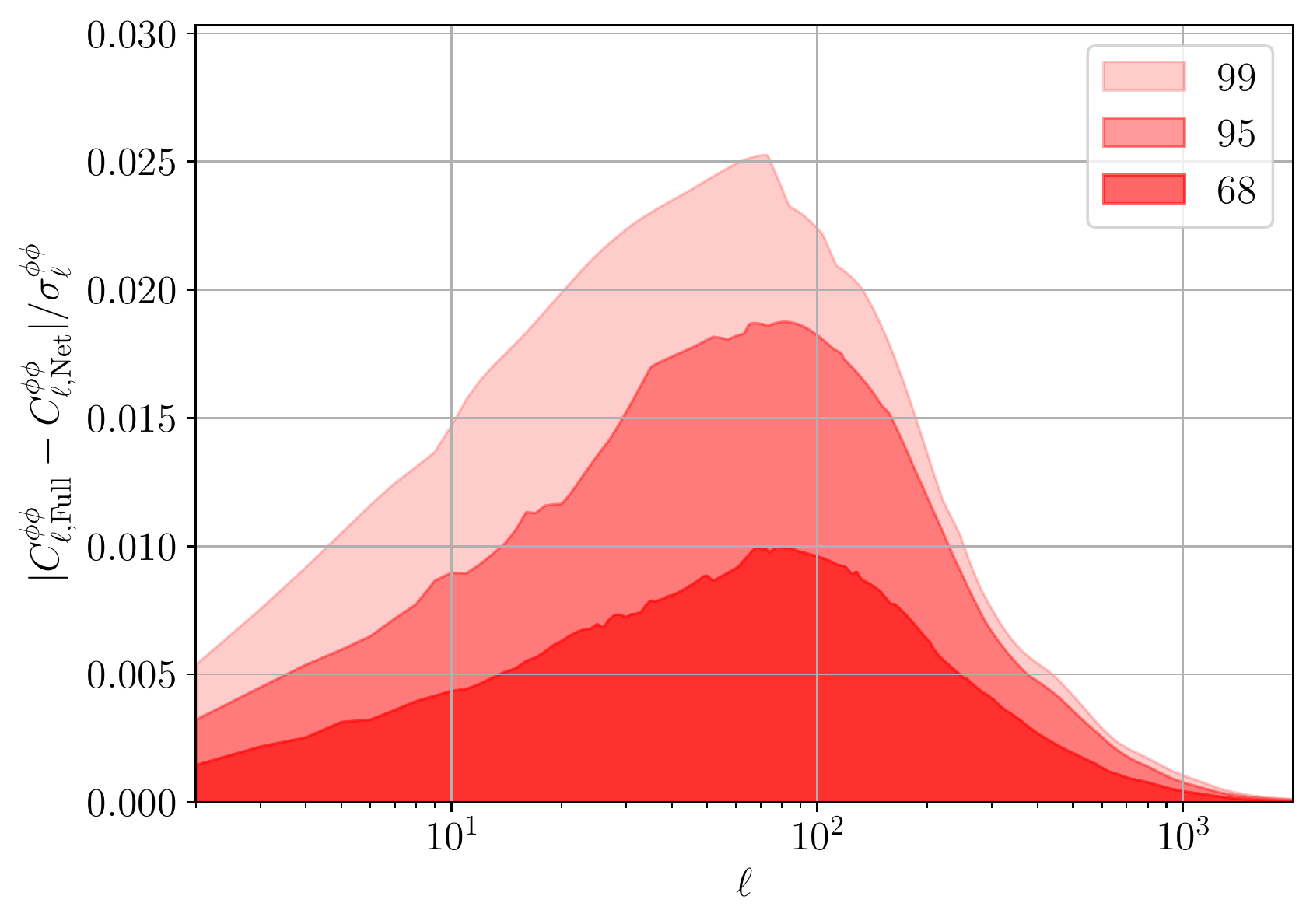}
	\end{subfigure}
	\vspace*{-1em}
	\caption{\label{fig:power_spectra_bbpp} \textbf{Left:} Same as \cref{fig:power_spectra_tt} but for the lensing potential auto-correlation $\phi \phi$. Shown logarithmically in $\ell$. \textbf{Right:} Relative errors on the BB spectra -- here we do not use the Planck noise spectrum, since Planck is noise-dominated for BB.}
\end{figure}

In \cref{fig:power_spectra_tt,fig:power_spectra_teee,fig:power_spectra_bbpp} we show the maximum deviation $\Delta_\ell$ on the TT, EE, TE, BB, and $\phi\phi$ spectrum within the best-fitting 68\%, 95\% and 99\% of models.
\Cref{fig:power_spectra_tt} shows that 95\% of TT spectra have $\Delta_\ell \leq 0.04$ at all multipoles -- only 1\% of the model actually reach $\Delta_\ell = 0.1$ for some multipoles. The most extreme deviation within our validation set is $\Delta_\ell = 0.13$ for a single $\ell$, but the average deviations across all models and $\ell$ values is found to be below $\Delta_\ell = 0.02$, showing that \classnet predicts really accurate temperature spectra. Regarding the EE spectrum, the highest deviation has a  similar magnitude and occurs at relatively low $\ell\sim 5$ (reionization peak).\footnote{As mentioned in \cref{ssec:source_functions}, this could be reduced by simply expanding the network [N5].} However, significant deviations are relatively rare, since we obtain $\Delta_\ell < 0.1$ for 94\% of cases, and for $\ell\geq10$ even the highest deviations are always below $0.06$. The average deviation is as low for the EE spectrum as for the TT spectrum. Next, we find similar results for the TE spectrum, which builds up from the same network predictions as the TT and EE spectrum. 

For the BB spectrum, we simply display relative errors on $C_\ell^\mathrm{BB}$ rather than $\Delta_\ell$s, due to the fact that the Planck noise is larger than the signal in the BB case. We see that the spectra are always predicted up to better than one per cent, which is sufficient even in the context of future CMB polarisation observations. Finally, for the lensing potential spectrum $C_\ell^{\phi \phi}$, we also achieve $\Delta_\ell \leq 0.01$ for almost all $\ell$, showing that \classnet can also safely be used for fitting Planck lensing data.

\begin{figure}[t]
	\centering
	\begin{subfigure}{0.45\linewidth}
		\includegraphics[width=\linewidth]{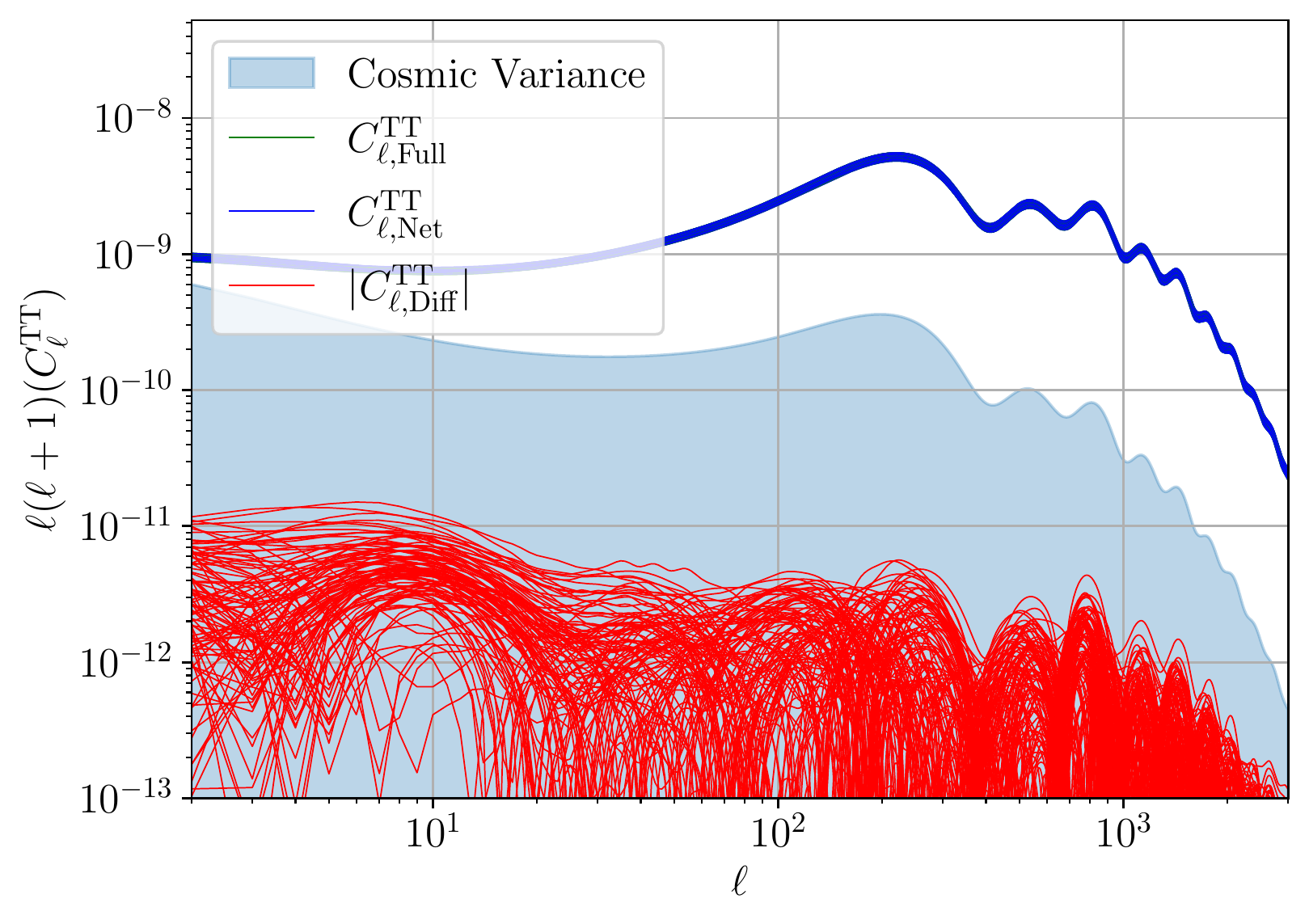}
		\caption{TT power spectrum}
	\end{subfigure}
	\begin{subfigure}{0.45\linewidth}
		\includegraphics[width=\linewidth]{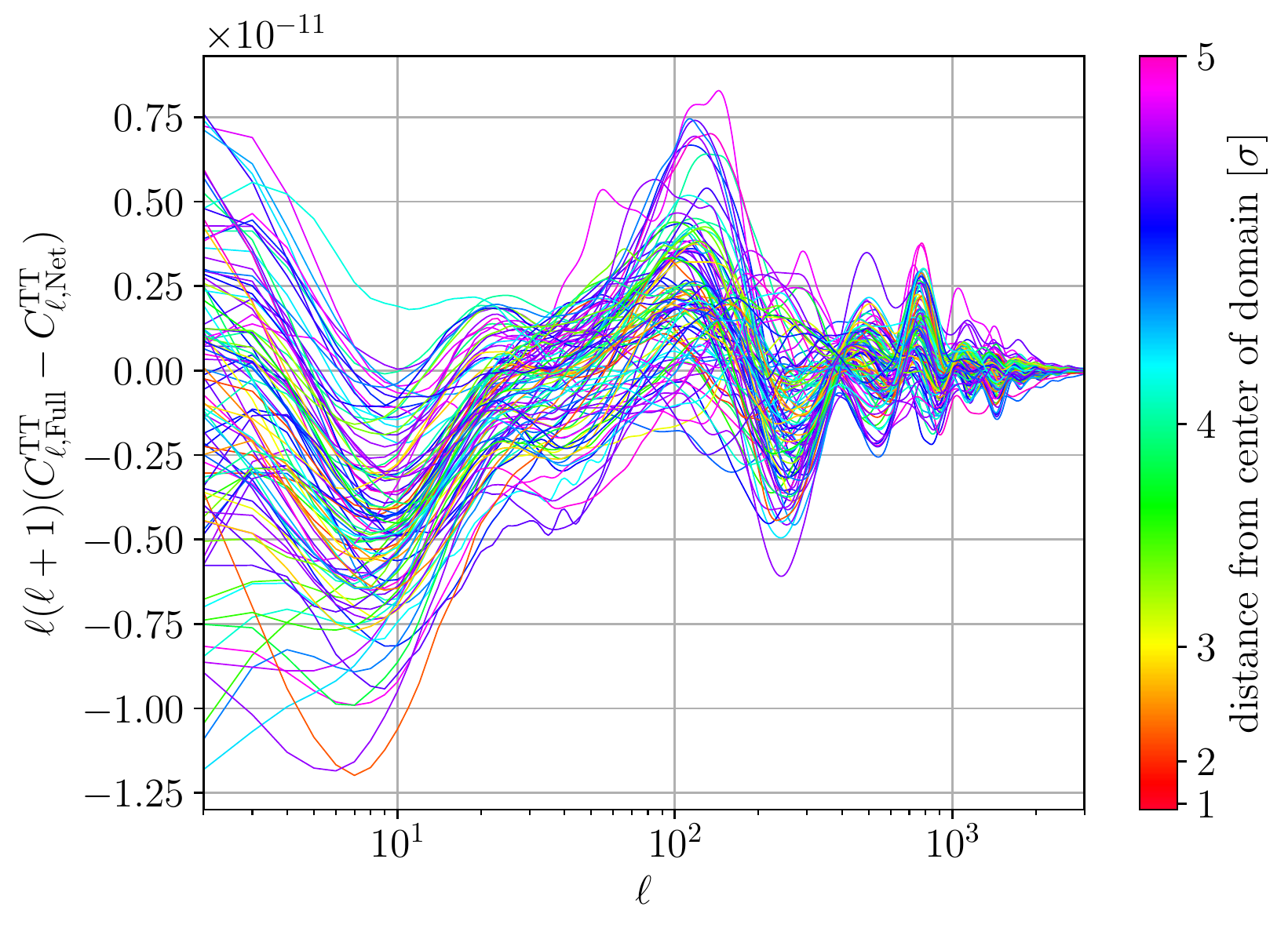}
		\caption{Difference in TT power spectrum}
	\end{subfigure}
	\caption{\label{fig:power_spectra_tt_lines}Temperature auto-correlation power spectra for the ${\cal O}(10^3)$ models of our validation set. Left: Power spectra predicted by \classfull (green) or \classnet (blue, almost masking the latter), and difference between them (red). In light blue we also show cosmic variance. Right: Zoom on the absolute differences in linear scale, colored by the distance of the cosmological model to the center of the ellipsoidal training region.}
\end{figure}

\begin{figure}[t]
\centering
\begin{subfigure}{0.45\linewidth}
	\includegraphics[width=\linewidth]{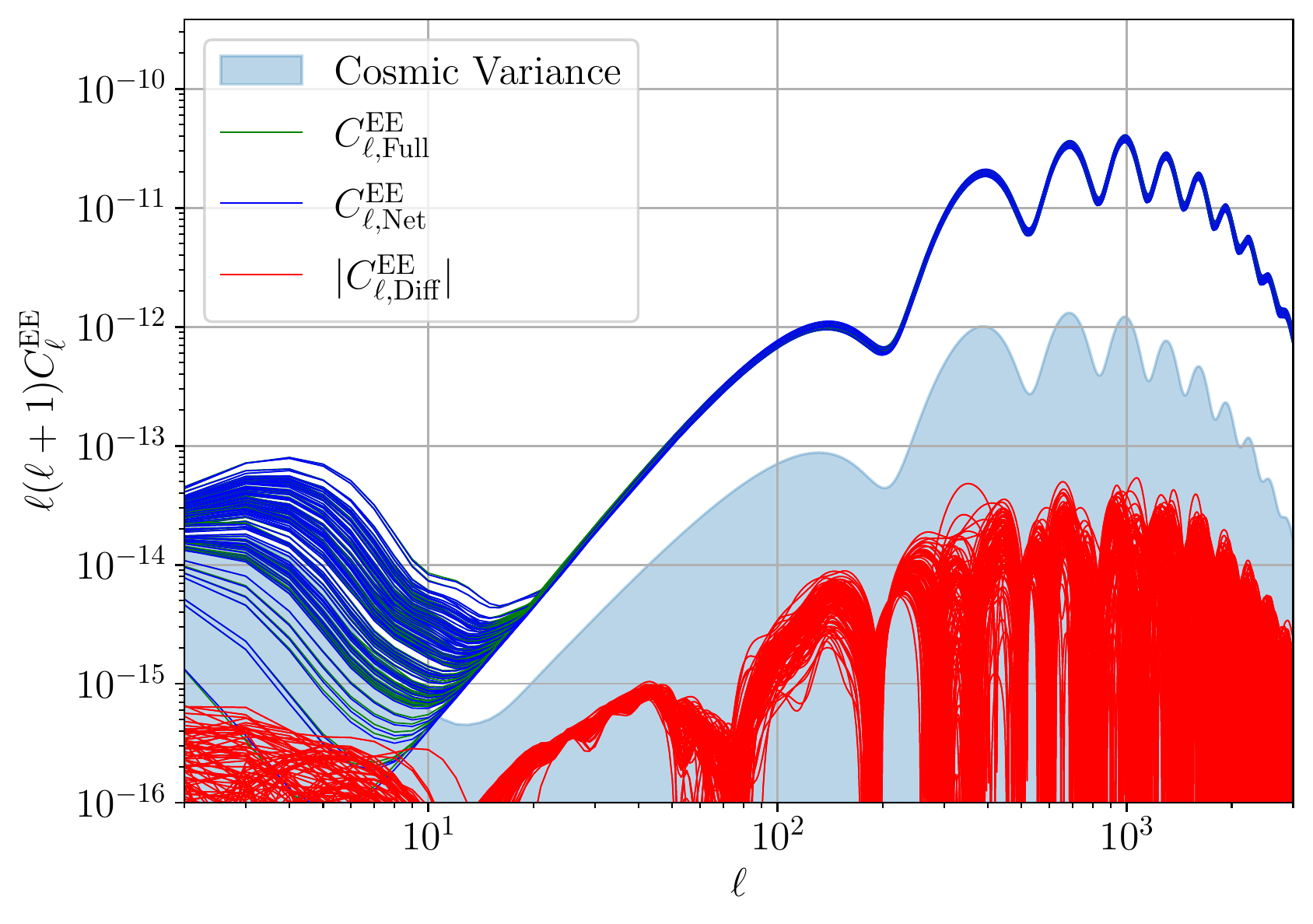}
	\caption{EE power spectrum}
\end{subfigure}
\begin{subfigure}{0.45\linewidth}
	\includegraphics[width=\linewidth]{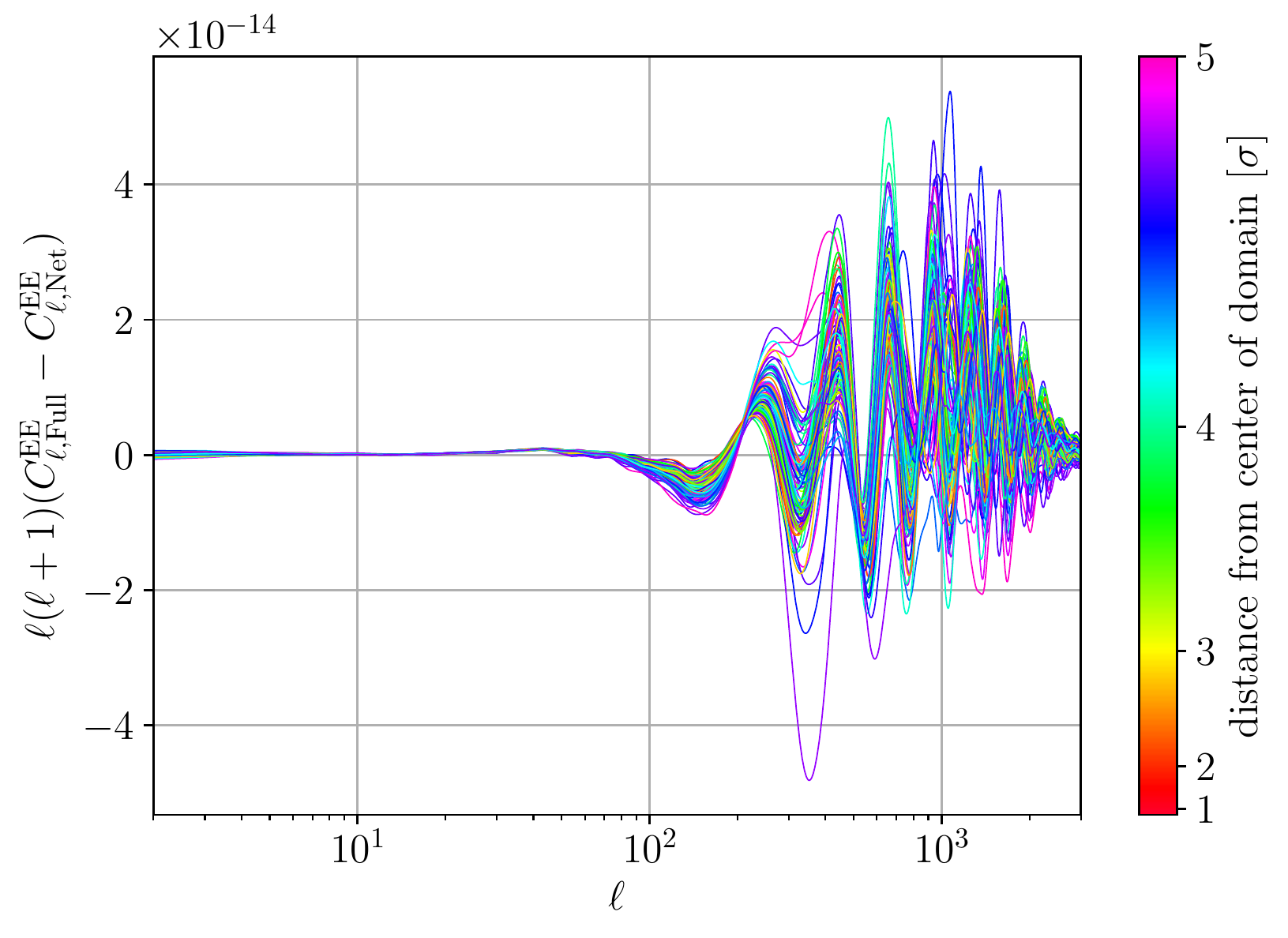}
	\caption{Difference in EE power spectrum}
\end{subfigure}
\caption{\label{fig:power_spectra_ee_lines} Same as \cref{fig:power_spectra_tt_lines} but for EE power spectra.}
\end{figure}

\begin{figure}[t]
\centering
\begin{subfigure}{0.45\linewidth}
	\includegraphics[width=\linewidth]{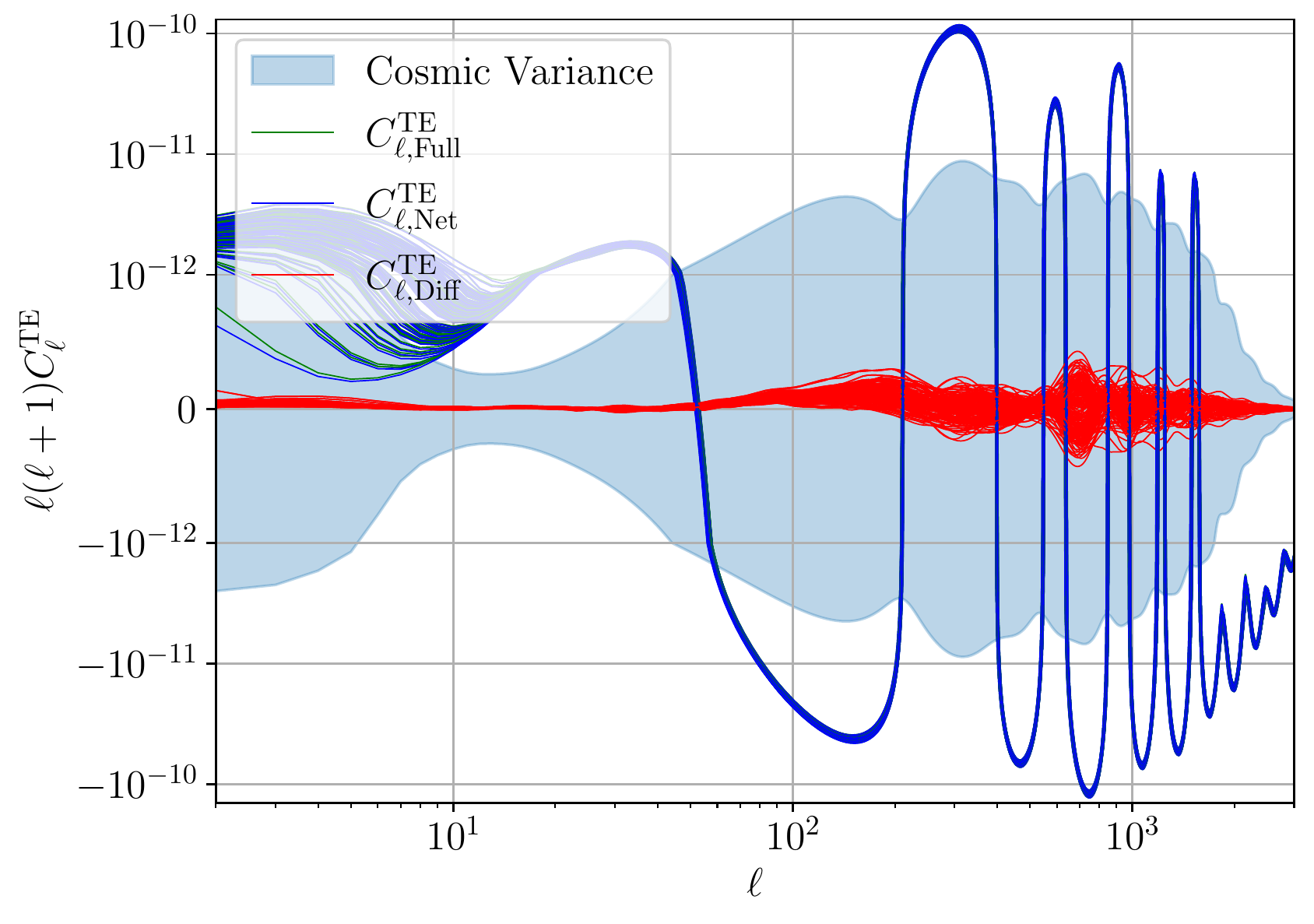}
	\caption{TE power spectrum}
\end{subfigure}
\begin{subfigure}{0.45\linewidth}
	\includegraphics[width=\linewidth]{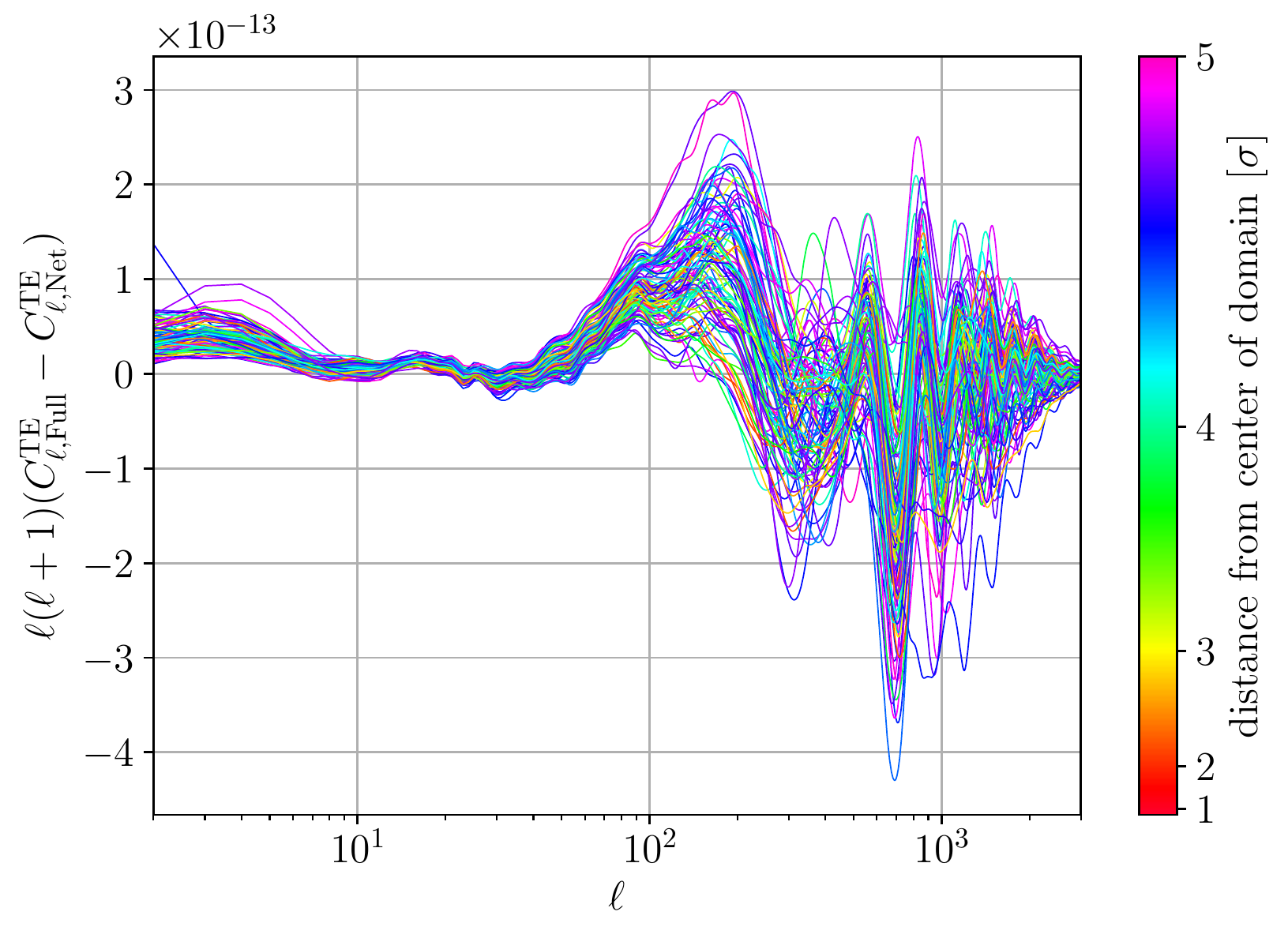}
	\caption{Difference in TE power spectrum}
\end{subfigure}
\caption{\label{fig:power_spectra_te_lines} Same as \cref{fig:power_spectra_tt_lines} but for TE power spectra. Note on the left that the scale is changing from a logarithmic to a linear representation.}
\end{figure}

\begin{figure}[t]
\centering
\begin{subfigure}{0.45\linewidth}
	\includegraphics[width=\linewidth]{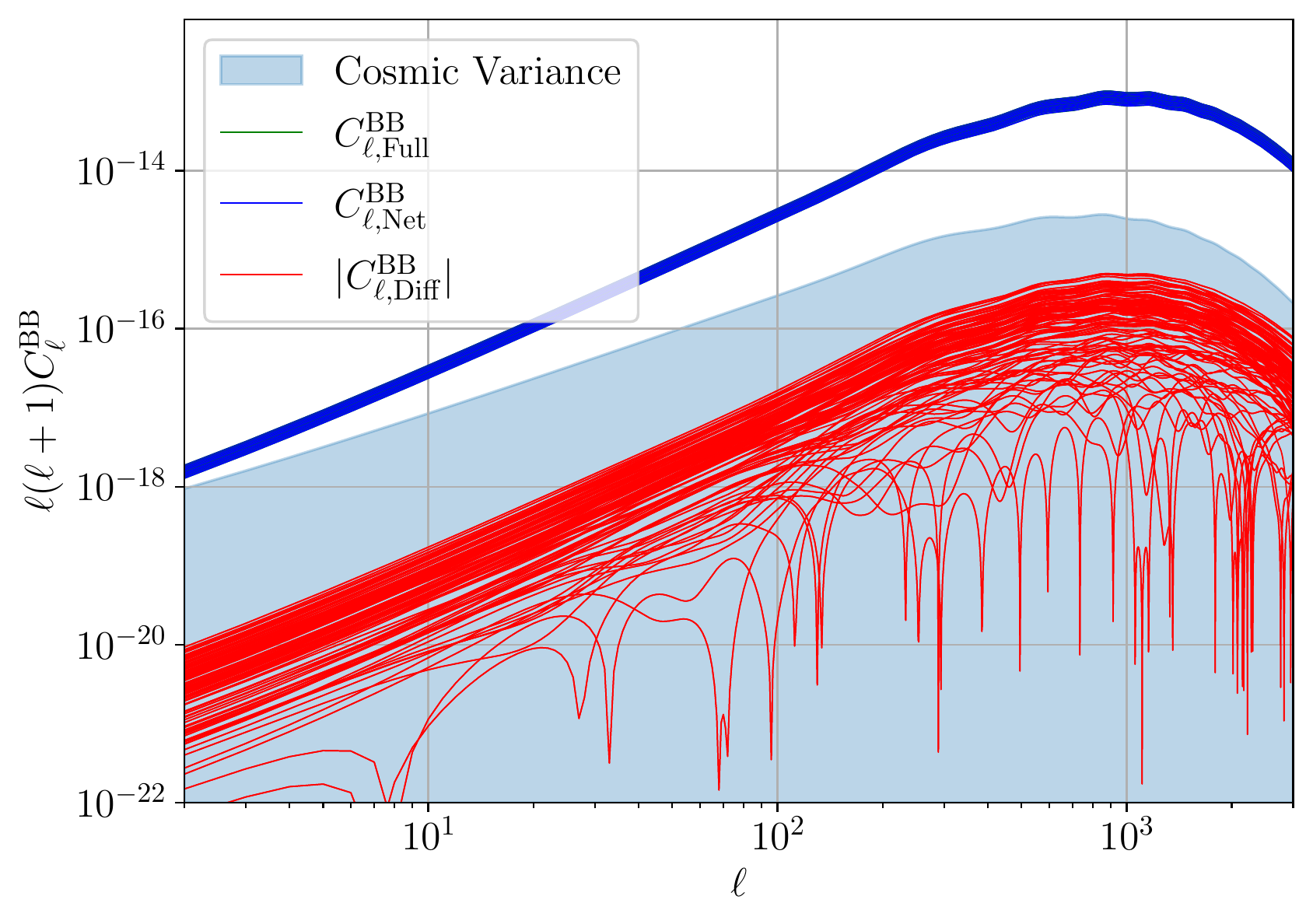}
	\caption{BB power spectrum}
\end{subfigure}
\begin{subfigure}{0.45\linewidth}
	\includegraphics[width=\linewidth]{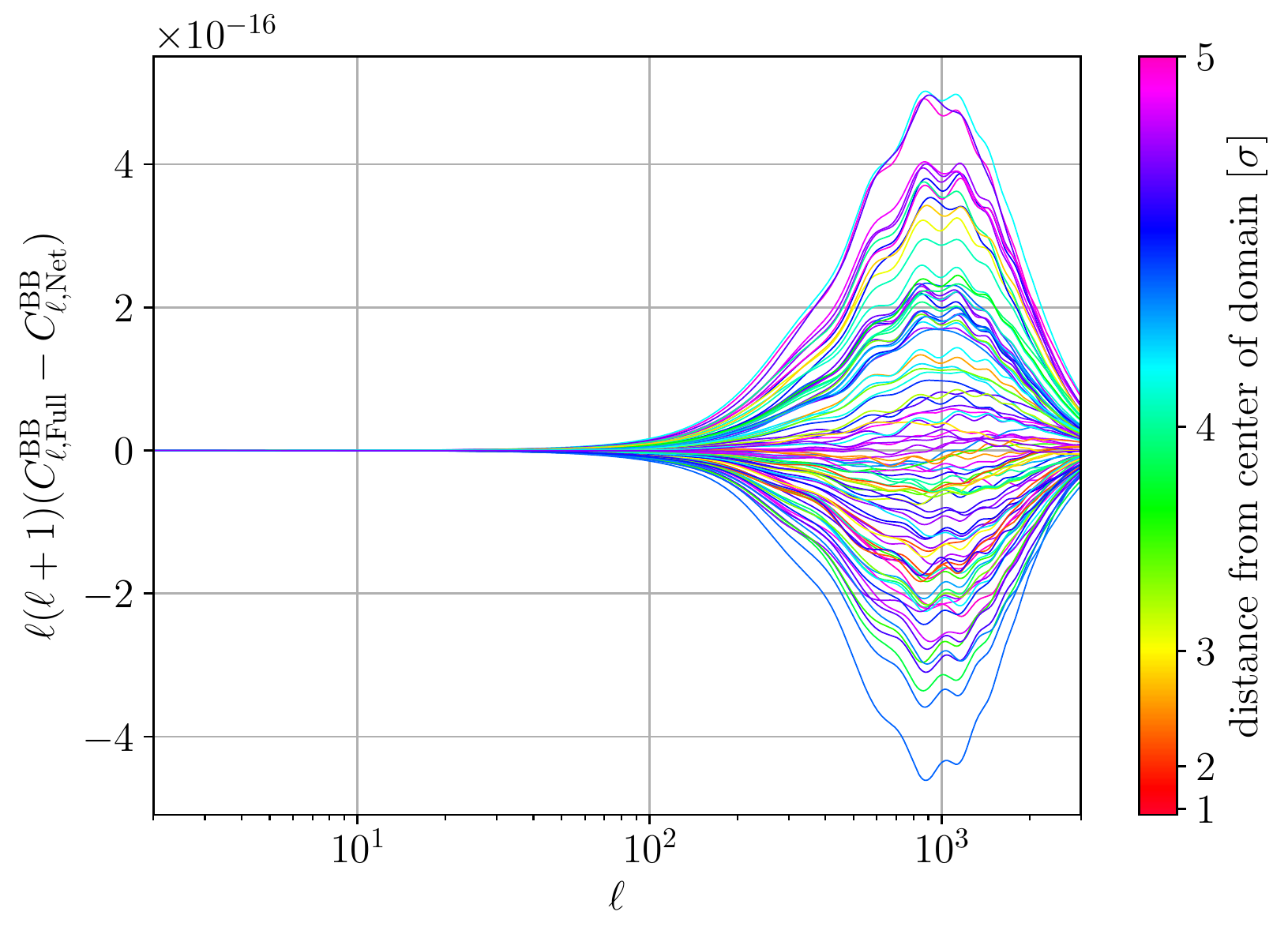}
	\caption{Difference in BB power spectrum}
\end{subfigure}
\caption{\label{fig:power_spectra_bb_lines} Same as \cref{fig:power_spectra_tt_lines} but for BB power spectra.}
\end{figure}

\begin{figure}[t]
\centering
\begin{subfigure}{0.45\linewidth}
	\includegraphics[width=\linewidth]{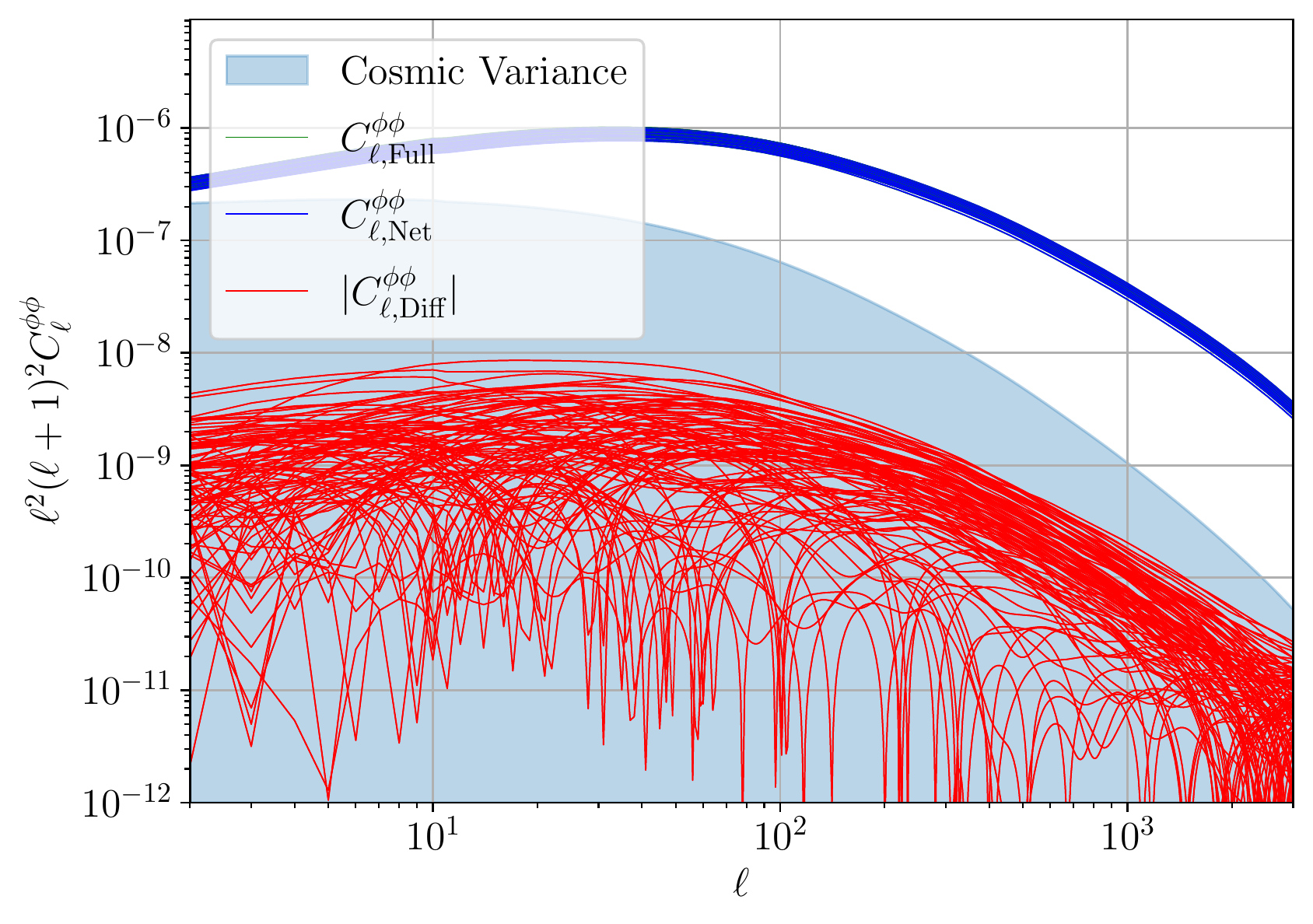}
	\caption{$\phi \phi$ power spectrum}
\end{subfigure}
\begin{subfigure}{0.45\linewidth}
	\includegraphics[width=\linewidth]{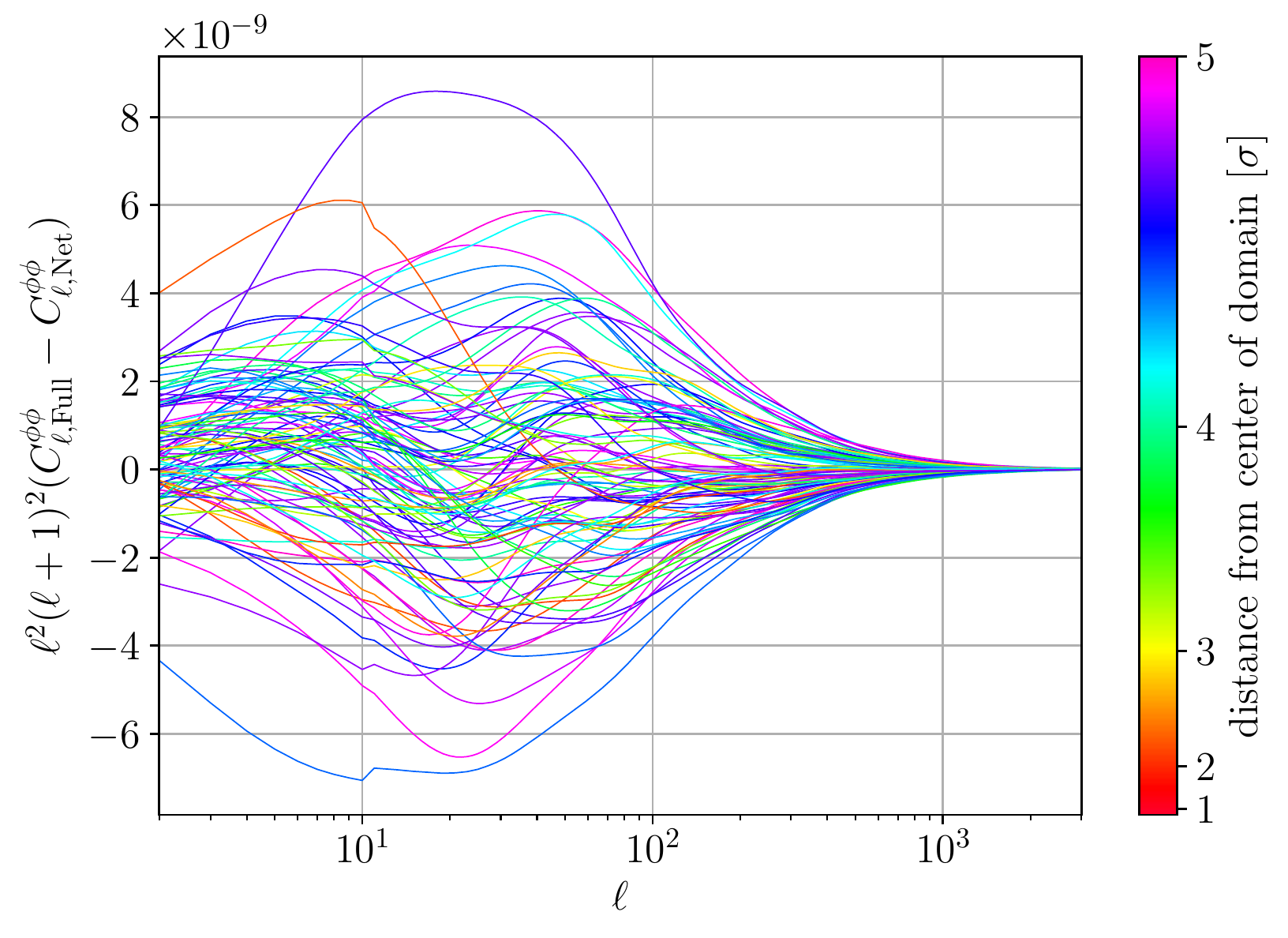}
	\caption{Difference in $\phi\phi$ power spectrum}
\end{subfigure}
\caption{\label{fig:power_spectra_pp_lines} Same as \cref{fig:power_spectra_tt_lines} but for $\phi \phi$ power spectra.}
\vspace{1\baselineskip}
\end{figure}

\begin{figure}[t]
	\centering
	\begin{subfigure}{0.45\linewidth}
		\includegraphics[width=\linewidth]{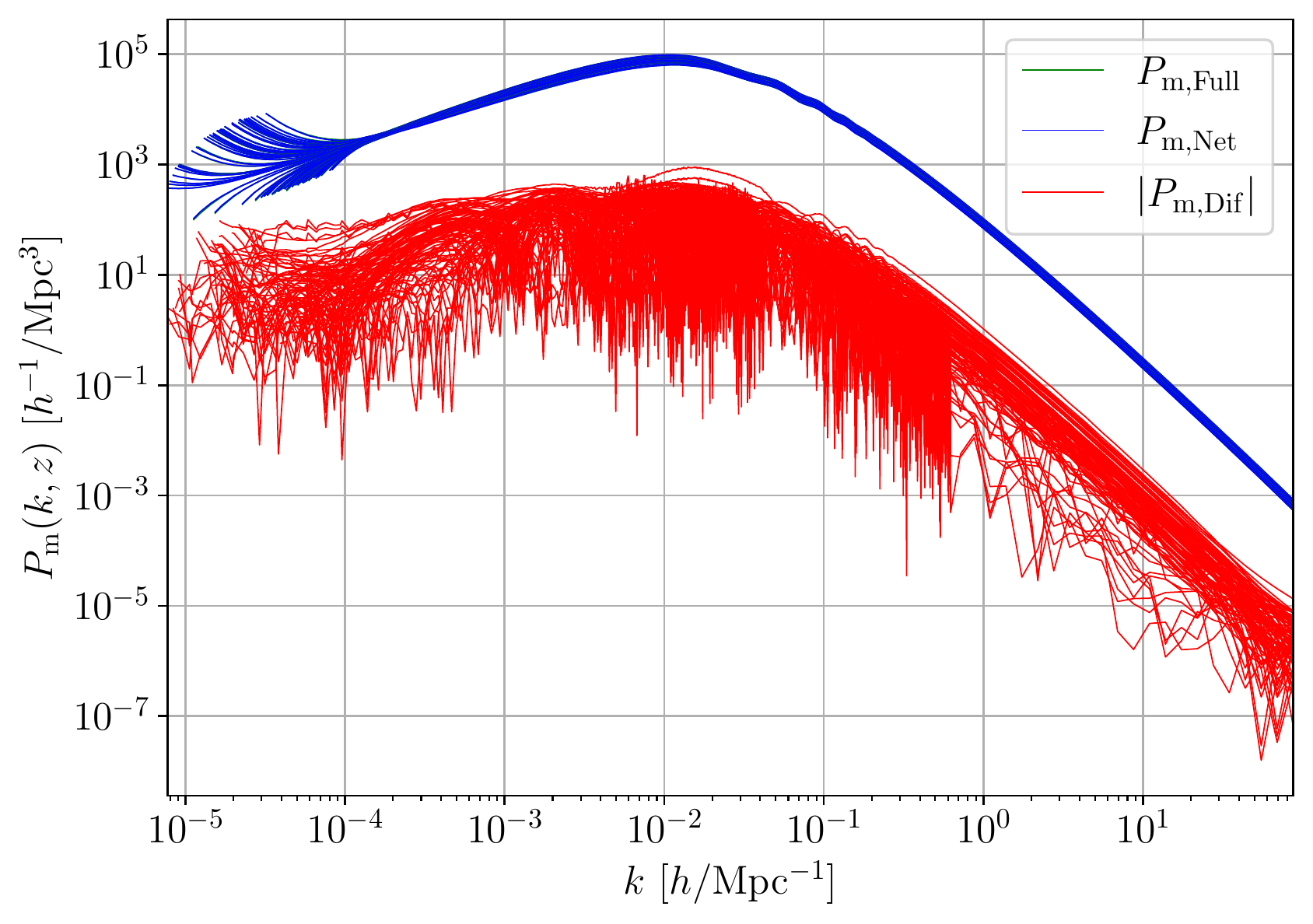}
		\caption{Matter power spectrum}
	\end{subfigure}
	\begin{subfigure}{0.45\linewidth}
		\includegraphics[width=\linewidth]{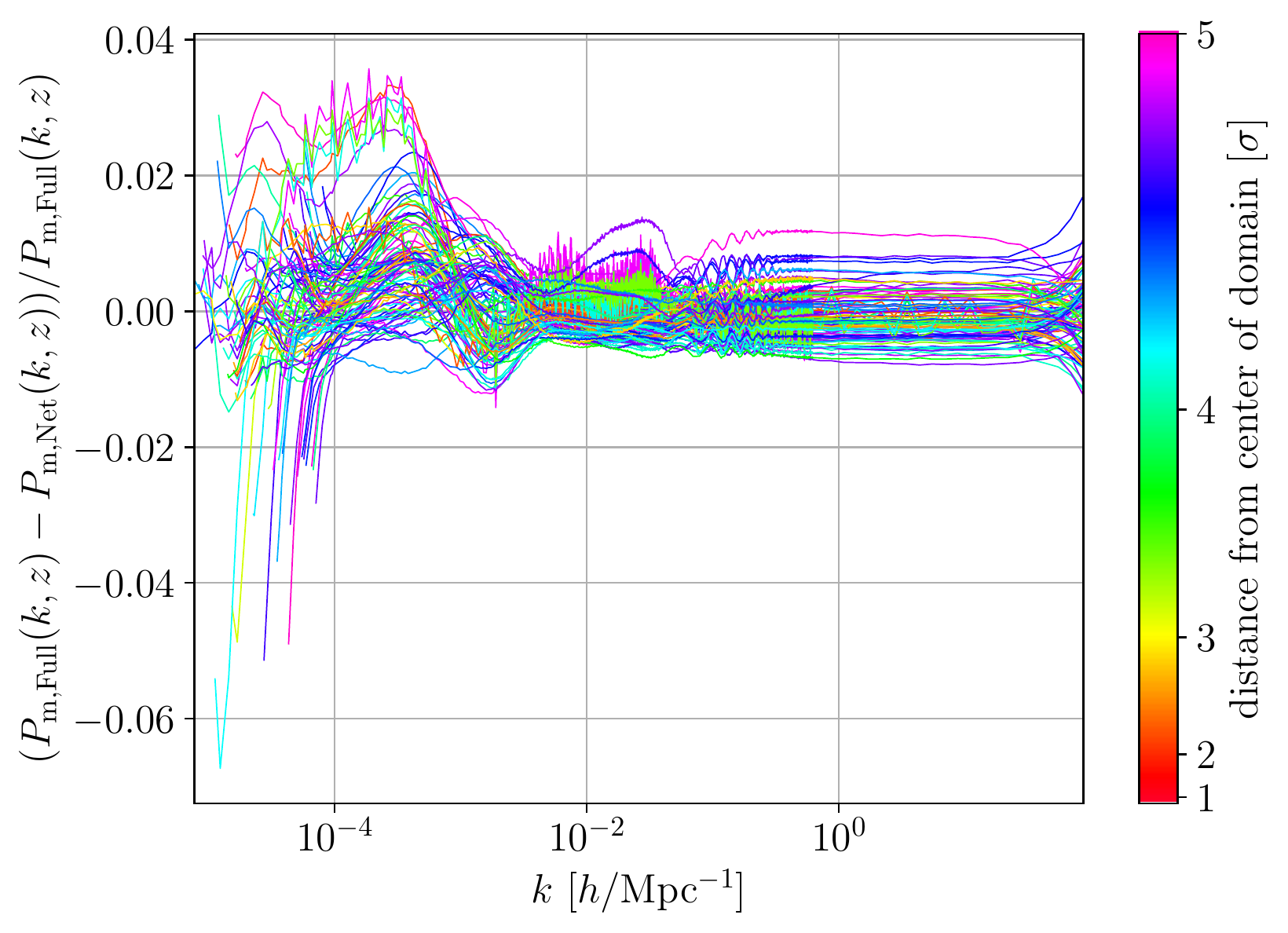}
		\caption{Matter power spectrum}
	\end{subfigure}
	\caption{\label{fig:matter_power_spectrum} Same as \cref{fig:power_spectra_tt}, but for the total matter power $P_\mathrm{m}(k,z)$ spectrum at redshift $z=0$ in units of Mpc$^3$, as a function of wavenumber $k$ in units of $h/$Mpc.}
\end{figure}

\begin{figure}[t]
	\centering
	\begin{subfigure}{0.45\linewidth}
		\includegraphics[width=\linewidth]{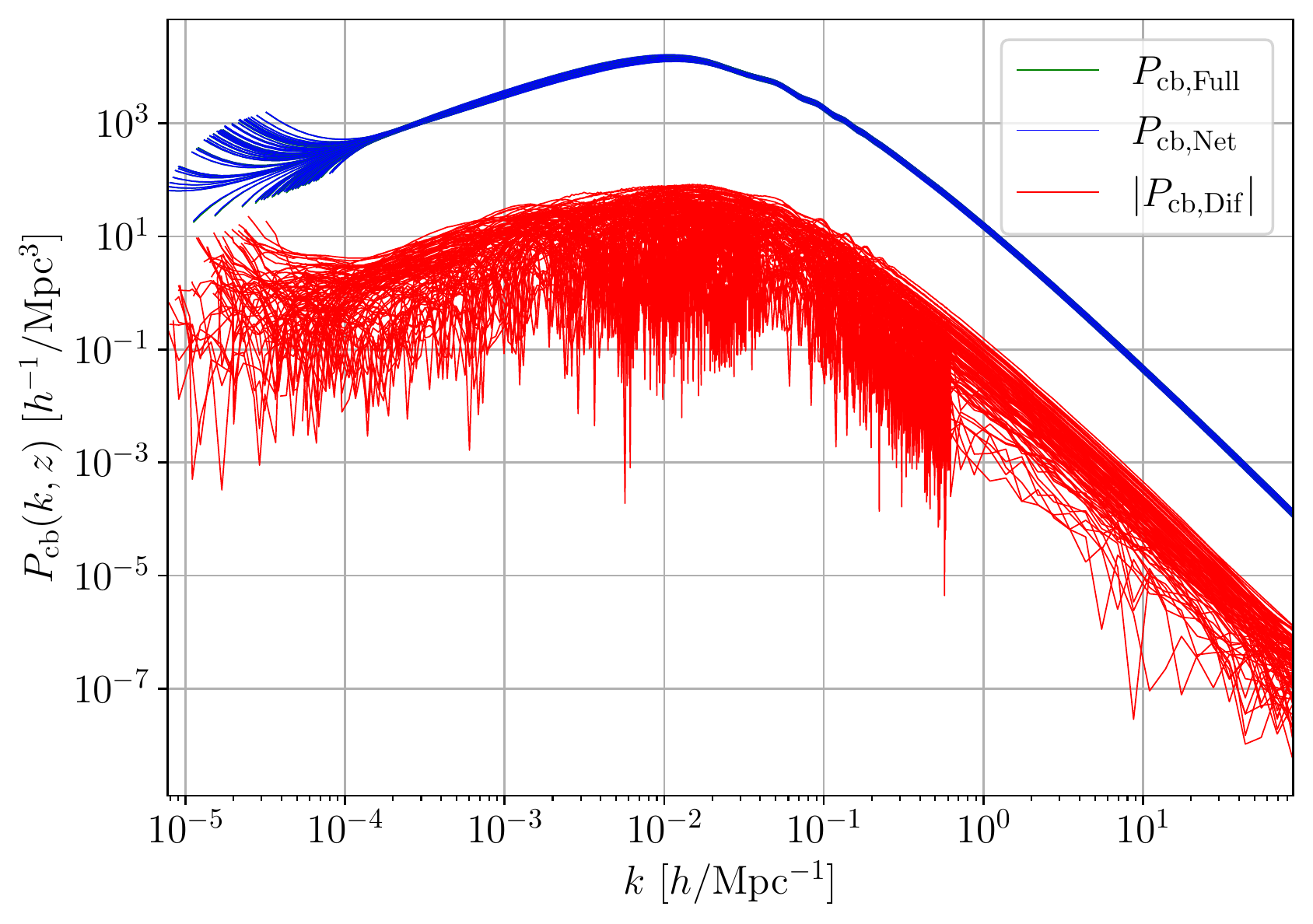}
		\caption{Matter power spectrum}
	\end{subfigure}
	\begin{subfigure}{0.45\linewidth}
		\includegraphics[width=\linewidth]{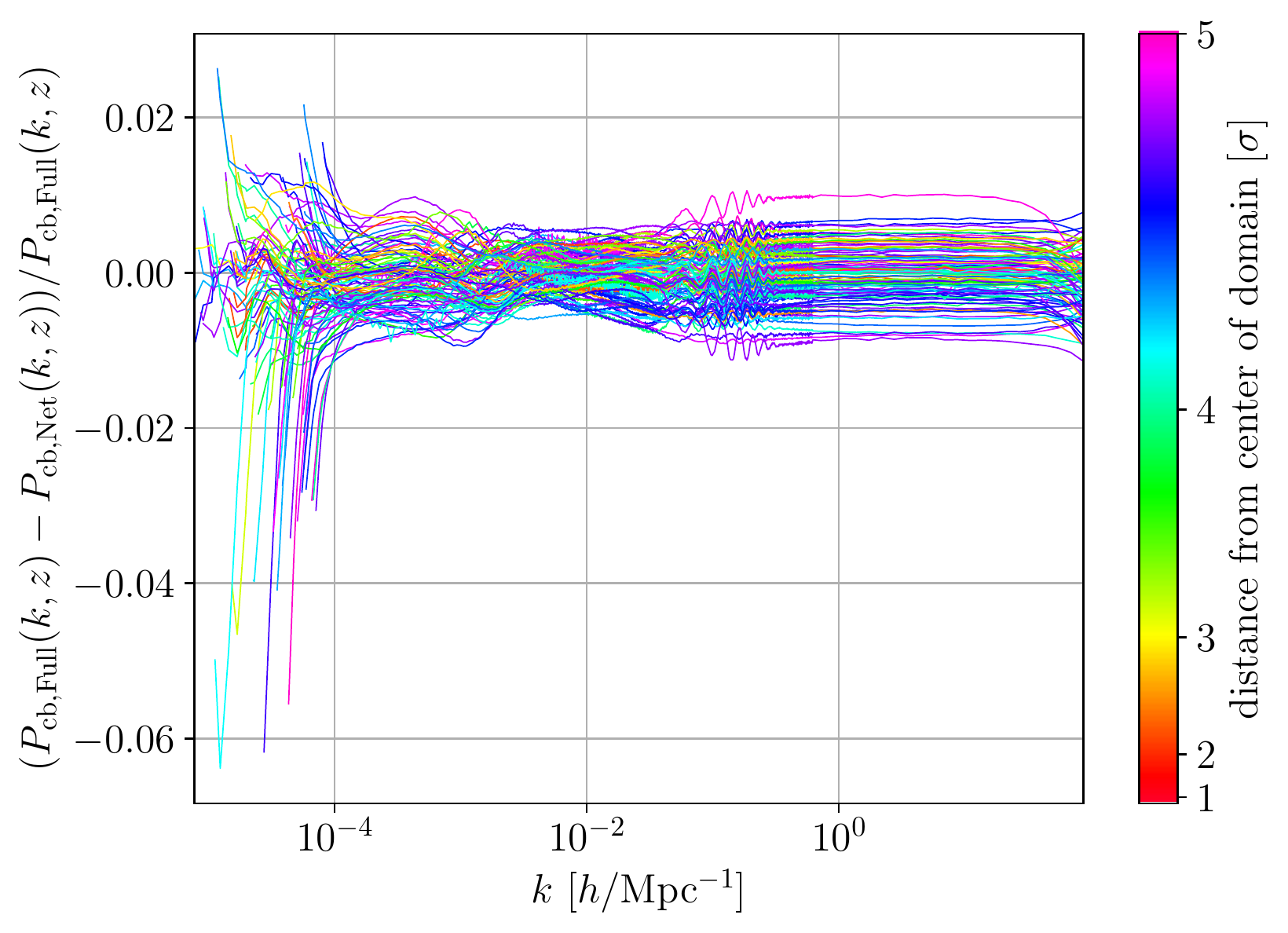}
		\caption{Matter power spectrum}
	\end{subfigure}
	\caption{\label{fig:cb_power_spectrum} Same as \cref{fig:power_spectra_tt}, but for the matter power spectrum of cold dark matter and baryons $P_\mathrm{cb}(k,z)$ at redshift $z=2$ in units of Mpc$^3$, as a function of wavenumber $k$ in units of $h/$Mpc.}
\end{figure}

In order to check for the presence of systematic deviations in the predictions, we also show in \cref{fig:power_spectra_tt_lines,fig:power_spectra_te_lines,fig:power_spectra_ee_lines,fig:power_spectra_bb_lines,fig:power_spectra_pp_lines,fig:matter_power_spectrum,fig:cb_power_spectrum} the $C_\ell$ and $P(k,z)$ residuals as individual lines. In the respective left panels, we show on a logarithmic scale the power spectra calculated with \classfull (green) or \classnet (blue),\footnote{Usually the blue lines are so close to the green that they mask the green lines.} and the absolute value of their difference in red. In the right panel we display the respective differences on a linear scale (note the scaling by two to three orders of magnitude in the $y$-axis), colored by the distance of the cosmological model to the center of the ellipsoidal training domain, to show that the residuals are randomly distributed across this domain, rather than smaller near the center (in which case outer lines would tend towards blue/violet).

In \cref{fig:power_spectra_tt_lines} we can check that for the TT spectrum the residuals are one to two orders of magnitude smaller than cosmic variance. In these residuals, we observe a slight mismatch of the oscillation phase/frequency (predominantly caused by network [N2] $T_0$ reco) causing small oscillations in the difference between full calculation and the network prediction. These become most dominant for small multipoles~$\ell$ where the time-saving small layout of the ISW network [N1] causes a small systematic deviation of the Sachs-Wolfe plateau. However, due to the large cosmic variance at small $\ell$ this deviation does not bias parameter inference, as observed in \cref{fig:power_spectra_tt} and in \cref{sec:parameter_estimation}.

A similar effect is seen in \cref{fig:power_spectra_ee_lines}. For the EE spectrum, the residuals are also very small and dominated by a small shift in the predicted phase of acoustic oscillations. Furthermore, a small difference is observable in the reionization peak at low $\ell$, which arises from the small size of the $T_2$ reio [N6] network -- A further increase of the size of this network could reduce such deviations but would also require longer evaluation times. Since we saw no observable biases on the cosmological parameters from this particular design choice (see \cref{sec:parameter_estimation}), we decided not to extend the network.

The cross-correlation of TE in \cref{fig:power_spectra_te_lines} does show a small residual trend at low multipoles, which is again comparatively small compared to cosmic variance. \Cref{fig:power_spectra_bb_lines,fig:power_spectra_pp_lines} show smoother residuals for the BB and $\phi\phi$ spectra: here the prediction error affects mostly the amplitude of the spectra, but it remains again well below cosmic variance.

\Cref{fig:matter_power_spectrum,fig:cb_power_spectrum} show the prediction for the matter and CDM+baryon power spectra from \classnet and \classfull at two arbitrary redshifts ($z=0$ and $z=2$). The agreement is at the level of $\sim$1\% for $k>5\times10^{-4} \mathrm{Mpc}^{-1}$. Below this, systematic deviations occur, as a consequence of the quadratic extrapolation of the matter density source function to low wavenumbers, explained further in \cref{sec:approx}.
However, wavenumbers below approximately $k \sim 5\cdot 10^{-4} \mathrm{Mpc}^{-1}$ have a negligible impact on all potentially observable quantities. The $P_\mathrm{m}(k,z)$ and $P_\mathrm{cb}(k,z)$ residuals are relatively flat, reflecting mainly a small error on the amplitude of the spectra. The residuals show only tiny oscillations on BAO scales, showing that BAO features are also accurately predicted by the networks.


We have demonstrated the excellent accuracy of \classnet at the level of the source function and the observable power spectra. Note that \classnet has not been designed for computing observables for a single model: in this case, \classfull should always be preferred, since a few seconds of computation are not a large price to pay. \classnet becomes vitally important in parameter inference pipelines, when thousands or even millions of models need to be evaluated. To judge the performance of \classnet in these cases, we will now look at its performance for Bayesian parameter estimation.

\subsection{Parameter Inference}
\label{sec:parameter_estimation}

Emulators such as \classnet are specially useful in the context of parameter inference, i.e.\ estimating a posterior distribution for a set of parameters of a specific cosmological model in the light of data. This is done by computing the likelihood of observables derived from experimental data, such as power spectra, given a theoretical prediction for these observables. The process involves evaluating an Einstein-Boltzmann Solver many times close to the maxima of the posterior following a Monte Carlo sampling algorithm. We are interested in measuring the accuracy of \classnet with respect to \classfull (i.e. to the full EBS computation) in the context of parameter inference on $\Lambda$CDM and its extensions with usual data sets.

\enlargethispage*{1\baselineskip}
We use the Planck 2018 data (TT, TE, EE + lensing), BAO data and Pantheon data already described in \cref{ssec:domain_cn}, in different combinations depending on the theoretical model under study. We use Cobaya \cite{cobaya}\cite{cobaya_code} as our Bayesian inference framework, and its Markov Chain Monte Carlo (MCMC) implementation \cite{Lewis:2002ah,Lewis:2013hha} as the sampler (including its \enquote{drag} option to treat nuisance parameters as fast parameters). 

The amount of evaluations required from the EBS depends on the desired accuracy of the sampled chains and the quality of prior knowledge -- like covariance matrices and reference distributions -- that can be supplied to the sampler, but it does not depend on whether \classnet or \classfull is used. As such, we will use the same setup for both codes. For each case, we run four chains in parallel, each being allowed to access eight cores for threading during the EBS computation. Both \classfull and \classnet runs use the same priors, same initial covariance matrices, and same starting points. The covariance matrix and starting points are determined by a \classfull run with lax convergence threshold (Gelman-Rubin $R-1=0.05$), the starting points of each of the chains in the final runs being the final points of the chains in the initial less-converged run.\footnote{The important thing is only that the point is the same for all newly started runs, the choice of the final point was arbitrary and we could have easily chosen the bestfit point instead.}
Starting all runs from a good guess of the best fit and covariance matrix removes the burn-in phase and reduces the amount of randomness in the amount of evaluations needed to converge in the final run.

In order to quantify the accuracy of \classnet with respect to \classfull, for each cosmological model to be tested (i.e.\ $\Lambda$CDM and possible modifications), we perform a separate MCMC run with each of \classnet and \classfull. We then compare these runs to quantify, for each parameter $x$, the bias on the mean value $m^x$, as well as the bias on the error (or more precisely on the confidence interval) $e^x$. For the means,
the bias for each model and each parameter $x$ is defined as

\begin{align}\label{eq:vvalue}
m^{x}_\mathrm{model} \equiv \frac{\left|\bar{x}^{\mathrm{Net}}-\bar{x}^{\mathrm{Full}}\right|}{\sigma_{x}^{\mathrm{Full}}}~,
\end{align}
where $\bar{x}$ denotes the posterior mean and $\sigma_x$ its 68\% confidence level interval half-length -- with the exception of the parameters $N_\mathrm{eff}$ and $\Omega_\nu h^2$ that are only constrained from one side and for which we use instead half of the 95\% upper confidence limit. For the errors, we simply compute the relative difference
\begin{align}\label{eq:bvalue}
e^{x}_\mathrm{model} \equiv \frac{\sigma_{x}^{\mathrm{Full}} - \sigma_{x}^{\mathrm{Net}}}{\sigma_{x}^{\mathrm{Full}}}~,
\end{align}
with $\sigma_x$ being again the 68\% confidence level interval half-length -- or the 95\% upper confidence limit for $N_\mathrm{eff}$ and $\Omega_\nu h^2$.

\begin{figure}[ht]
	\centering
	\begin{subfigure}{0.99\linewidth}
		\includegraphics[width=\linewidth]{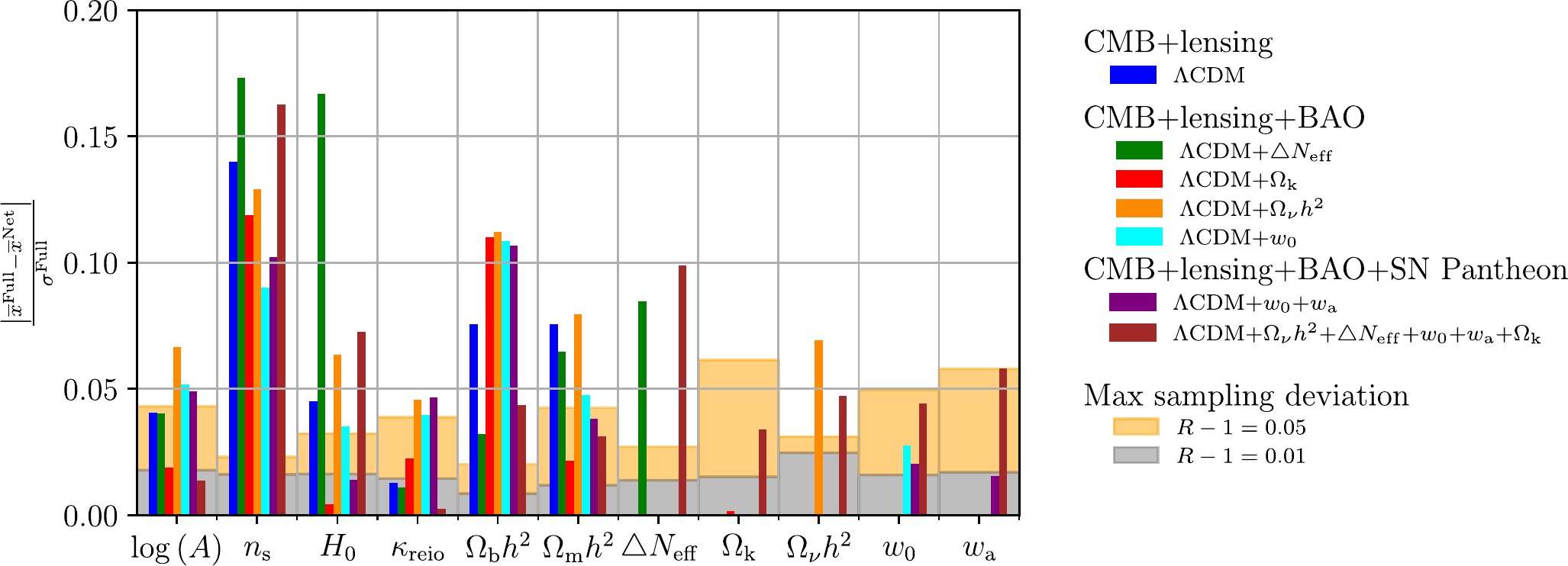}
	\end{subfigure}
	\caption{Deviation of means between \classnet and \classfull measured in units of 68\% CL widths (respectively half of the 95 \% upper CL for $\triangle N_\mathrm{eff}$, $\Omega_\nu h^2$) for each parameter in each model. For each parameter the colored bar shows the observed deviation, while the grey/orange filled contour in the background shows the expected deviation purely from sampling uncertainties for a Gelman-Rubin convergence of $|R-1|<0.01$/$|R-1|<0.05$\,, respectively. See also \cref{tab:vvalues} for the precise values.}
	\label{fig:precision_plot}
\end{figure}

\begin{figure}[ht]
	\centering
	\begin{subfigure}{0.99\linewidth}
		\includegraphics[width=\linewidth]{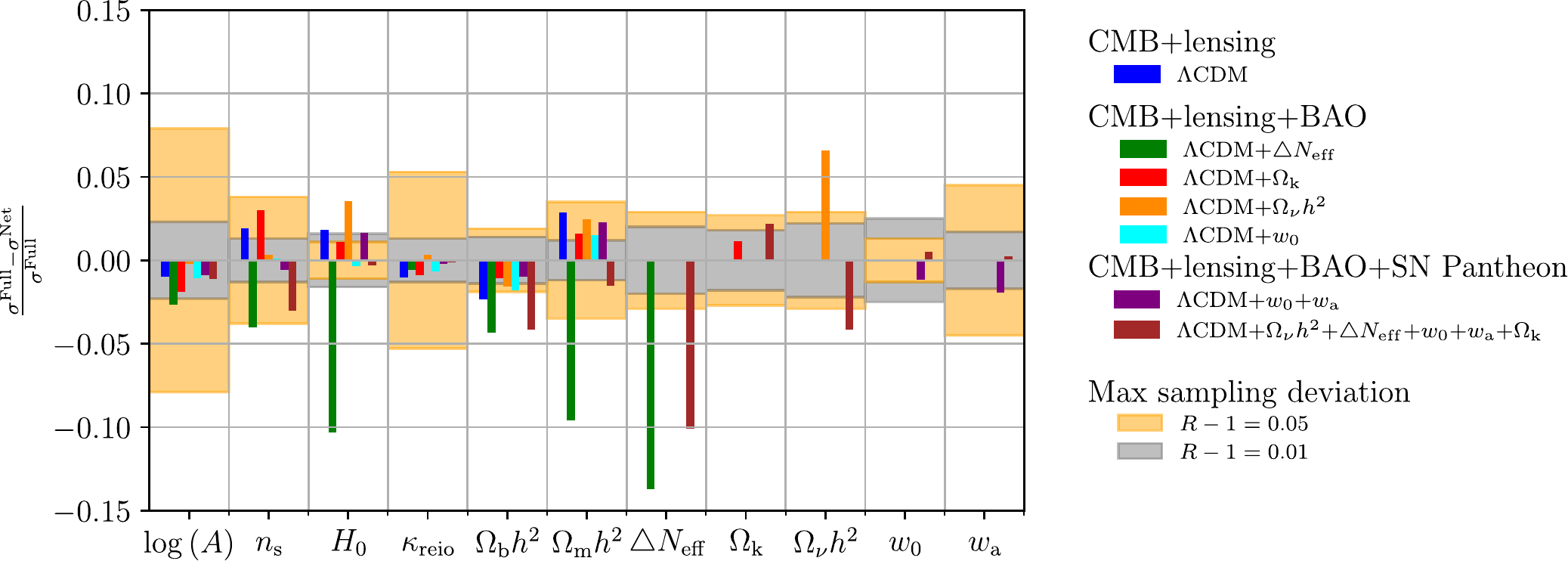}
	\end{subfigure}
	\caption{Same as \cref{fig:precision_plot} but for the relative deviation of the 68\% CL (or respectively half of the 95 \% upper CL for $\triangle N_\mathrm{eff}$, $\Omega_\nu h^2$) between \classnet and \classfull. See also \cref{tab:bvalues} for the precise values.}
	\label{fig:bias_plot}
\end{figure}

Notice that the \classfull (and \classnet) mean and covariance vary between different individual MCMC runs (and in particular this means that the \classfull result does not represent the exact true posterior). This implies that deviations of parameters (measured with $m_\mathrm{model}^x$) that are below the intrinsic sampling variance between individual runs are irrelevant. To ensure that this sampling variance is very small, we set the Gelman-Rubin convergence criterion to $R-1=0.01$.
To assess the size of the intrinsic sampling variance, we perform five identical\footnote{Except for the inherent randomness of the MCMC chain acceptance or rejection of proposed steps.} Cobaya runs with \classfull in the full 11 dimensional extended model described in \cref{eq:cosmo_inputs}. In these runs we obtain an average parameter deviation over all parameters $x$ of $\bar{m}^{\mathrm{ref}}_\mathrm{extended}=0.0078$ ($|\bar{e}|^{\mathrm{ref}}_\mathrm{extended}=0.019$\footnote{We state the mean of the absolute deviation from the averaged value since we are interested in typical deviation scales.}) and a maximum deviation of $m^{\mathrm{ref},\mathrm{max}}_\mathrm{extended}=0.025$ ($|e|^{\mathrm{ref},\mathrm{max}}_\mathrm{extended}=0.025$). In \cref{fig:precision_plot,fig:bias_plot} we show how big the deviations $m^x_\mathrm{model}$ of the mean and $e^x_\mathrm{model}$ of the confidence limit are between \classnet and for a variety of different models, and directly compare the corresponding numbers to the maximum sampling deviations (shown as grey/orange contours in the background). If the deviations are below this level, they could have been caused by chance and will be considered irrelevant.

\begin{figure}[ht]
  \centering \includegraphics[width=0.9\linewidth]{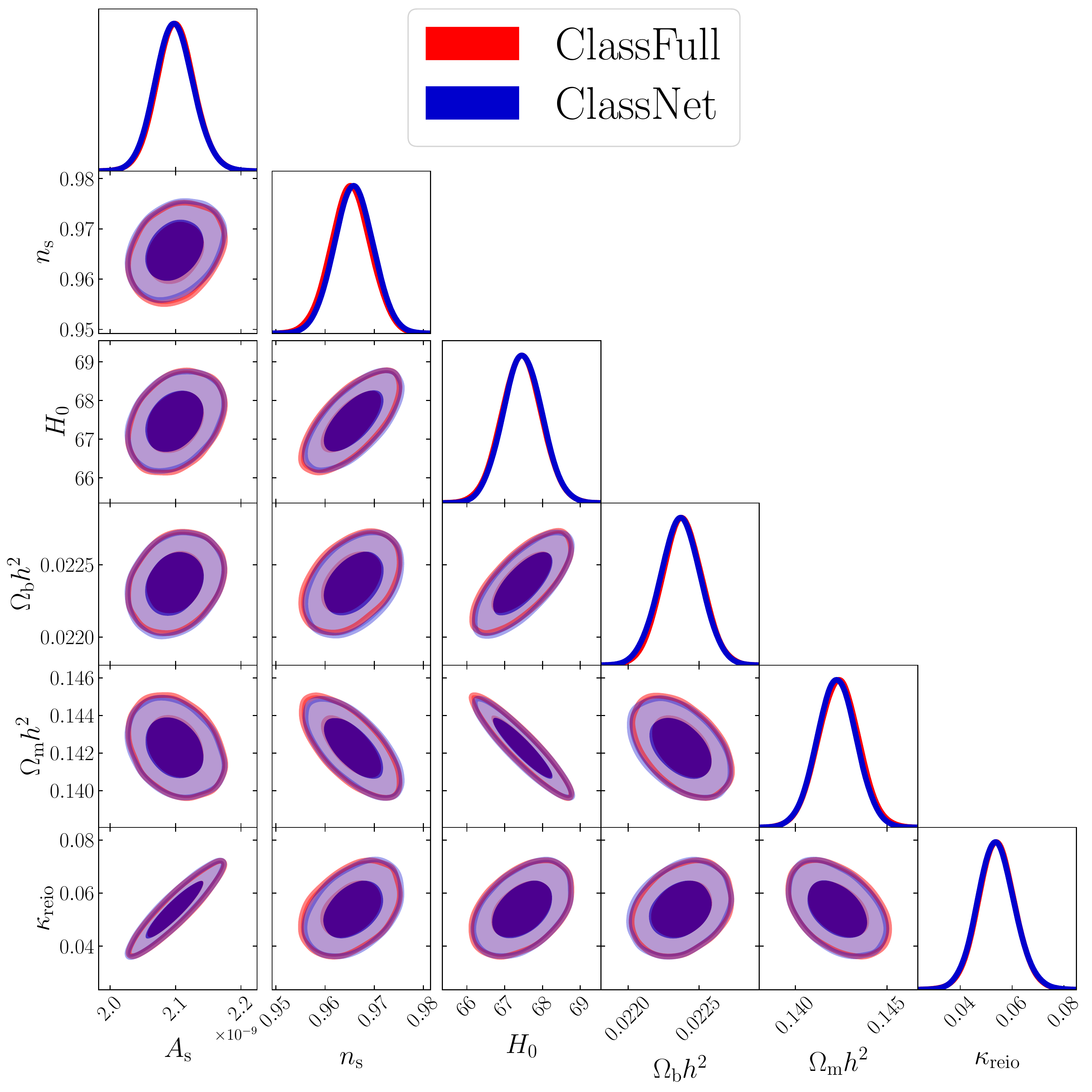}
  \caption{1- and 2-dimensional posterior contours for $\Lambda$CDM with Planck 2018 data with either \classfull (standard CLASS, in red) or \classnet (in blue). Differences are only barely visible for $n_s$ (see text for quantification). We only show here the contours for cosmological parameters, although our run also included the usual Planck nuisance parameters.}
  \label{fig:parameter_estimation_lcdm}
\end{figure}

\begin{figure}[t]
	\centering
	\includegraphics[width=0.95\linewidth]{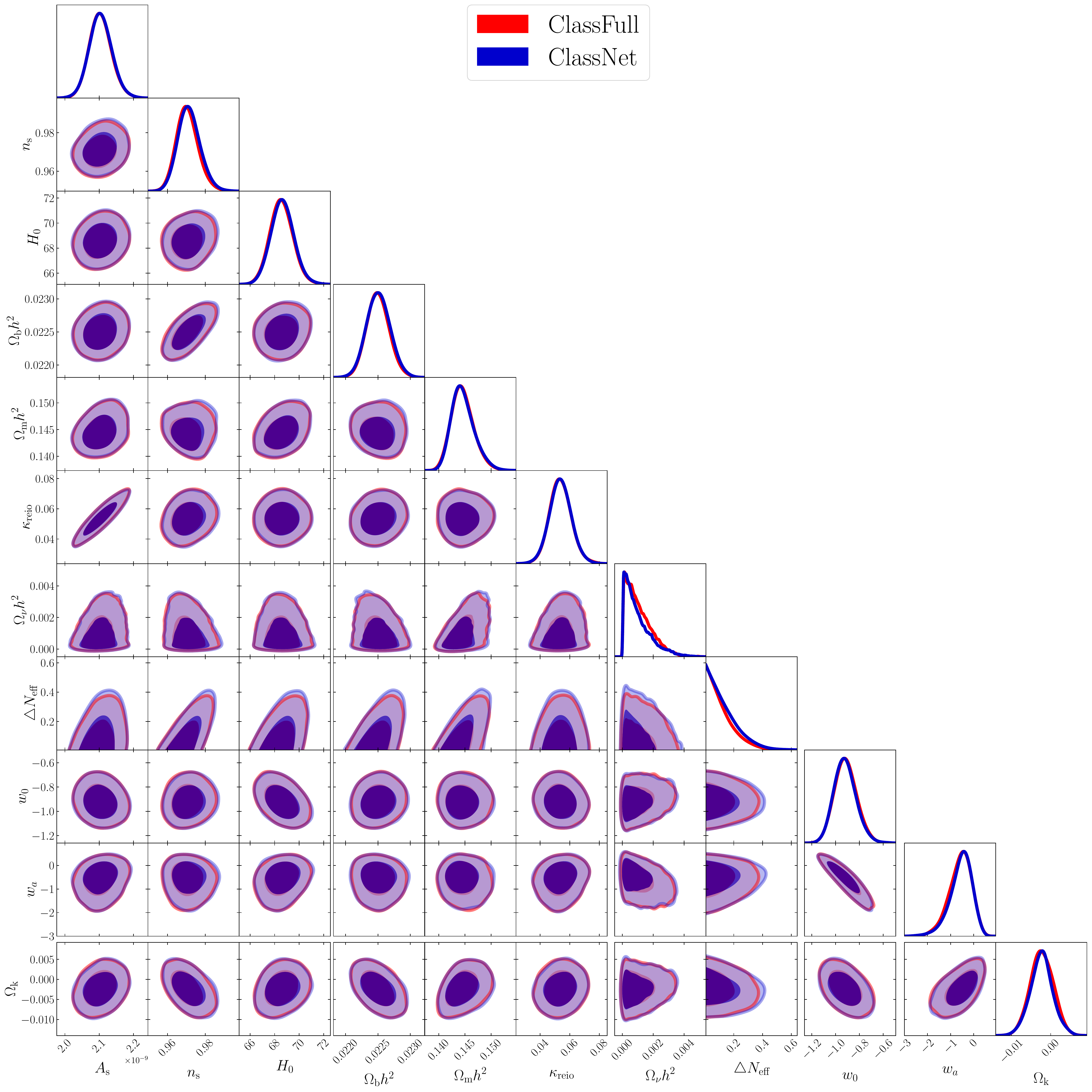}
	\caption{1- and 2-dimensional posterior contours for the fully-extended, 11-parameter $\Lambda$CDM model (with dynamical dark energy, varying neutrino mass, effective relativistic degrees of freedom and curvature) with the full set of CMB+BAO+SN data, with both ClassFull (standard CLASS, in red) and ClassNet (in blue). We only show here the contours for cosmological parameters, although our run also included the usual Planck and Pantheon nuisance parameters.}
	\label{fig:parameter_estimation_11p}
\end{figure}

Below we discuss the accuracy of \classnet, first, for the minimal $\Lambda$CDM model introduced in \cref{ssec:domain_cn}, then, for simple one- or two-parameter extensions, and finally, for the extended 11-parameter model. We show all biases on the means $m^{x}_\mathrm{model}$ for each model in \cref{fig:precision_plot,tab:vvalues} and all biases on the errors $e^{x}_\mathrm{model}$ in \cref{fig:bias_plot,tab:bvalues}. We also present the triangle plots obtained with \classfull or \classnet in \cref{fig:parameter_estimation_lcdm} for the minimal $\Lambda$CDM model, in \cref{fig:parameter_estimation_11p} for the extended 11-parameter model and in \cref{ssec:triangles} for other models.


\begin{table}
	\centering
	\resizebox{0.95\textwidth}{!}{
	\begin{tabular}{c | c c c c c c c c c c c}
		Model & $\log(10^{10}A_s)$ & $n_s$ & $H_0$ & $\kappa_\mathrm{reio}$ & $\Omega_b h^2$ & $\Omega_m h^2$ &$\Delta N_\mathrm{eff}$ & $\Omega_k$ &$\Omega_\nu h^2$ 
		& $w_0$ & $w_a$ \\ \hline
		$\Lambda$CDM & 0.041 & 0.140 & 0.045 & 0.013 & 0.076 & 0.076 & - & - & - & - & -\\
		$\Lambda$CDM+$\Delta N_\mathrm{eff}$ & 0.040 & 0.173 & 0.167 & 0.011 & 0.032 & 0.065 & 0.085 & - & - & - & -\\
		$\Lambda$CDM+$\Omega_k$ & 0.019 & 0.119 & 0.004 & 0.022 & 0.110 & 0.022 & - & 0.001 & - & - & -\\
		$\Lambda$CDM+$\Omega_\nu h^2$ & 0.066 & 0.129 & 0.063 & 0.045 & 0.112 & 0.079 & - & - & 0.069 & - & -\\
		$\Lambda$CDM+$w_0$ & 0.052 & 0.090 & 0.035 & 0.040 & 0.108 & 0.047 & - & - & - & 0.028 & -\\
		$\Lambda$CDM+$w_0$+$w_a$ & 0.049 & 0.102 & 0.014 & 0.046 & 0.107 & 0.038 & - & - & - & 0.020 & 0.016 \\
		$\Lambda$CDM+all & 0.013 & 0.162 & 0.073 & 0.003 & 0.044 & 0.031 & 0.099 & 0.034 & 0.047 & 0.044 & 0.058 \\ \hline \\[-0.3cm]
		$|R-1| < 0.05$ (max) & 0.043 & 0.023 &  0.032 & 0.039 & 0.020 & 0.042 & 0.027 & 0.037 &0.031 & 0.050 & 0.058 \\ 
		$|R-1| < 0.01$ (max) & 0.018 & 0.016 & 0.016 & 0.014 & 0.008 & 0.012 & 0.014 & 0.015 &0.025 & 0.017 & 0.016
	\end{tabular}
	}
	\caption{\label{tab:vvalues} Bias on the mean $m^x_\mathrm{model}$ according to \cref{eq:vvalue} for all models and parameters $x$ under consideration. The lower two lines show for comparison the maximum $m^x$ expected from finite MCMC sampling. See also \cref{fig:precision_plot} for a graphical representation.}
\end{table}

\begin{table}
	\centering
	\resizebox{0.95\textwidth}{!}{
		\begin{tabular}{c | c c c c c c c c c c c}
			Model & $\log(10^{10}A_s)$ & $n_s$ & $H_0$ & $\kappa_\mathrm{reio}$ & $\Omega_b h^2$ & $\Omega_m h^2$ &$\Delta N_\mathrm{eff}$ & $\Omega_k$ &$\Omega_\nu h^2$ 
			& $w_0$ & $w_a$ \\ \hline
			$\Lambda$CDM & -0.010 & 0.019 & 0.018 & -0.010 & -0.023 & 0.029 & - & - & - & - & -\\
			$\Lambda$CDM+$\Delta N_\mathrm{eff}$ & -0.027 & -0.040 & -0.103 & -0.006 & -0.044 & -0.096 & -0.137 & - & - & - & -\\
			$\Lambda$CDM+$\Omega_k$ & -0.019 & 0.030 & 0.011 & -0.009 & -0.011 & 0.016 & - & 0.012 & - & - & -\\
			$\Lambda$CDM+$\Omega_\nu h^2$ & -0.002 & 0.003 & 0.035 & 0.003 & -0.016 & 0.025 & - & - & 0.066 & - & -\\
			$\Lambda$CDM+$w_0$ & -0.011 & 0.001 & -0.003 & -0.006 & -0.018 & 0.015 & - & - & - & $<$0.001 & -\\
			$\Lambda$CDM+$w_0$+$w_a$ & -0.009 & -0.006 & 0.017 & -0.002 & -0.010 & 0.023 & - & - & - & -0.011 & -0.019 \\
			$\Lambda$CDM+all & -0.011 & -0.031 & -0.003 & -0.001 & -0.042 & -0.015 & -0.101 & 0.022 & -0.041 & 0.005 & 0.002 \\ \hline \\[-0.3cm]
			$|R-1| < 0.05$ (max) & 0.079 & 0.038 &  0.011 & 0.053 & 0.019 & 0.035 & 0.029 & 0.027 &0.029 & 0.013 & 0.045 \\ 
			$|R-1| < 0.01$ (max) & 0.023 & 0.013 & 0.016 & 0.013 & 0.014 & 0.012 & 0.020 & 0.018 &0.022 & 0.025 & 0.017
		\end{tabular}
	}

	\caption{\label{tab:bvalues} Bias on the error $e^x_\mathrm{model}$ according to \cref{eq:bvalue} for all models and parameters. The lowest two lines show for comparison the average $e^x$ expected from finite MCMC sampling. See also \cref{fig:bias_plot} for a graphical representation.}

\end{table}


\paragraph{$\boldsymbol{\Lambda}$CDM model: }
In \cref{fig:parameter_estimation_lcdm} we can see a comparison between a \classnet and \classfull run for the case of the $\Lambda$CDM model. Differences are barely visible, with the possible exception of the primordial tilt $n_s$\,, for which the deviation is $m_{\Lambda \mathrm{CDM}}^{n_s}=0.14$, while  the maximum sampling error is expected to be only around $0.025$ at a convergence of $|R-1|<0.05$. Thus, \classnet biases the result on $n_s$ by $0.14\sigma$, which is still acceptable for parameter inference. The reason for this small biasing is further detailed in \cref{ssec:trouble} -- in a nutshell,
our systematic differences in the prediction of the $C_\ell$ happen to mimick a slightly more tilted TT angular power spectrum, in particular due to the [N2] ($T_0$~reco) network, as visible also in \cref{fig:power_spectra_tt_lines}. The second most biased parameter is the baryon density, with a deviation $m^{\Omega_\mathrm{b}h^2}_{\Lambda \mathrm{CDM}}\sim 0.08$ which is about 3 times bigger than sampling variance. This is likely caused by the intrinsic degeneracy between $\Omega_\mathrm{b}h^2$ and $n_s$\,. For other parameters, the biasing is even lower, with a parameter-averaged difference $\bar{m}_{\Lambda \mathrm{CDM}}=0.059$. All parameter correlations are very well captured by the \classnet run. The confidence interval scales are well recovered with a maximum deviation of $e_{\Lambda \mathrm{CDM}}^{\Omega_\mathrm{m}}=0.029$. This underestimation is within the expected absolute sampling deviations of $\sim0.035$.

We checked a posteriori that in our runs in \classnet mode, only a fraction of $5.8 \cdot 10^{-4}$ of the models in the MCMC chains lay outside of the validation domain and required a \classfull evaluation. Thus, despite being defined using the Planck+BAO+SN likelihood and the 11-parameter model, our training domain encompasses the well-fitting region for Planck 2018 alone and the 6-parameter $\Lambda$CDM model. This shows that, in the definition of the training domain, the role of  BAO+SN data was mainly to restrict the range of the extended model parameters $\{\Omega_k, w_0, w_a\}$ rather than that of $\Lambda$CDM parameters. Our networks are thus found to be extremely efficient even for Planck-only fits, as long as spatial curvature and dynamical dark energy are not included.

\pagebreak[20]
\paragraph{$\boldsymbol{\Lambda}$CDM+$\Delta N_\mathrm{eff}$ model: }
We repeat the analysis with a free number of effective relativistic degrees of freedom $\Delta N_\mathrm{eff}$ and using the Planck+BAO likelihoods (see \cref{tab:vvalues}, \cref{tab:bvalues} as well as \cref{fig:parameter_estimation_Nur} in \cref{ssec:triangles}). The 95 \% upper confidence limit on $\Delta N_\mathrm{eff}$ rises when using \classnet by a factor of $\sim 14\%$ which exceeds the reference sampling deviations of $\bar{|e|}^{\mathrm{ref},\mathrm{\triangle N_\mathrm{eff}}}_\mathrm{extended}=0.019$. 
We analyse the origin of this bias in \cref{ssec:trouble}.
Through parameter degeneracies, this bias propagates also to $H_0$ and $n_s$ ($m^{H_0} \sim m^{n_s} \sim 0.17$ and $e^{H_0}\sim-0.10$). These values, which are still small, correspond to the largest biases observed across all our runs. 
Also in this case, we find that only a fraction of $\sim 10^{-4}$ of the sampled points lay outside the validation region of \classnet and were obtained with \classfull evaluations.

\paragraph{$\boldsymbol{\Lambda}$CDM+$\Omega_k$ model: }
Next, we investigate an extension of the $\Lambda$CDM model with spatial curvature, still with the Planck+BAO likelihoods (see \cref{tab:vvalues}, \cref{tab:bvalues} as well as \cref{fig:parameter_estimation_Nur} in \cref{ssec:triangles}). The constraints are all well recovered in this case, with only minor deviations in $m^{n_s}$ of order $0.12$ and $m^{\Omega_b h^2}$ of $0.11$. The deviation in the mean value of $\Omega_k$ is well below sampling deviation and thus unbiased. This also holds true for the estimates of the confidence limit for all parameters, as they do not exceed the reference sampling deviations. 
Additionally we find that none of the sampled points in the MCMC chains were required to be evaluated by \classfull.

\enlargethispage*{1\baselineskip}
\paragraph{$\boldsymbol{\Lambda}$CDM+$\Omega_\nu h^2$ model: }
We now allow the previously-fixed neutrino mass to vary, with a flat prior on the parameter $\Omega_\nu h^2$ (\texttt{omega\_ncdm} in \class, with three degenerate massive neutrinos). We constrain the parameter space again with the Plack+BAO likelihoods. Also in this case, most parameter contours are recovered almost perfectly (see \cref{tab:vvalues}, \cref{tab:bvalues} as well as \cref{fig:parameter_estimation_Omnu} in \cref{ssec:triangles}). The parameter $\Omega_\nu h^2$ shows a slight deviation of $m^{\Omega_\nu h^2}\sim 0.11$\,, which propagates to a deviation on $H_0$ of $m^{H_0}\sim 0.06$. The same holds true for the confidence limits which both slightly exceed the sampling variation with $e^{\Omega_\nu h^2}=0.07$ and $e^{H_0}=-0.04$. 
In this case, all sampled points were obtained with \classnet.

\paragraph{$\boldsymbol{\Lambda}$CDM+$w_0$ model: }
We introduce dark energy gradually, first by allowing for a floating equation of state parameter $w\neq 1$, assumed to be constant over time, while sticking to the Planck+BAO likelihoods (see \cref{tab:vvalues}, \cref{tab:bvalues} as well as \cref{fig:parameter_estimation_Omnu} in \cref{ssec:triangles}). The additional parameter $w_0$ has negligible bias (comparable to the mean sampling deviation),  and the other parameters are well recovered as well. All uncertainty estimates turn out to be unbiased. 
In this case, a small a fraction of $\sim 10^{-3}$ samples in the MCMC chains were obtained with \classfull. This can be attributed to the fact that the training domain included the supernovae likelihood while this run did not.

\paragraph{$\boldsymbol{\Lambda}$CDM+$w_0$+$w_a$ model: }
For dynamical dark energy following the CPL parametrization, we add the parameters $w_0+w_a$\, and switch to the full Planck + BAO + Pantheon likelihood combination. All contours in this case align very nicely (see \cref{tab:vvalues}, \cref{tab:bvalues} as well as \cref{fig:parameter_estimation_w0wa} in \cref{ssec:triangles}). Again, we only observe a slight bias of the posterior mean deviations for $n_s$ and $\Omega_b h^2$ of $\sim 0.1\sigma$, and confidence intervals are all accurately predicted. 
In this run, the chains were sampled without any \classfull evaluations.

\paragraph{$\boldsymbol{\Lambda}$CDM+all model: }
Finally, we put everything together: we sample on all modifications of $\Lambda$CDM described in section \cref{sec:architecture}, creating a cosmological model with 11 free parameters. We use the full Planck + BAO + Pantheon likelihood combination. As shown in \cref{tab:vvalues}, we find a small bias of $\sim0.16\sigma$ for the posterior mean prediction of $n_s$ as well as a bias of $\sim0.07\sigma$ for $H_0$. The 95\% upper bounds on the parameters $\Delta N_\mathrm{eff}$ and $\Omega_\nu h^2$ are shifted by $e^{\Delta N_\mathrm{eff}}\sim-0.1$ and $e^{\Omega_\nu h^2} \sim -0.04$, as can be seen in \cref{fig:parameter_estimation_11p} and \cref{tab:bvalues}. Apart from these biases, all parameter estimations are well recovered and follow the known correlations. As already stated above, only a fraction of $5.8\cdot10^{-4}$ of the sampled chains were evaluated with \classfull.

\vspace*{1\baselineskip}
We have shown in this section that \classnet can predict source functions well enough to be used for parameter inference from current data sets while assuming $\Lambda$CDM or its most common extensions. The deviations we found between \classnet and \classfull are slightly above what we found for the sampling variance in a small number of cases, but still not very significant. \Cref{fig:precision_plot} provides a visual representation of the deviations in all cases, and \cref{tab:vvalues,tab:bvalues} a numerical summary. We should keep in mind that we have applied a very strong convergence criterion of $|R-1|<0.01$ when measuring the sampling variance.

\section{Conclusions\label{sec:conclusion}}

With the new efficient and accurate network architecture presented in this work, we believe that we have reached a number of significant objectives:
\enlargethispage*{4\baselineskip}
 \begin{itemize}
 \item When using our new set of neural networks instead of a full integration of the system of linear cosmological perturbation equations, we fully remove the calculation of source functions $S_x$ from the list of bottlenecks in an EBS. This has been checked with our implementation of the networks in a version of \class called \classnet, released publicly together with this work. It applies independently of the number of cores and of the computed observables.
 \item When \classnet is used for computing only the matter and baryon + CDM power spectrum, it becomes faster than the usual \class code by a factor of about 55 on 4 CPUs (and even more on less CPUs). Then, the total execution time falls below the order of ${\cal O}(0.1s)$, that is well below the execution time of the typical likelihood of a survey. This can lead to a considerable speed up in a parameter inference run, that will then be limited only by the execution time of the likelihood.
 \item When \classnet is used for computing the CMB power spectra, it becomes faster than the usual \class code by a factor of about 2.7 on 4 CPUs (and even more on less CPUs). This is sufficient for speeding up a parameter inference run from e.g. Planck data by approximately the same factor.
 \item Our neural networks are trained for an extended cosmology, which includes five additional free parameters compared to the minimal $\Lambda$CDM model. The ease at which such new parameters can be included hints at the flexibility of the network to treat any non-standard cosmology. Moreover, thenetworks are compatible with any assumption regarding the primordial power spectrum of scalar adiabatic perturbations, allowing for a huge flexibility of possible models of inflation. 
 \item We have demonstrated that \classnet reaches sufficient accuracy for fitting current data sets such as Planck, BAOs from BOSS, and supernovae data from Pantheon, with a biasing of the reconstructed parameters in the range of $0.01\sigma$ to $0.1\sigma$ in most cases, or up to $0.17\sigma$ in only very few cases well documented in our Results section (mostly biasing $n_s$ or $\Delta N_\mathrm{eff}$).
 \item Thanks to a new network splitting and a more efficient architecture, which relies on passing some exact background/thermodynamical functions (as well as some analytical approximations) as inputs to the network, the size of our network could be kept very small. Then the networks are both fast to evaluate and fast to retrain for cosmologies other than those considered in this work. Together with \classnet, we also release some scripts to guide a new network training by the user.
 \end{itemize}
Summarizing, the multitude improvements to the \classnet code have allowed us to eliminate the {\tt perturbation} module as a bottleneck of the EBSs, while remaining highly accurate for individual evaluations and parameter inference runs. Additionally, the extension of the parameter space far beyond $\Lambda$CDM, with up to five additional parameters, allows much broader applicability of \classnet, and shows the promising future for other non-standard cosmologies.
 
This release already offers an opportunity for considerable savings of CPU time and electricity costs for cosmology groups. Still, we intend to keep working on the {\sc CosmicNet} project with multiple goals: We will continue improving the design of the network, the hyperparameters, and the training strategy in order to seek for ever more robust, accurate, and efficient networks. Furthermore, we will optimize the use of neural networks for the purpose of parameter inference, with the possibility of using a parameter inference to guide the training like in reference \cite{Nygaard:2022wri}. Moreover, we plan to set up a centralized database of trained network weights, allowing for an efficient exchange of information between different groups, in order to avoid duplicate efforts and in order to increase reproducibility. Finally, we will explore different ways to speed up the remaining bottlenecks in EBSs, such as the line-of-sight integration.

We conclude that the {\sc CosmicNet} project based on the highly efficient and accurate {\sc classNet} branch has a promising future and will soon significantly reduce computational and economic costs for its users around the world.

\section*{Acknowledgements}
Sven G\"unther acknowledges support from the DFG grant LE 3742/6-1.
Nils Sch\"oneberg acknowledges support at different stage of this work from the DFG grant LE 3742/4-1 and from the Maria de Maetzu fellowship grant: Esto publicaci\'on es parte de la ayuda CEX2019-000918-M, financiado por MCIN/AEI/10.13039/501100011033.
Simulations were performed with computing resources granted by RWTH Aachen University under project jara0184.

\clearpage
\appendix
\section{Analytic approximations for network N7 \label{sec:approx}}

\subsection{Approximations for $\Lambda$CDM+$M_\nu$}

The network [N7] is designed to predict relative differences between the true transfer functions -- usually called source functions in the rest of this paper -- $S_{\phi+\psi}(k,\tau)$, $S_{\delta_\mathrm{m}}(k,\tau)$, $S_{\delta_\mathrm{cb}}(k,\tau)$ and some analytic approximation. For $\Lambda$CDM cosmologies extended to massive neutrinos, our analytic approximation can be inferred from the transfer functions of Hu \& Eisenstein.\footnote{We use the C subroutine {\tt TFmdm\_onek\_mpc()} available at \url{http://background.uchicago.edu/~whu/transfer/transferpage.html}} (HE)\cite{hu_eisenstein,Eisenstein:1997jh} We implemented these functions in \classnet. We take advantage of the fact that the \class background and thermodynamics module contain accurate calculations of quantities that are usually approximated analytically within the HE algorithm, such as: the scale-independent growth factor $D(\tau)$, the Hubble rate $H(\tau)$, the redshift at baryon drag time, the sound horizon at that time, the redshift at equality and the wavenumber crossing the Hubble scale at this time. In our implementations, these quantities are read directly from \class.

A difficulty comes from the normalisation of these transfer functions. In \class (and other modern EBSs), the source functions $S_x(k,\tau)$ are normalised in the early universe to the condition ${\cal R}(k,\tau_\mathrm{ini})=1$ for all $k$, where ${\cal R}$ is the comoving curvature perturbation. This is true in particular for $S_x(k,\tau)=\delta_x(k,\tau)$ with $x\in \{\mathrm{m}, \mathrm{cb}\}$. Instead, the HE transfer functions $T_\mathrm{m}^\mathrm{HE}(k,z)$ and $T_\mathrm{cb}^\mathrm{HE}(k,z)$\footnote{These transfer functions are called respectively {\tt tf\_cbnu} and {\tt tf\_cb} in the code of Hu \& Eisenstein.} are, by convention, the same quantities first divided by $k^2$ and then normalised to one in the large wavelength limit,
\begin{equation}
  T^\mathrm{HE}_x(k,z(\tau)) = \left[ \frac{\delta_x(k,\tau)}{k^2} \right] / \lim_{\tilde{k} \rightarrow 0} \left[ \frac{\delta_x(\tilde{k},\tau)}{\tilde{k}^2} \right] \qquad \mathrm{for}~~x\in \{\mathrm{m}, \mathrm{cb}\}~.
\end{equation}
In the HE scheme, the perturbations $\delta_x$ are implicitly expressed in the synchronous gauge, for which $[k^{-2} \delta_x(k,\tau)]$ is approximately independent of $k$ in the small $k$ limit (more precisely, for scales crossing the Hubble radius after radiation domination, $k\ll k_\mathrm{eq}$). \class and the trained networks of \classnet predict a gauge-independent version of the density perturbations, $\delta_x = \delta_x^{(g)} + 3 H \theta_x^{(g)}/k^2$, where ($\delta_x^{(g)}$, $\theta_x^{(g)}$) are the density and velocity perturbations of $x$ in an arbitrary gauge $g$. In the synchronous gauge, the second term is always subdominant: thus, irrespectively of the fact that \class is being run in the synchronous or newtonian gauge, we can think of $\delta_x$ in \class as being the synchronous gauge density perturbations. So the relation between the \class sources for $\delta_\mathrm{m,cb}$ and the HE transfer functions reads
\begin{equation}
  T^\mathrm{HE}_x(k,z(\tau))= \left[ \frac{S_x(k,\tau)}{k^2} \right] / \lim_{\tilde{k} \rightarrow 0} \left[ \frac{S_x(\tilde{k},\tau)}{\tilde{k}^2} \right] \qquad \mathrm{for}~~x\in \{\mathrm{m}, \mathrm{cb}\}~.
 \end{equation}
 Then, in order to predict $S_x(k,\tau)$ from $T^\mathrm{HE}_x(k,z(\tau))$, we need to predict \mbox{$S_x(\tilde{k},\tau)=\delta_x(\tilde{k},\tau)$}, that is, the linear growth factor of large-scale density fluctuations (with very small wavenumber $\tilde{k}$) between an early time at which  we impose the normalisation condition ${\cal R}(\tilde{k},\tau_\mathrm{ini})=1$ and a later time $\tau$ chosen during matter or dark energy domination (the network [N7] is not trying to accurately predict transfer functions during radiation domination).

In this context, since we are considering the limit of $\tilde{k} \to 0$, we can easily assume that the comoving wavelength $2\pi/\tilde{k}$ is even much larger than the current Hubble horizon $c/H_0$\,, and thus also much larger than the Hubble horizon at intermediate times, summarized by $\tilde{k} \ll \mathcal{H}$ for all times with $\mathcal{H} = a(\tau)H(\tau)$. An investigation into the synchronous adiabatic initial conditions (ICs) and the equations of motion on super-Hubble scales tells us that we simply have $\delta_\mathrm{c} = \delta_\mathrm{b} = \delta_\mathrm{m} = \frac{3}{4} \delta_\gamma = \frac{3}{4} \delta_\nu = -\frac{1}{2} h$ at all times of interest. Then, using the Einstein equations
\begin{align}
	\frac{1}{2} \mathcal{H} h' = k^2 \eta + \frac{3}{2} a^2 (\rho_\mathrm{r} \delta_\mathrm{r} + \rho_\mathrm{m} \delta_\mathrm{m})~,\label{eq:app:einsteinh} \\
	k^2 \eta' \propto \frac{4}{3} \rho_\mathrm{r} \theta_\mathrm{r} + \rho_\mathrm{m} \theta_\mathrm{m} \ll 1~,
\end{align}
and the normalisation condition $\eta = {\cal R} = 1$, one finds that for adiabatic ICs the growing mode for $\delta_\mathrm{m}$  -- during matter and radiation domination and on super-Hubble scales -- is approximately of the form\footnote{Since when $a \to f a$ also $k \to f k$, this result is invariant under a rescaling of the scale factor $a(t)$.}
\begin{equation}\label{eq:app:deltam}
	\frac{-\delta_\mathrm{m}(\tilde{k},\tau)}{\tilde{k}^2} \simeq \frac{2}{5} \frac{a/a_0^3}{\Omega_\mathrm{m} H_0^2}
\end{equation}
with $a_0$ being the scale factor today. A more precise approximation is discussed in \cref{app:omegak_approx}, where we also discuss the impact of curvature. To eliminate the scale factor (which is not an input of the network in our current implementation, but could be in the future), we adopt the well-known expression for the scale factor during radiation and matter domination,$a(\tau)/a_0 = \frac{1}{4} (\Omega_m a_0^2 H_0^2) \tau^2\left[1 + 2 (\sqrt{2}+1) \frac{\tau_\mathrm{eq}}{\tau}\right]$. Taking then some arbitrary intermediate time $\tau_\mathrm{md} = \tau(z_\mathrm{md}=50)$ during matter domination, we know that
\begin{equation}
\delta_\mathrm{m}(\tilde{k},\tau_\mathrm{md}) \simeq -\frac{(\tilde{k} \tau_\mathrm{md})^2}{10} \left[ 1+2(\sqrt{2}+1) \frac{\tau_\mathrm{eq}}{\tau_\mathrm{md}} \right]~.
\end{equation}
At later times, the $\Lambda$ or Dark Energy begins to suppress this growth even on super-Hubble scales. Then the  growth is captured by the scale-independent growth factor $D(\tau)$. This leads to a factor of $D(\tau)/D(\tau_\mathrm{md})$ between $\tau_\mathrm{md}$ and a later redshift. Since $D(\tau) \propto a(\tau)$ at high redshift, we can use this approximation for all relevant redshifts.\footnote{At very high redshift $D/a$ is not constant during radiation domination, but this $\sim 20\%$ difference in amplitude is well absorbed by the prediction of the network [N7].}

Putting everything together, we can use the approximation
\begin{equation}
  S_x(k,\tau) \simeq - T^\mathrm{HE}_x(k,z(\tau)) \, \frac{(k \tau_\mathrm{md})^2}{10}\left[1+2(\sqrt{2}+1)\frac{\tau_\mathrm{eq}}{\tau_\mathrm{md}}\right] \frac{D(\tau)}{D(\tau_\mathrm{md})} \quad \mathrm{for}~~x\in \{\mathrm{m}, \mathrm{cb}\}~.
  \end{equation}
 Since $\tau_\mathrm{eq}$ and $D(\tau)$ are computed by \class for each cosmology in the background module, this approximation can be implemented in \classnet. In this work, for simplicity, we used a slightly simplified version, 
\begin{equation}
  S_x(k,\tau) \simeq -T^\mathrm{HE}_x(k,z(\tau)) \, \frac{(k \tau_\mathrm{md})^2}{7.8}\frac{D(\tau)}{D(\tau_\mathrm{md})} \quad \mathrm{with} \quad \tau(z_\mathrm{md}=50)~, \label{eq:approxx}
  \end{equation}
obtained by fixing $\tau_\mathrm{eq}$ to its Planck best-fit value. We actually ask the network to predict corrections between the true and approximate ratios $S_x(k,\tau) / D(\tau)$, which depend only weakly on time, and we multiply the final prediction by $D(\tau)$.  
 
 Finally, for metric fluctuations, the Einstein equation gives two usueful relations in the sub-Hubble limit: $\phi=\psi$ and the Poisson equation $-k^2 \phi = \frac{3}{2} a^2 H^2 \Omega_\mathrm{m} \delta_\mathrm{m}$. With the gauge-invariant expression of $\delta_\mathrm{m}$ used in \class and \classnet, this relation remains true even on super-Hubble scales, such that we can always use
 \begin{equation}
  S_{\phi+\psi}(k,\tau) \simeq T^\mathrm{HE}_\mathrm{m}(k,z(\tau)) \, \left\{ \frac{3 a^2(\tau) H^2(\tau) \Omega_\mathrm{m}(\tau)  \tau_\mathrm{md}^2}{10}\left[1+2(\sqrt{2}+1)\frac{\tau_\mathrm{eq}}{\tau_\mathrm{md}}\right]  \frac{D(\tau)}{D(\tau_\mathrm{md})}\right\}~. \label{eq:approxpp}
 \end{equation}  
 The factor between curly brackets is independent of $k$, and also approximately independent of time deep inside the matter dominated regime. We actually let the network predict this factor, which means that the role of the network [N7]$_{\phi+\psi}$ is to predict the ratio $S_{\phi+\psi}(k,\tau) /T^\mathrm{HE}_\mathrm{m}(k,z(\tau))$.
 
\subsection{Approximations with curvature, dark radiation and dynamical dark energy}

The Hu \& Eisenstein transfer functions are designed to predict observables deep inside the sub-Hubble limit but, for CMB calculations, $S_{\phi+\psi}$ is needed up to scales approaching the Hubble scale. The HE transfer functions account for the effect of a non-zero spatial curvature parameter $K$ on $\delta_\mathrm{m,cb}$ up to factors of $s_2=\left( 1 - 3 K/k^2\right)^{1/2}=\left( 1 - 3 \Omega_k a_0^2 H_0^2 /k^2\right)^{1/2}$ that become irrelevant deep in the sub-Hubble limit. To take them into account, we need to multiply the approximation (\ref{eq:approxx}) by a factor $s_2^2$.  The same factor appears in the Poisson equation and cancels out, such that (\ref{eq:approxpp}) does not need to be modified in presence of curvature.

The HE transfer functions assume no relativistic relics beyond photons and ordinary neutrinos. To account for the effect of $N_\mathrm{eff}$ on the source functions $S_{\phi+\psi}$, $S_{\delta_\mathrm{m}}$, $S_{\delta_\mathrm{cb}}$, we use the existence of a well-known approximate degeneracy at the level of such observables between variations of $N_\mathrm{eff}$ and $H_0$. In particular, any $\Lambda$CDM+$\Omega_k$+$N_\mathrm{eff}$ model with $\Delta N_\mathrm{eff}\neq 0$ is nearly degenerate with another model featuring $\Delta N_\mathrm{eff} = 0$, a different value of $H_0$ and the same value of fractional densities $\Omega_x$. This degeneracy is explained in more details in \cite{Rossi:2014nea}, section IV.C and equations (6-8).  We use it to generalise our approximations to the case of the $\Lambda$CDM+$\Omega_k$+$N_\mathrm{eff}$ model.

Finally, the effect of dynamical dark energy on $S_{\phi+\psi}$, $S_{\delta_\mathrm{m}}$, $S_{\delta_\mathrm{cb}}$ is essentially contained in the expression of the growth factor $D(\tau)$. Since \classnet reads this factor from the \class background module for each cosmology, our approximations do cover the most general case studied in this paper.

\subsection{Large-scale approximation scheme}\label{app:omegak_approx}

The full derivation of the evolution of $\delta_\mathrm{m}$ in the synchronous gauge following the Einstein equations~\ref{eq:app:einsteinh} actually gives a differential equation in matter/radiation domination. Using the Meszaros variable $y = a/a_\mathrm{eq}$ the Hubble function can be put into a nice form \\$\mathcal{H}^2 = (aH)^2 = \frac{\Omega_m H_0^2 a_0^3}{a_\mathrm{eq}} \cdot (y^{-1} + y^{-2})$\,. Transforming the derivative $\delta_m' = \mathcal{H} \mathrm{d}\delta_m/\mathrm{d}\ln y$ allows us to rephrase the whole differential equation in terms of mostly dimensionless factors\footnote{Note that this term is invariant to a re-scaling of the scale factor $a$, since any transformation $a \to f a$ also rescales $k \to f k$.}
\begin{equation}
	-\frac{\mathrm{d}\Delta}{\mathrm{d}\ln y} = \frac{3}{2} \frac{y+\frac{4}{3}}{1+y} \Delta + \frac{a_\mathrm{eq} a_0^{-3}}{\Omega_m H_0^2} \frac{1}{y^{-1}+y^{-2}}~,
\end{equation}
\enlargethispage*{2\baselineskip}
with $\Delta = \delta_m/k^2$\,. This first order differential equation can trivially be solved and one obtains
\begin{equation}
	-\delta_m(a)/k^2 = A \frac{\sqrt{1+y}}{y^2} + \frac{a_\mathrm{eq} a_0^{-3}}{\Omega_m H_0^2} f(y) \qquad \qquad \mathrm{with}~~ f(y) = \frac{2 y^3 - 4 y^2 + 16 y +32}{5 y^2}~.
\end{equation}
One can quickly determine that at late times the homogeneous solution only grows as $y^{-3/2}$, while the particular solution grows as $y^1$, quickly dominating the overall term. However, if one wants to be very strict, one can also neglect terms of order $y^0$ or $y^{-1}$ and expand $f(y)$ to first order, obtaining $f(y) \approx 2/5 \cdot y$ in this limit. In this limit, one then finds\\ $-\delta_m(a)/k^2 \approx 2/5 \cdot (a/a_0^3)/(\Omega_m H_0^2)$ as in \cref{eq:app:deltam}. However, it is trivial to keep the non-leading terms of $f(y)$ as well, vastly improving the accuracy of the approximation from around $3\%$ to around $0.4\%$. Note, that this is true even when neglecting the homogeneous solution. It is determined from the initial conditions (e.g. $A = \left[\delta^\mathrm{ini}_m/k^2 - f(y_\mathrm{ini}) a_\mathrm{eq} a_0^{-3}/(\Omega_m H_0^2)\right] \frac{y_\mathrm{ini}^2}{\sqrt{1+y_\mathrm{ini}}}$), but does not contribute significantly. The initial condition for $\delta_m^\mathrm{ini}$ is usually taken to be $-k^2 \tau_\mathrm{ini}^2/4 \cdot \left[1-\frac{4\tau_\mathrm{ini}/\tau_{eq}}{10 (1+\sqrt{2})}\right]$, but is of little relevance here.

The addition of curvature turns out to be a rather simple modification, since all the initial conditions and equations of motion are unaffected except for the Einstein \cref{eq:app:einsteinh}, where a pre-factor of $(1-3K/k^2)$ appears in front of the $k^2 \eta$ term, leading directly to a factor of $(1-3K/k^2)$ for the dominant particular solution of the differential equation. Overall, the curvature can thus be dealt with by simply multiplying the solution with $(1-3K/k^2)$ where $K = -\Omega_k H_0^2$.

However, in addition to changing the solution of the $\delta_m$ for $k \to 0$, the curvature also introduces a minimal $k$ in both open $K < 0$ and closed $K > 0$ universes. In an open universe, one has $k_\mathrm{min} = \sqrt{-K}$, while in a closed universe one has $k_\mathrm{min} = \sqrt{(8-m) K}$ with $m=0$ for scalar perturbations. As such, the network cannot rely on always receiving the same $k_\mathrm{min}$ for every cosmology. However, dynamically adapting the $k$-grid of the network during evaluation is not possible as the number of the nodes is fixed in the architecture. As such, we choose to instead predict always the full $N_k$ nodes, and simply discard those nodes $i$ with $k_i < k_\mathrm{min}$\, and only compute the loss on these nodes. Additionally, for the final output the three nodes, one slightly below $k_\mathrm{min}$ and two slightly above $k_\mathrm{min}$, are fitted quadratically in order to interpolate at $k_\mathrm{min}$\,. 
\section{Triangle Plots}
\label{ssec:triangles}
In this section we show the remaining triangle plots for the various extensions of the $\Lambda$CDM model investigated in the main text. We always compare the $1 \sigma$ and $2 \sigma$ contours for either \classnet or \classfull, which show in general an excellent agreement.

\begin{figure}[H]
	\centering
	\begin{subfigure}{0.95\linewidth}
		\includegraphics[width=0.5\linewidth]{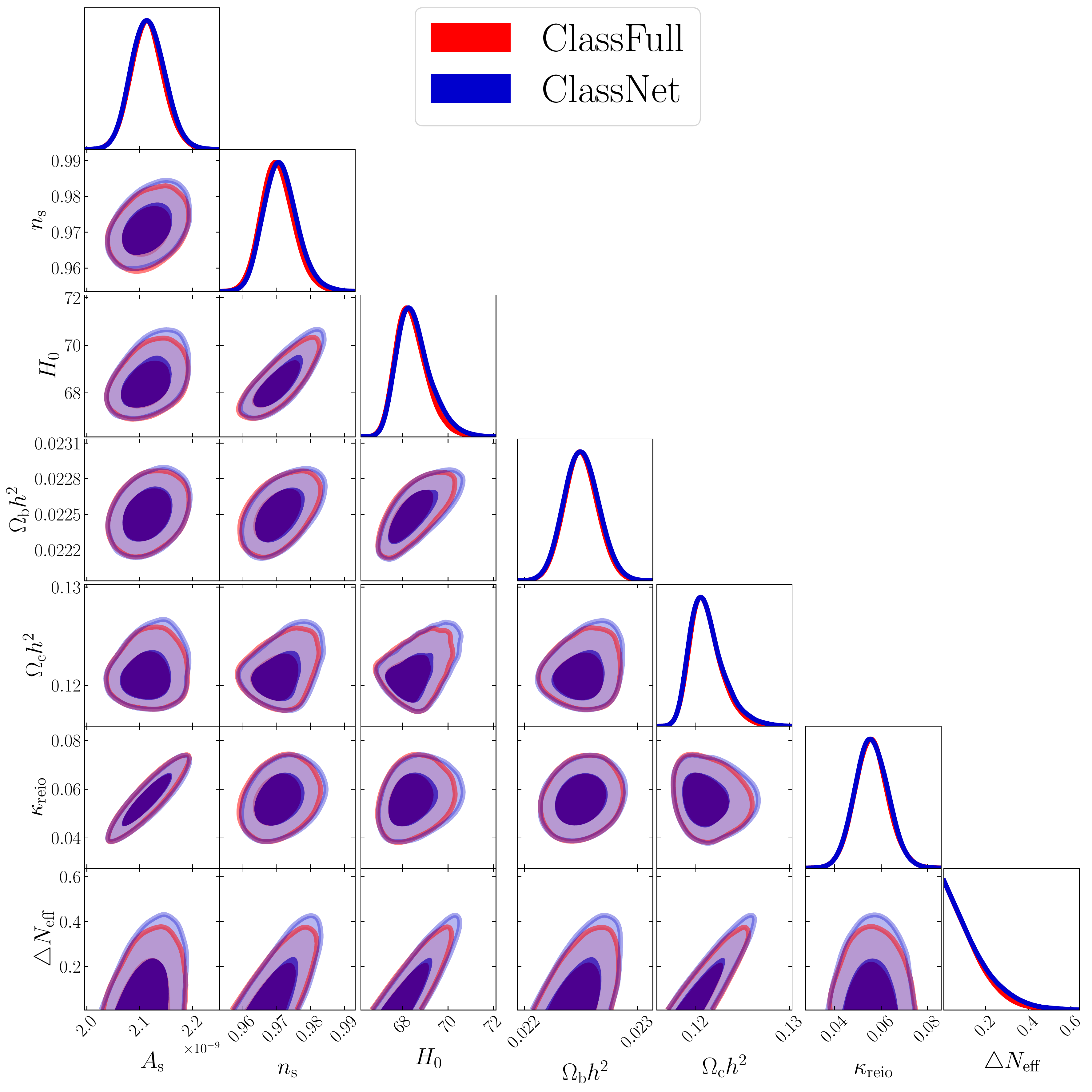}
		\includegraphics[width=0.5\linewidth]{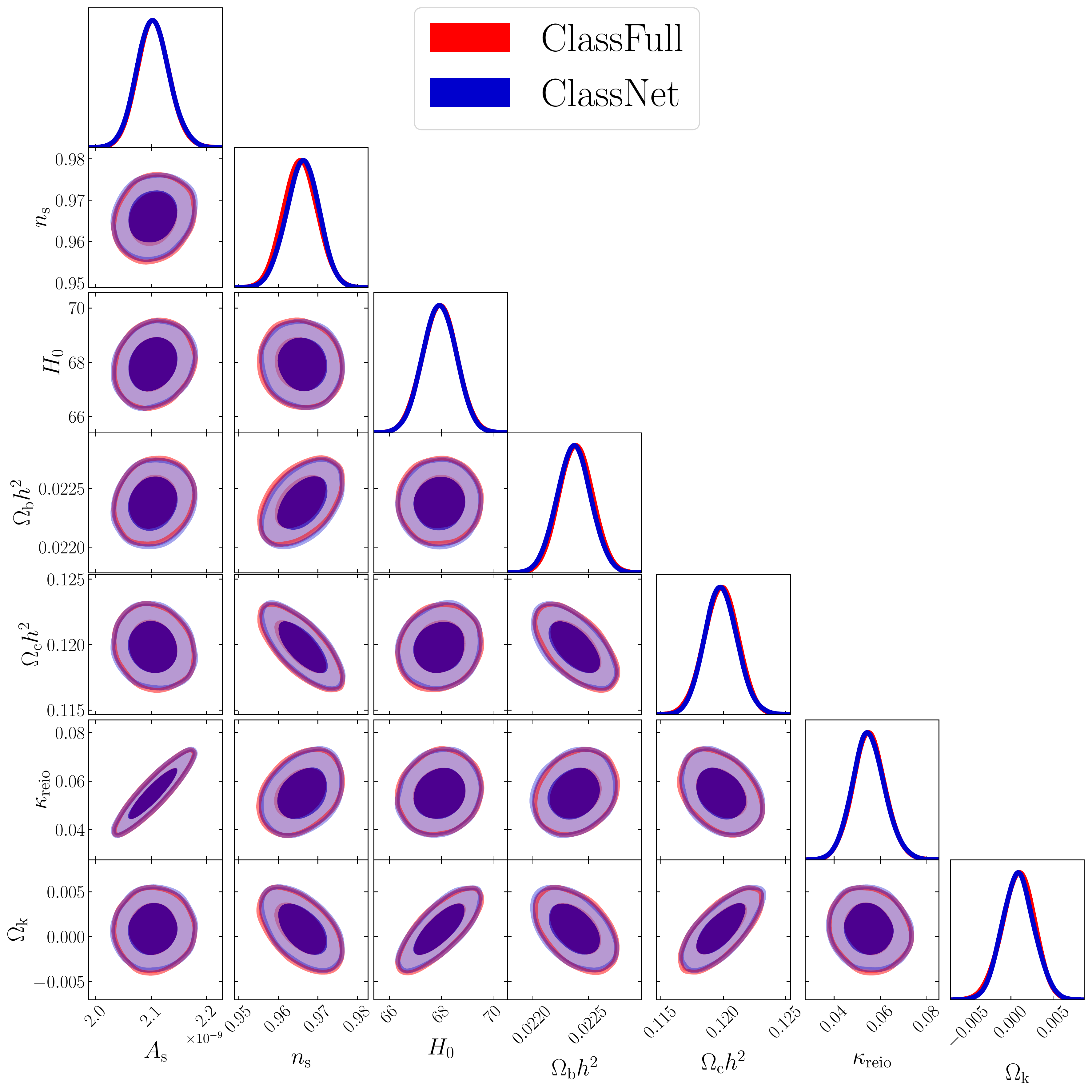}
	\end{subfigure}
	
	\caption{1 and 2 $\sigma$ contours in the case of extended $\Lambda$CDM with the Planck+BAO likelihoods. \textbf{Left:} Extended model with $\Delta N_\mathrm{eff}$\,. \textbf{Right:} Extended model with $\Omega_k$\,.}
	\label{fig:parameter_estimation_Nur}
\end{figure}

\begin{figure}[H]
	\centering
	\begin{subfigure}{0.95\linewidth}
		\includegraphics[width=0.5\linewidth]{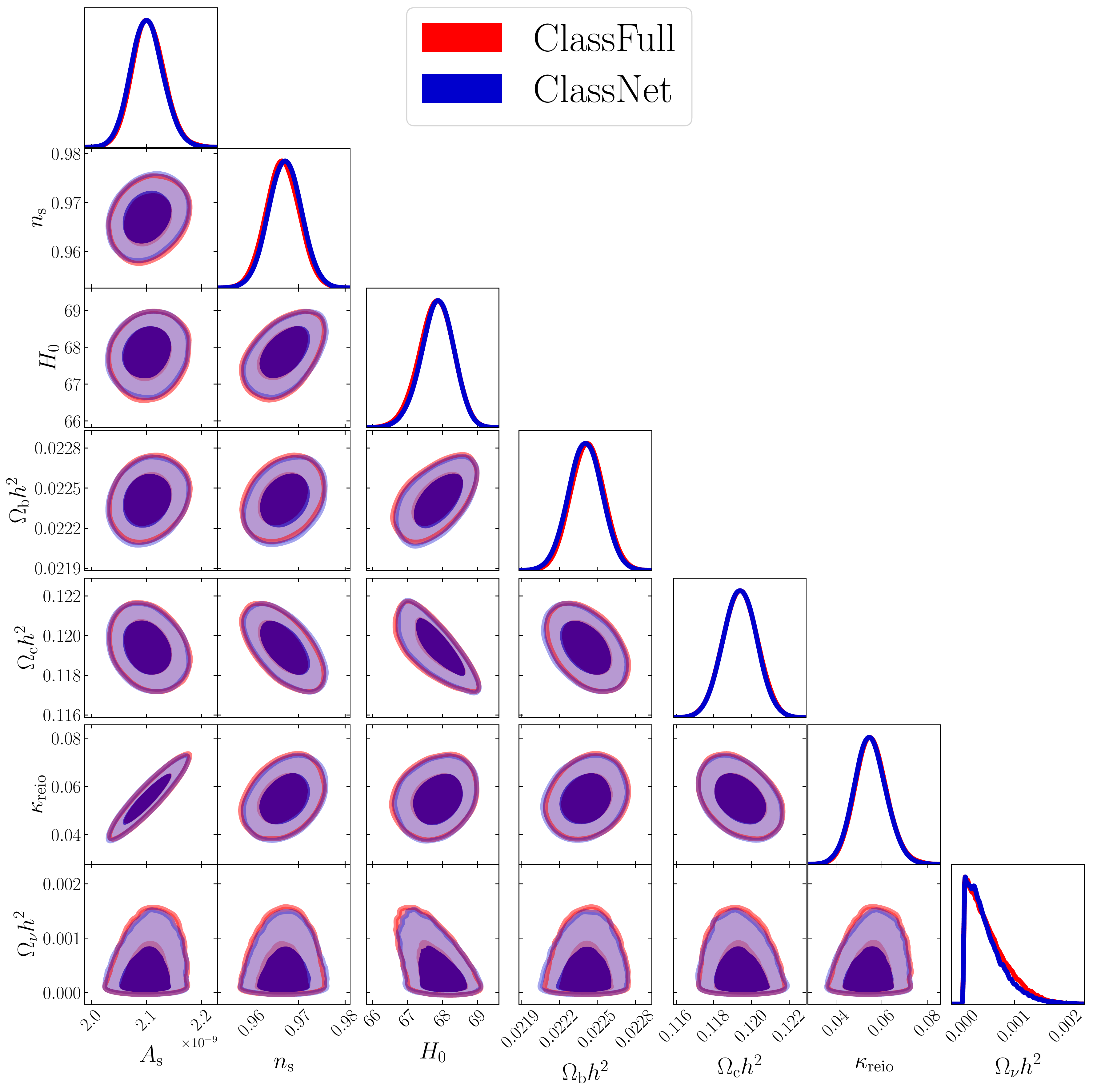}
		\includegraphics[width=0.5\linewidth]{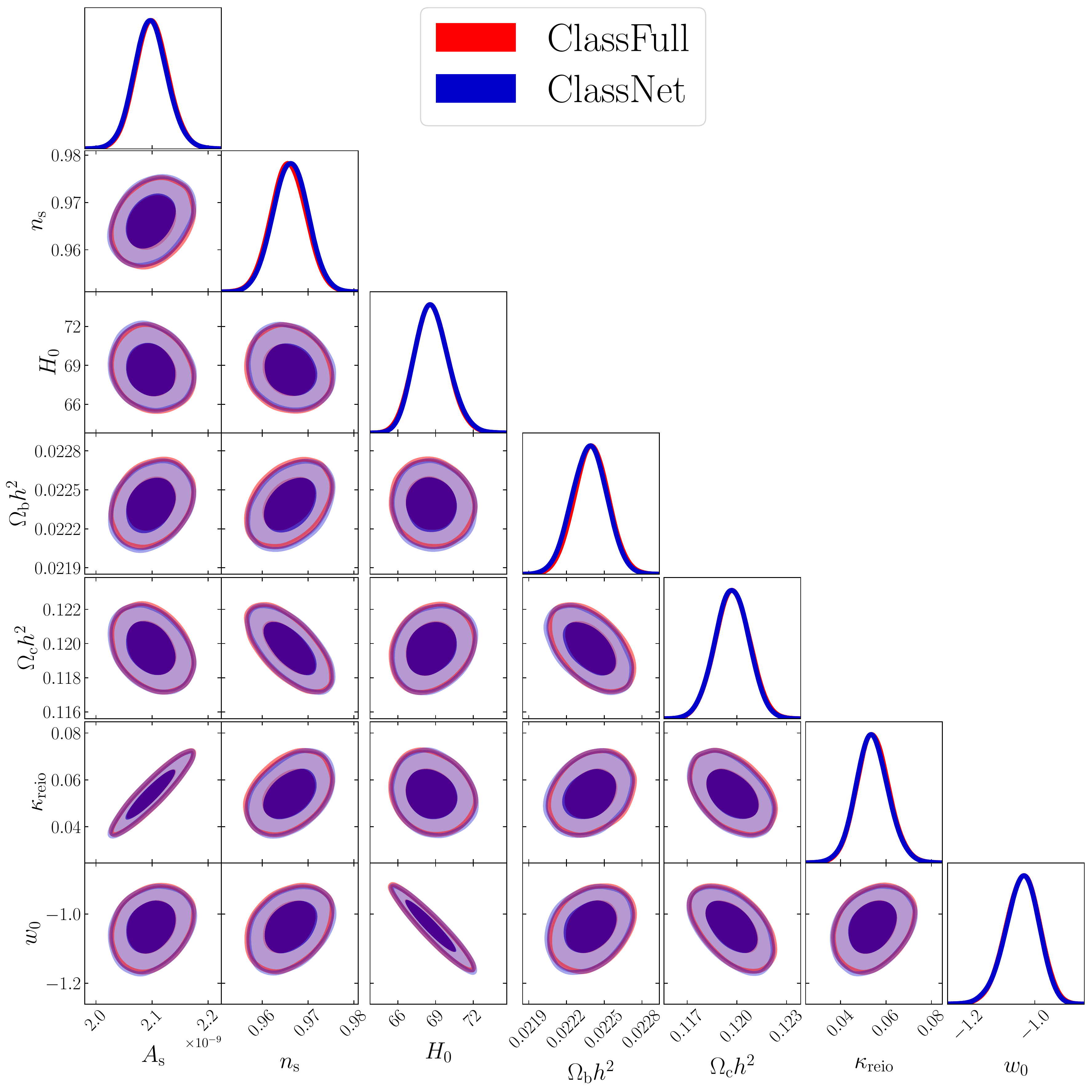}
	\end{subfigure}
	
	\caption{1 and 2 $\sigma$ contours in the case of extended $\Lambda$CDM with the Planck+BAO likelihoods. \textbf{Left:} Extended model with $\Omega_\nu h^2$\,. \textbf{Right:} Extended model with $w_0$\,.}
	\label{fig:parameter_estimation_Omnu}
\end{figure}

\begin{figure}[H]
	\centering
	\begin{subfigure}{0.6\linewidth}
		\includegraphics[width=\linewidth]{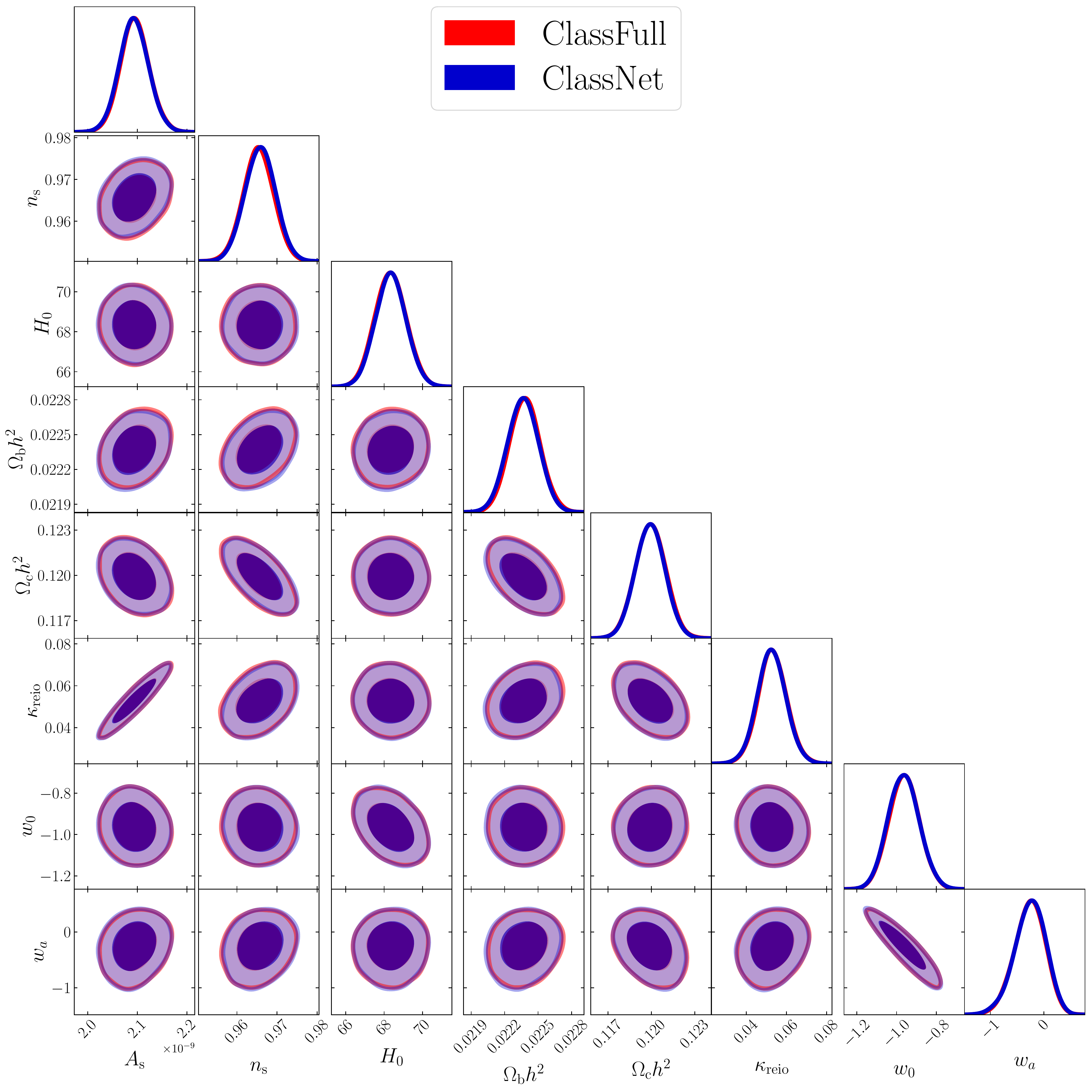}
	\end{subfigure}
	
	\caption{1 and 2 $\sigma$ contours in the case of $\Lambda$CDM + $w_0+w_a$\,,  with the Planck+BAO+SN likelihoods.}
	\label{fig:parameter_estimation_w0wa}
\end{figure}
\section{Origin of parameter bias} \label{fig:sysdev}

We have seen in \cref{sec:parameter_estimation} that \classnet reaches sufficient accuracy for parameter estimation from current data. However, there are small biases on certain cosmological parameters. We found that the bias on the mean and/or error exceeds the MCMC sampling variance mainly for $n_s$, $\Omega_\mathrm{b} h^2$ and $\Delta N_\mathrm{eff}$ -- up to a level of 10\% to 17\%, which is acceptable, but still worth investigating in view of future progress. To a lesser extent, the means and/or errors for $H_0$, $\Omega_\mathrm{m} h^2$ and $\Omega_\nu h^2$ are also slightly biased, but at an even smaller level, depending on which data set is used. For other parameters, the bias is below the MCMC sampling variance.

Our goal in this appendix is to understand better the origin of these small biases. This hints at what should be our priority in future versions of the \classnet networks.

\subsection{CMB spectrum residuals}
\label{sec:resi_cor}

\enlargethispage*{2\baselineskip}
If the residuals of the \classnet $C_\ell$ spectra (compared to the \classfull calculations) had completely random shapes for each given cosmological model and were randomly distributed over parameter space, we would expect that network errors get averaged out during parameter estimation, first, when summing over $\ell$ values in each likelihood, and second, when moving around in parameter space. Conversely, a bias on the mean and/or error of a cosmological parameter can appear in two cases:
\begin{description}
\item[a)] When the average shape of the power spectra residuals shows a systematic trend that can be compensated by a shift in a cosmological parameter;\footnote{It has recently been proposed in \cite{Grandon:2022gdr} that such a shift could be compensated by using the known systematic deviations within the validation set by absorbing the shifts within the likelihood underlying the inference.} 
\item[b)] When the shape or amplitude of the residuals has a correlated dependence on the value of a cosmological parameter.
\end{description}

The shape of the residuals for $C_\ell^{TT}$ and $C_\ell^{EE}$ was already displayed in \cref{fig:power_spectra_tt_lines,fig:power_spectra_ee_lines}, colour-coded by the distance to the center of the training domain. However, by looking at these figures, it is very difficult to identify effects of type {\bf a)}, and state, for instance, whether the average shape of the residuals can be compensated by a small shift in $n_s$ or in $\Omega_\mathrm{b}h^2$.

Effects of type {\bf b)} are easier to highlight. For this purpose, it is enough to plot the residuals colour-coded by the value of each cosmological parameter in the basis passed to the networks, that is, ($\Omega_\mathrm{b}h^2$, $\Omega_\mathrm{m}h^2$, $H_0$, $\kappa_\mathrm{reio}$, $\Omega_\nu$, $\Delta N_\mathrm{eff}$, $\Omega_k$, $w_0$, $w_a$). In the following figures, we will plot the residuals corresponding to a uniform sampling of our ellipsoidal training domain, with all previous parameters being varied, while the primordial parameters ($A_s$, $n_s$) are fixed to their Planck best-fit value.

\begin{figure}[H]
	\centering
	\begin{subfigure}{0.45\linewidth}
		\includegraphics[width=\linewidth]{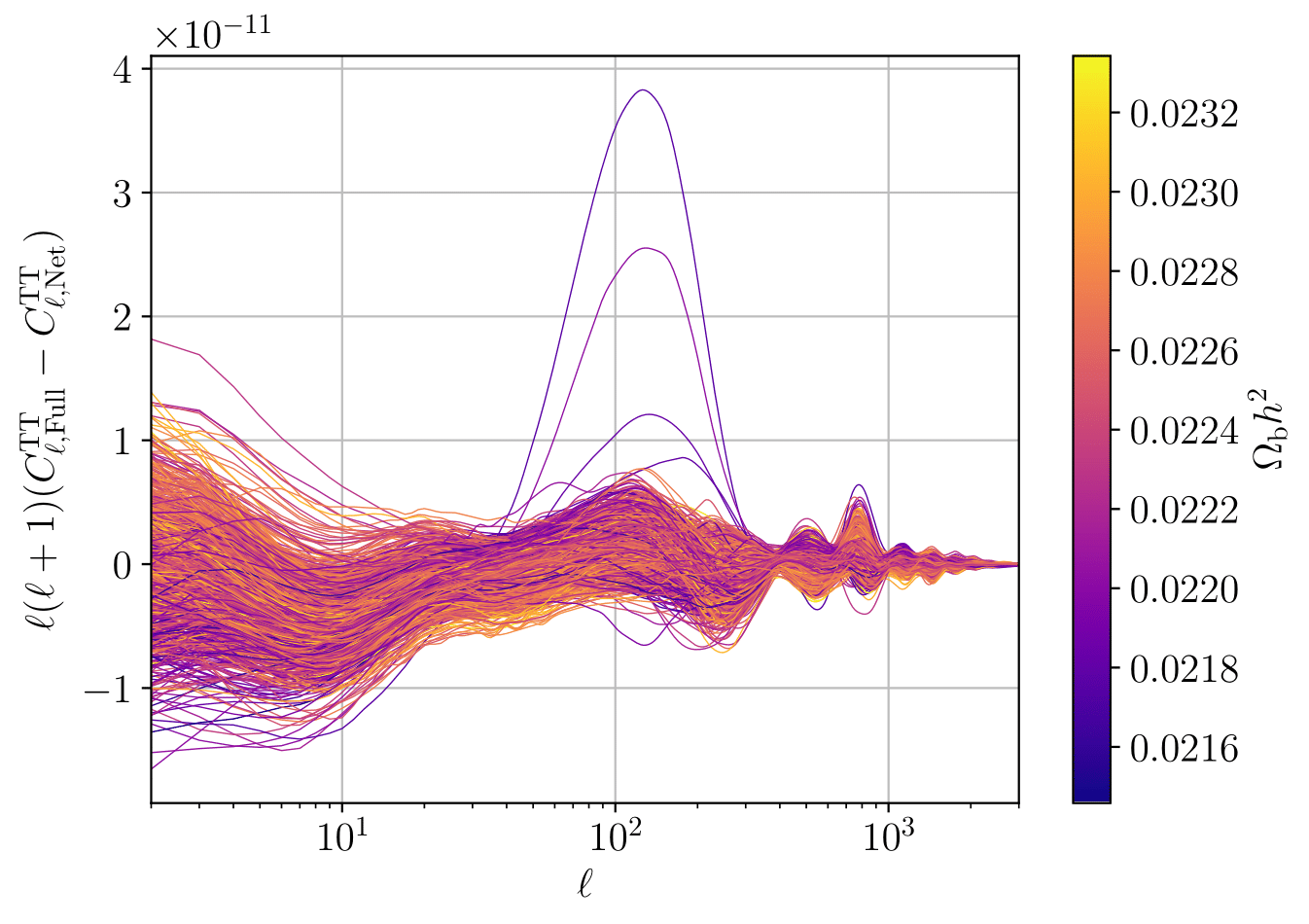}
	\end{subfigure}
	\begin{subfigure}{0.45\linewidth}
		\includegraphics[width=\linewidth]{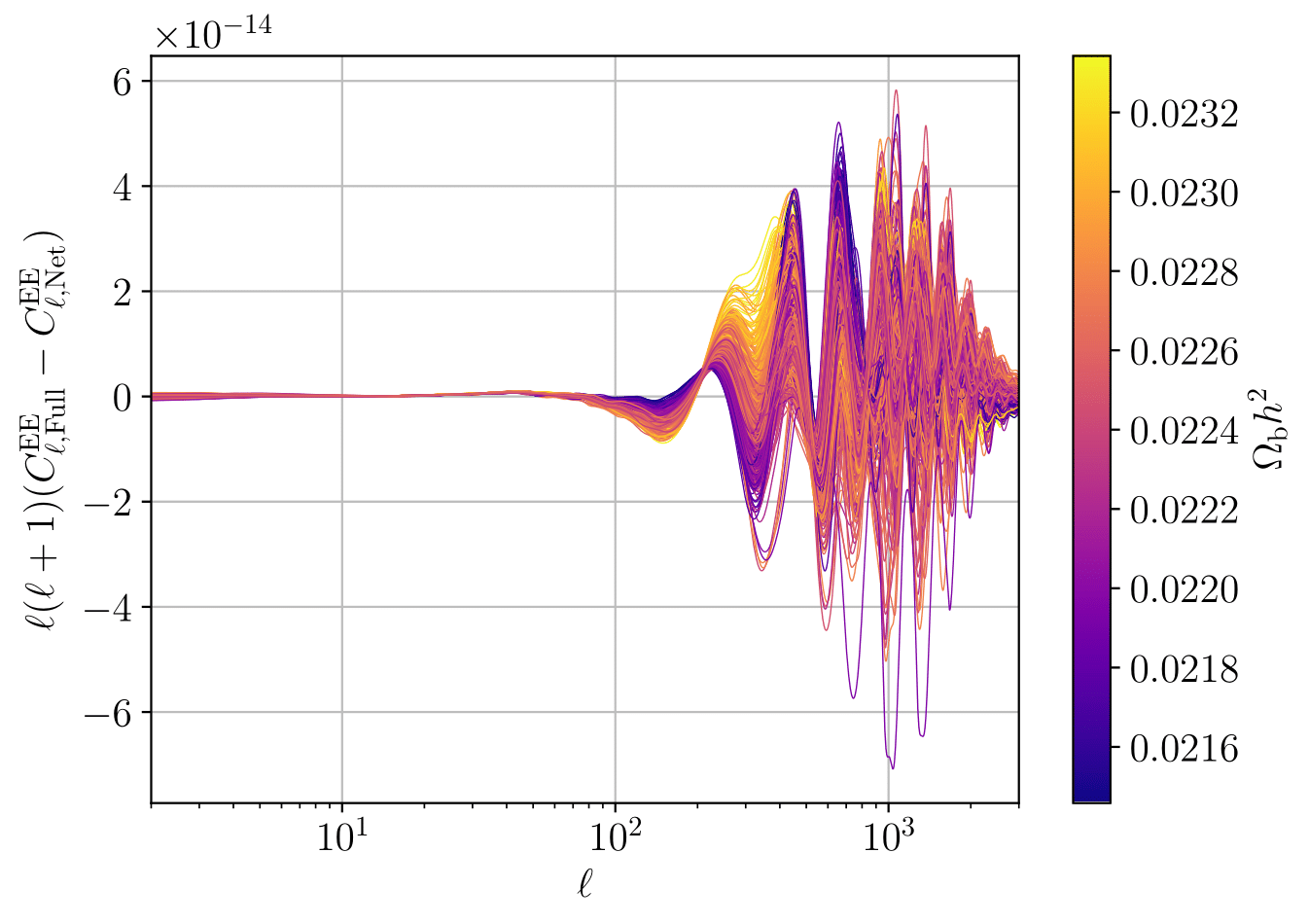}
	\end{subfigure}
	\begin{subfigure}{0.45\linewidth}
		\includegraphics[width=\linewidth]{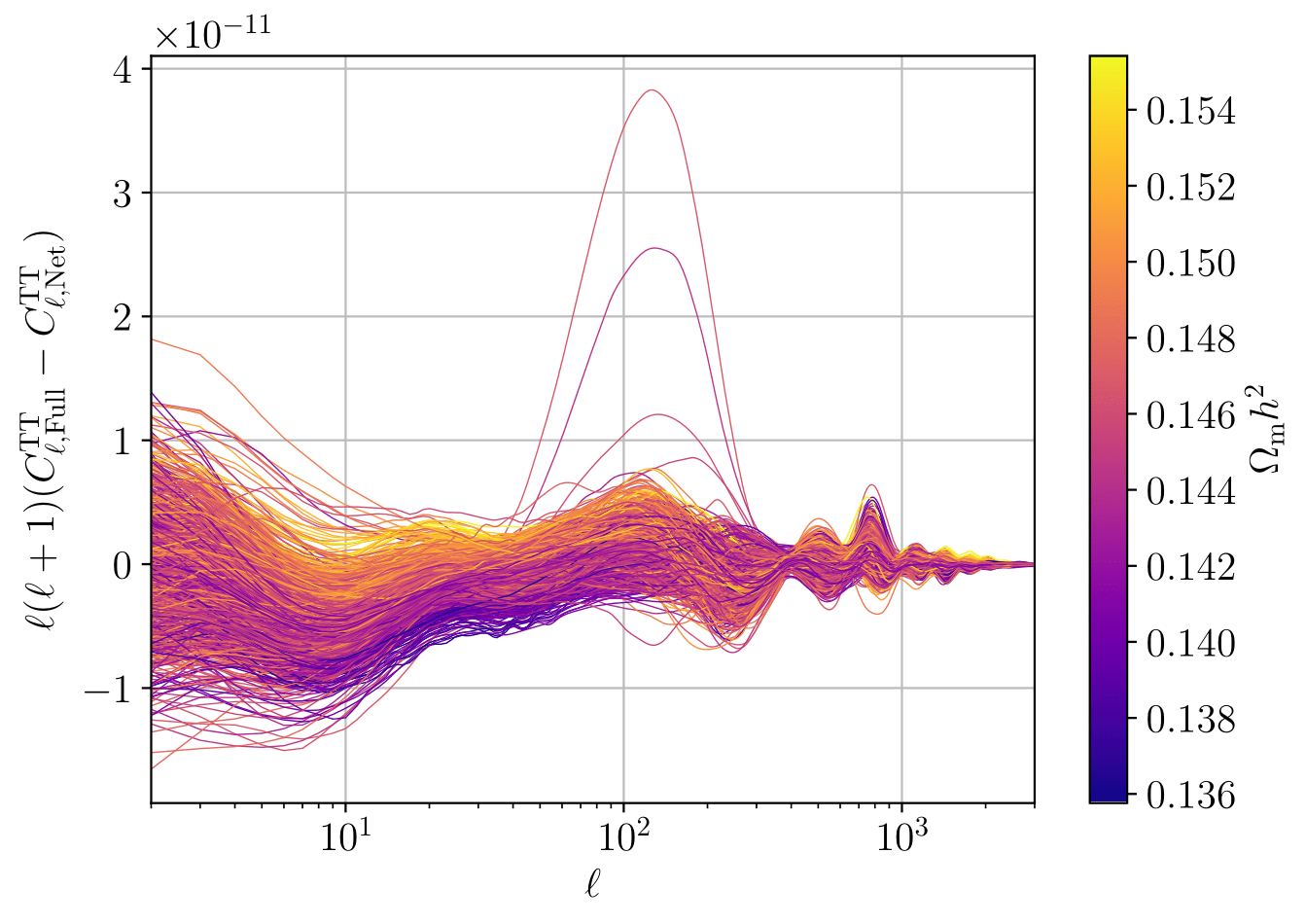}
	\end{subfigure}
	\begin{subfigure}{0.45\linewidth}
		\includegraphics[width=\linewidth]{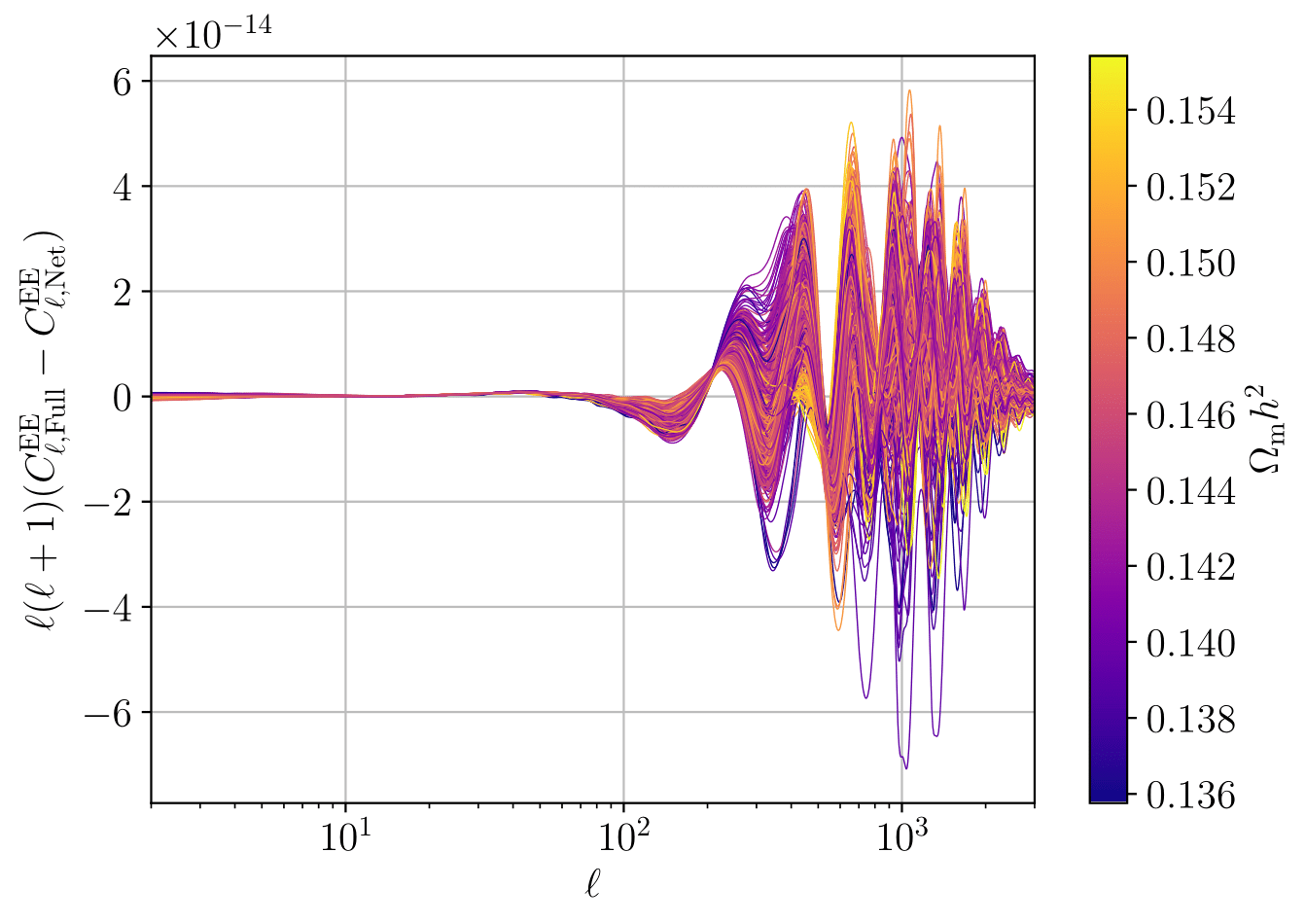}
	\end{subfigure}
	\begin{subfigure}{0.45\linewidth}
		\includegraphics[width=\linewidth]{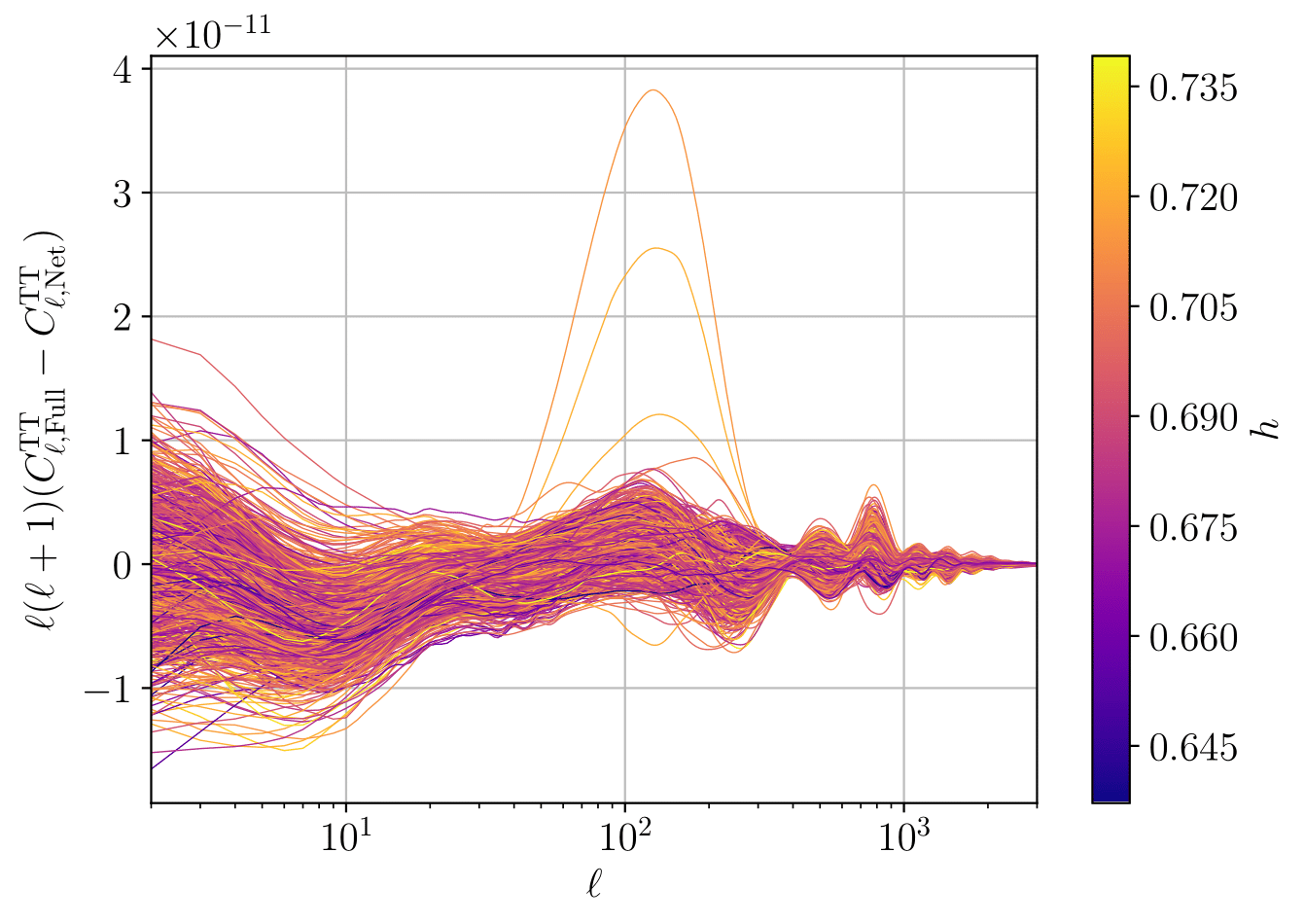}
	\end{subfigure}
	\begin{subfigure}{0.45\linewidth}
		\includegraphics[width=\linewidth]{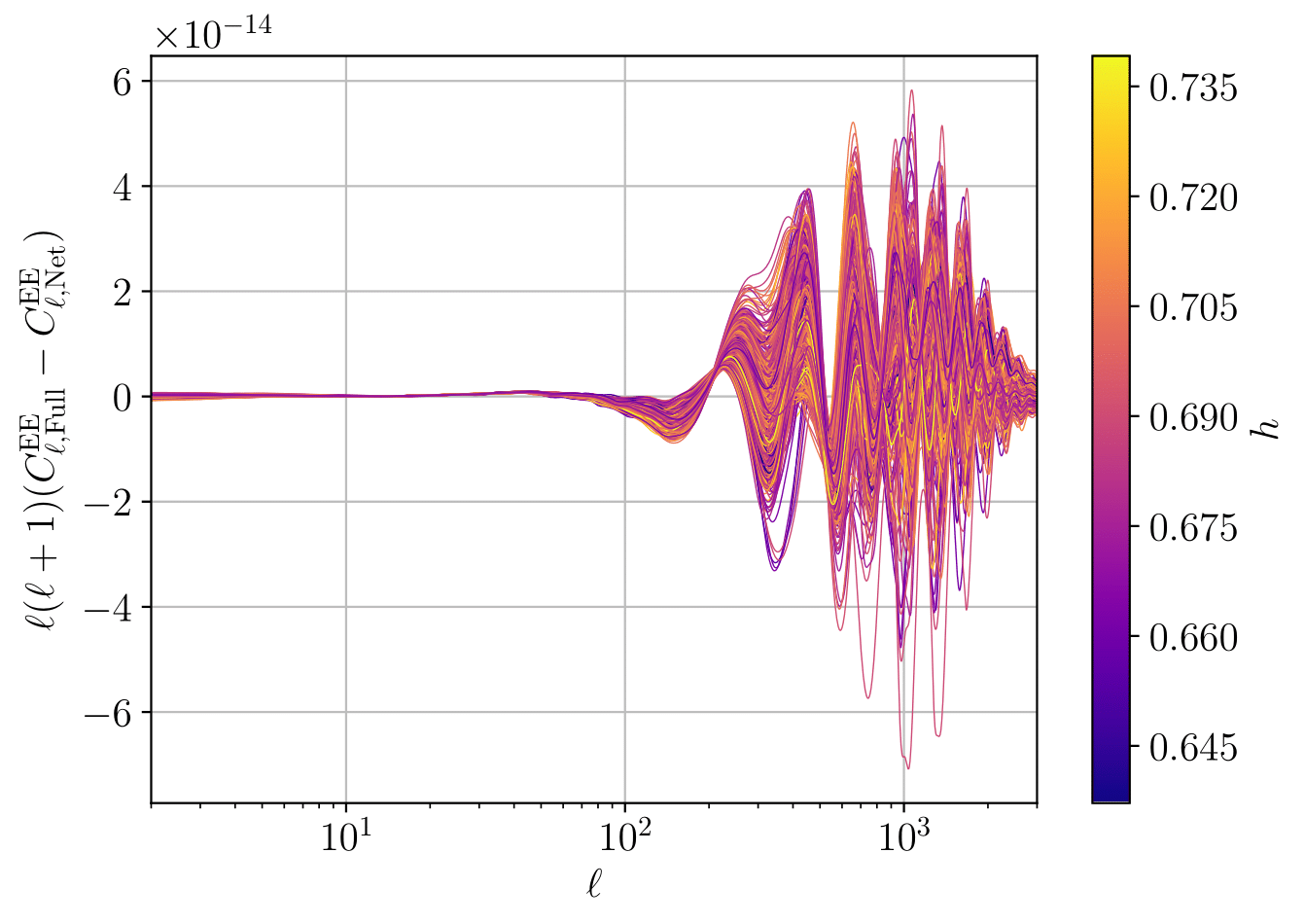}
	\end{subfigure}
	\begin{subfigure}{0.45\linewidth}
		\includegraphics[width=\linewidth]{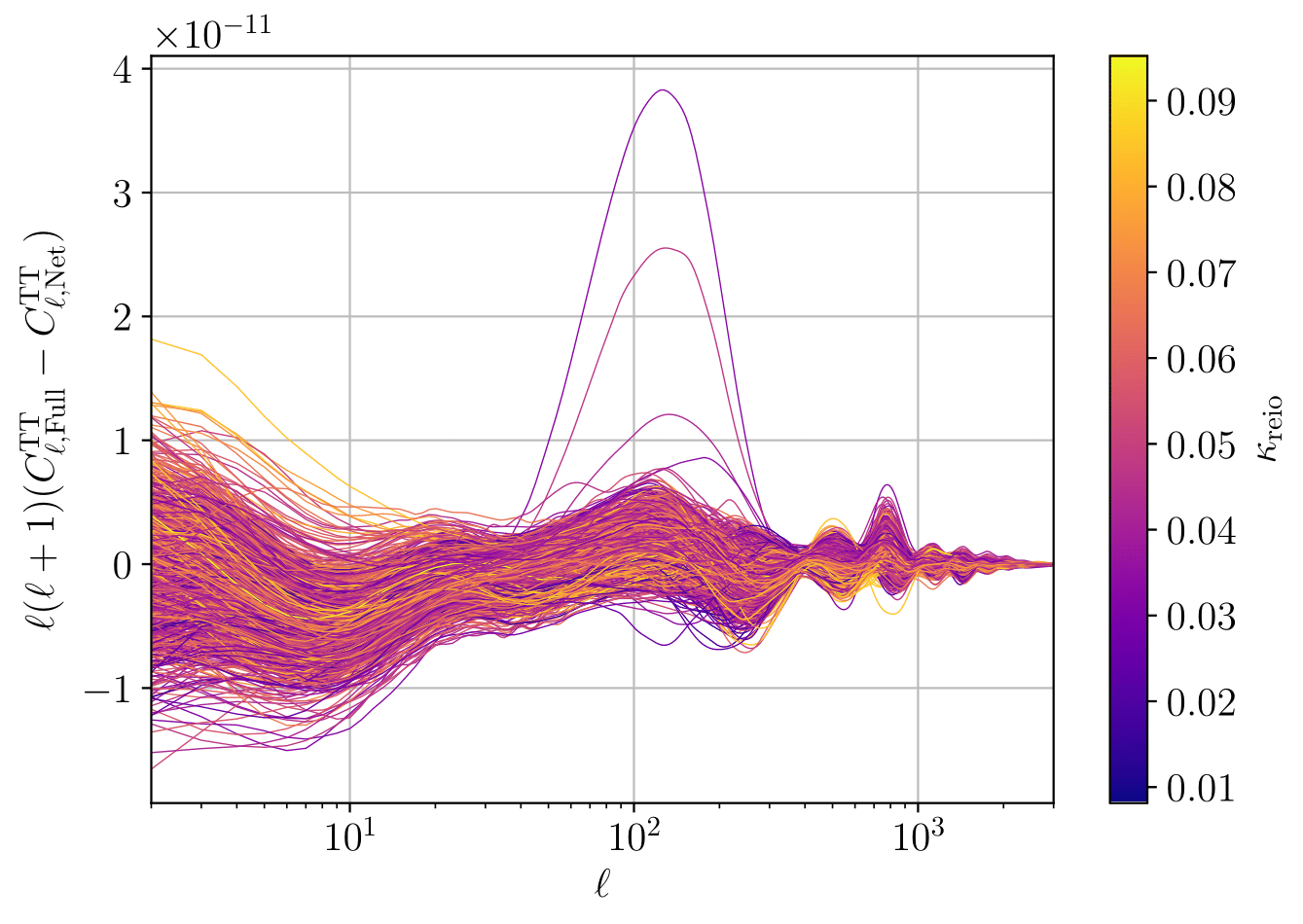}
	\end{subfigure}
	\begin{subfigure}{0.45\linewidth}
		\includegraphics[width=\linewidth]{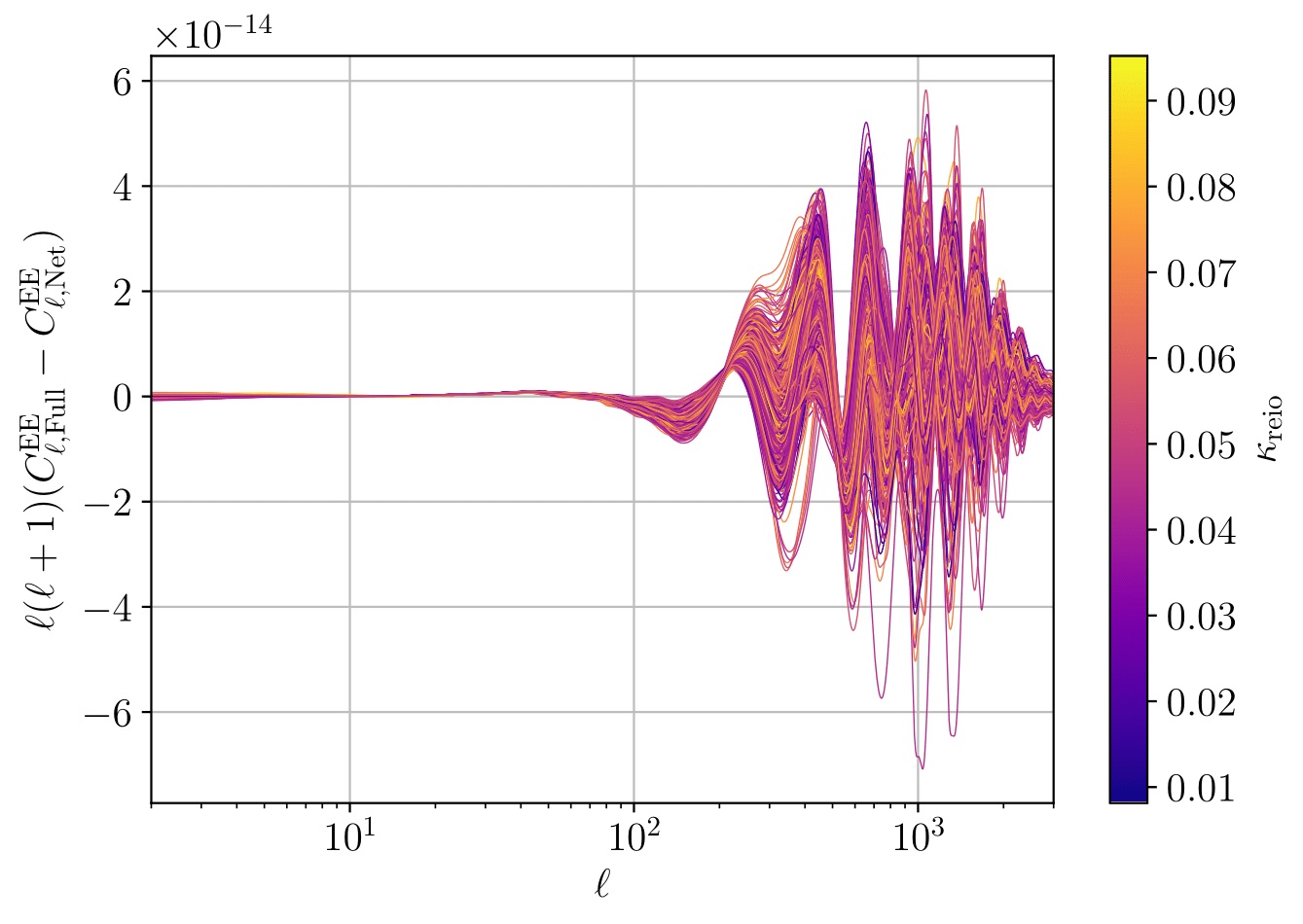}
	\end{subfigure}
	\caption{Absolute deviation between the \classfull and \classnet power spectrum of temperature (left) and polarisation (right) across a set of 994 cosmologies uniformly sampled from the test domain, colour-coded as a function of individual parameters: from top to bottom, $\Omega_\mathrm{b}h^2$, $\Omega_\mathrm{m}h^2$, $h$ and $\kappa_\mathrm{reio}$\,.}
	\label{fig:pardep_lcdm}
\end{figure}

In \cref{fig:pardep_lcdm}, we plot the  $C_\ell^{TT}$ and $C_\ell^{EE}$ residuals color-coded as a function of the value of the $\Lambda$CDM parameters ($\Omega_\mathrm{b}h^2$, $\Omega_\mathrm{m}h^2$, $H_0$, $\kappa_\mathrm{reio}$). For $C_\ell^{TT}$, no clear correlation appears, apart from a dependence of the residuals in the range $20 \leq \ell \leq 100$ on the value of $\Omega_\mathrm{m}h^2$. Since $\Omega_\mathrm{m}$ (or more precisely $\Omega_\mathrm{bc} = \Omega_\mathrm{m}-\Omega_\nu$) determines the value of the redshift of equality, $z_\mathrm{eq}$, and thus the magnitude of the early ISW effect, we conclude that the accuracy of the network [N1]  -- in charge of most of this effect -- could be improved in order to reduce the bias on $\Omega_\mathrm{m} h^2$. For $C_\ell^{EE}$, we see a small correlation between the amplitude and phase of the oscillating part of the residuals with $\Omega_\mathrm{b}h^2$ and $\Omega_\mathrm{m}h^2$. This suggests that the network [N5] lacks a bit of accuracy when estimating the phase and amplitude of the acoustic oscillations in the photon multipoles $G_0$, $G_2$, $F_2$, which depend on the photon-to-baryon ratio $R=\frac{3 \rho_b}{4 \rho_\gamma}$ and on the redshift of equality $z_\mathrm{eq}$, and thus, on $\Omega_\mathrm{b}h^2$ and $\Omega_\mathrm{m}h^2$.

\begin{figure}[t]
	\centering
	\begin{subfigure}{0.45\linewidth}
		\includegraphics[width=\linewidth]{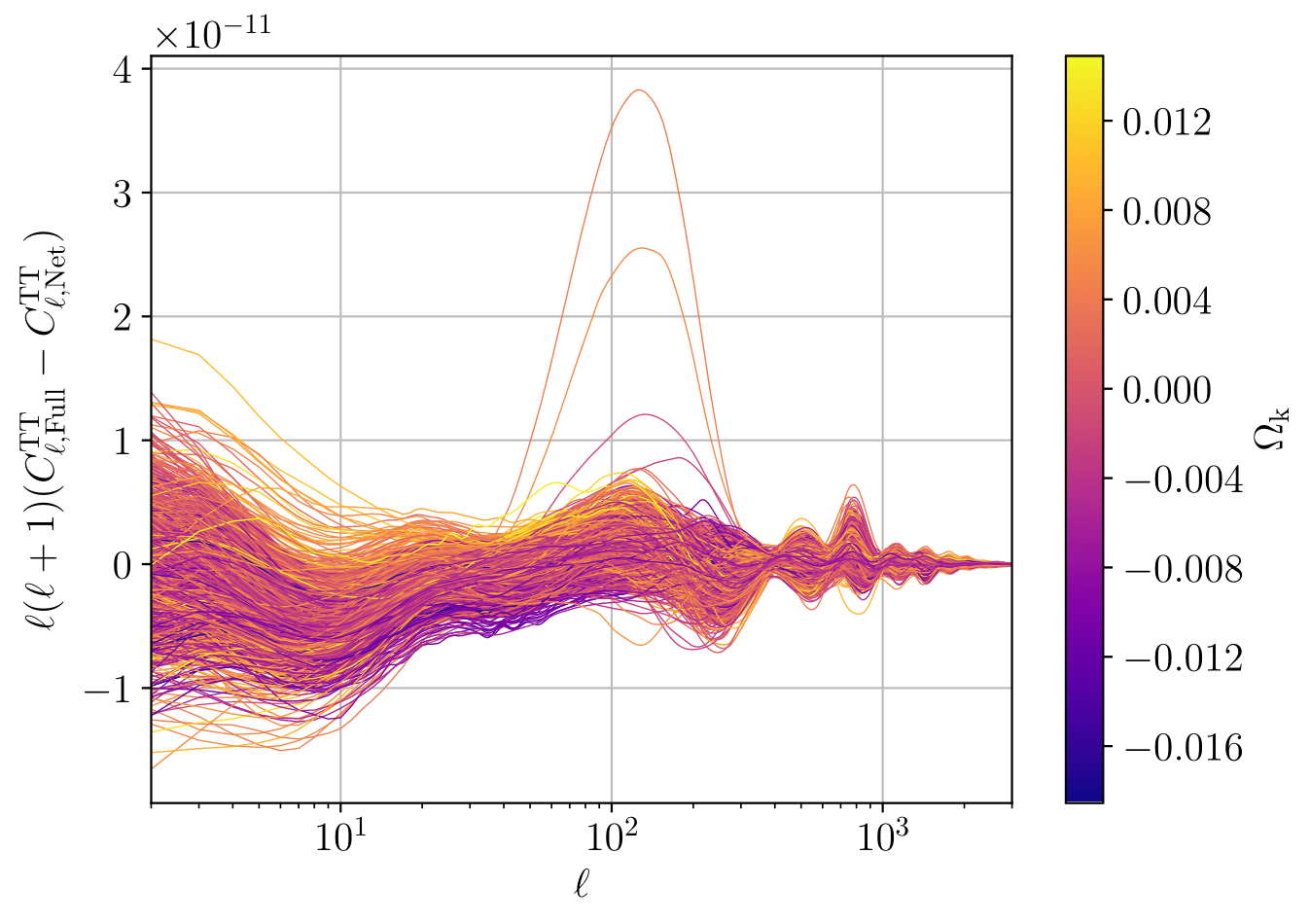}
	\end{subfigure}
	\begin{subfigure}{0.45\linewidth}
		\includegraphics[width=\linewidth]{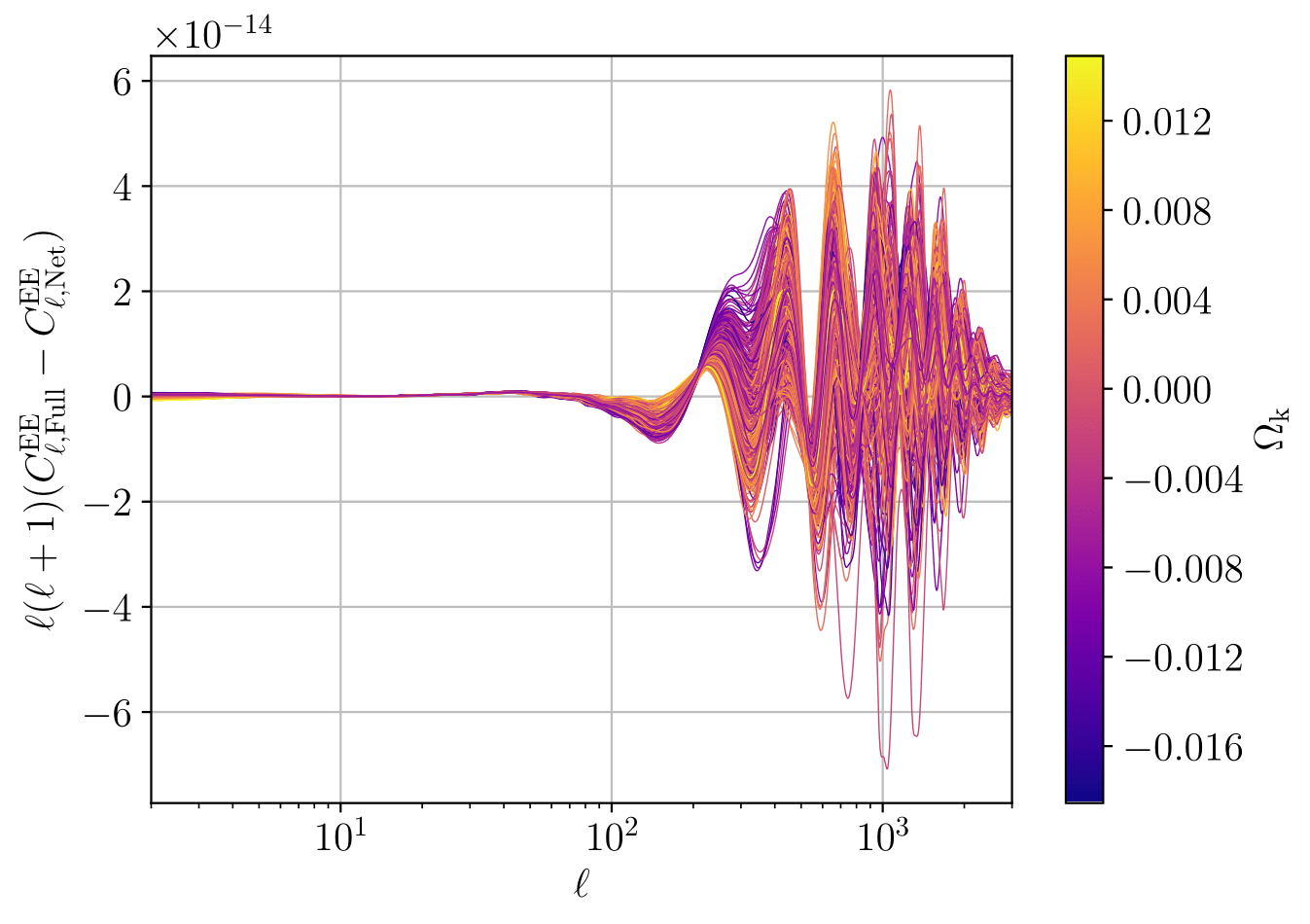}
	\end{subfigure}
	\begin{subfigure}{0.45\linewidth}
		\includegraphics[width=\linewidth]{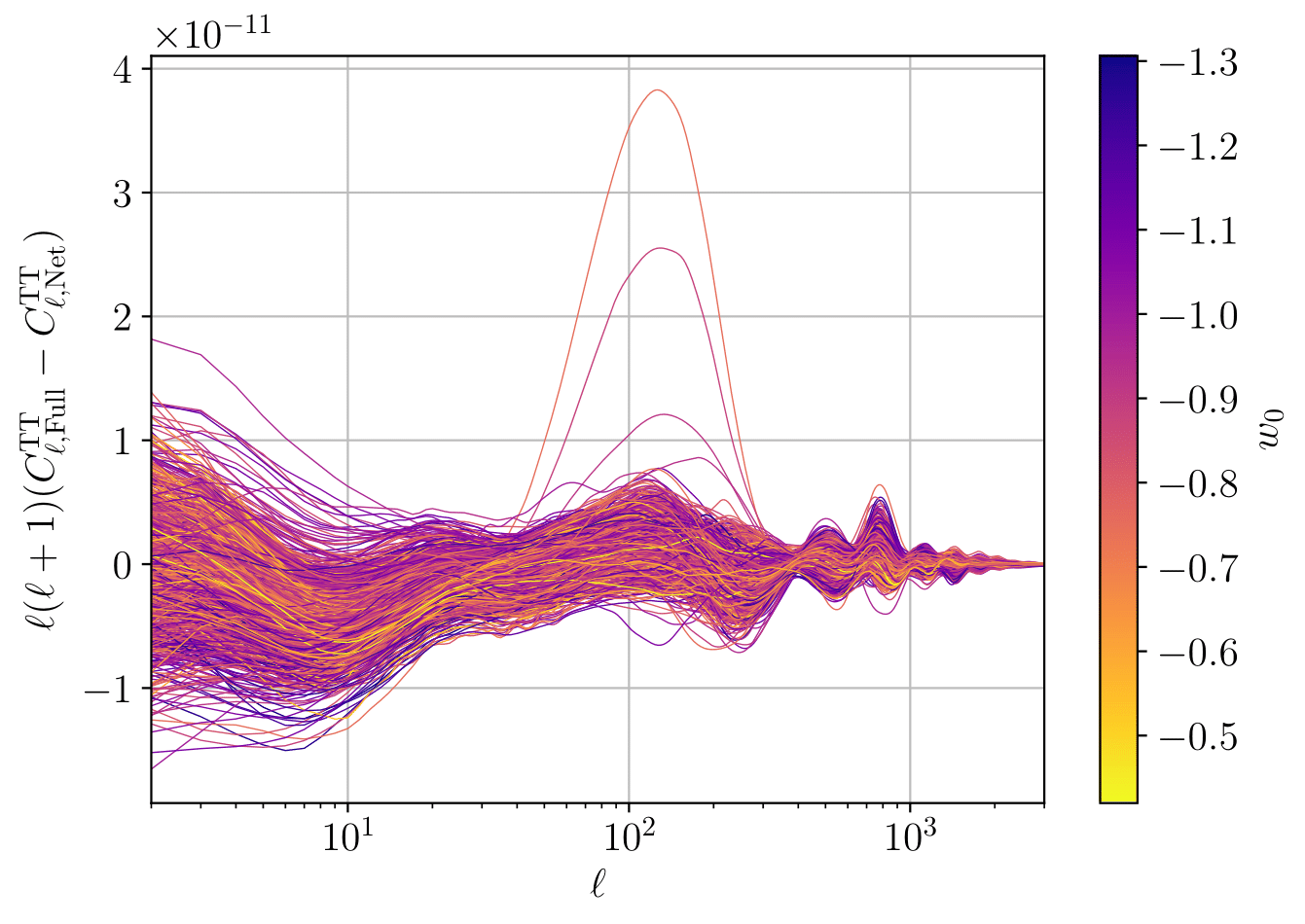}
	\end{subfigure}
	\begin{subfigure}{0.45\linewidth}
		\includegraphics[width=\linewidth]{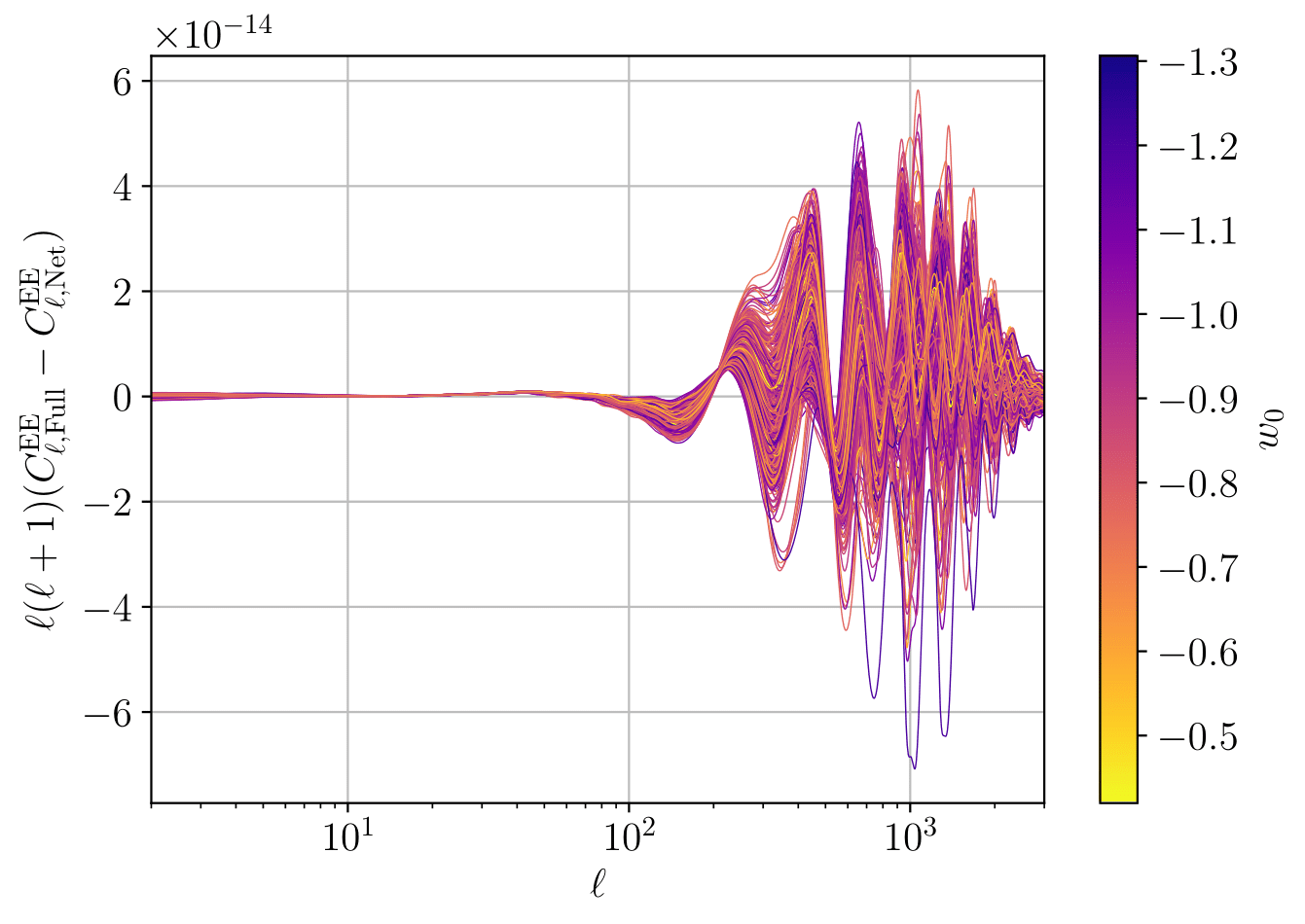}
	\end{subfigure}
	\begin{subfigure}{0.45\linewidth}
		\includegraphics[width=\linewidth]{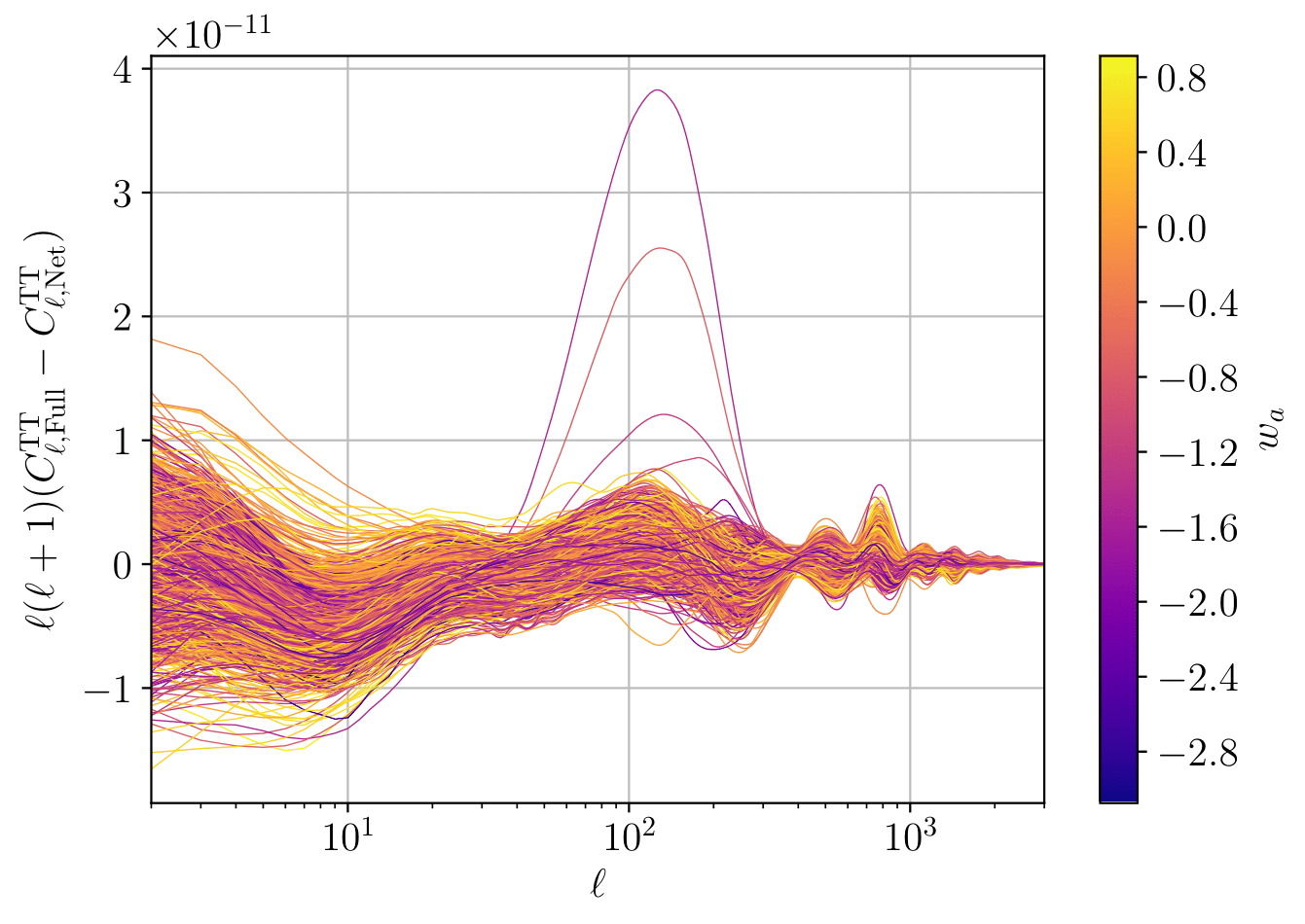}
	\end{subfigure}
	\begin{subfigure}{0.45\linewidth}
		\includegraphics[width=\linewidth]{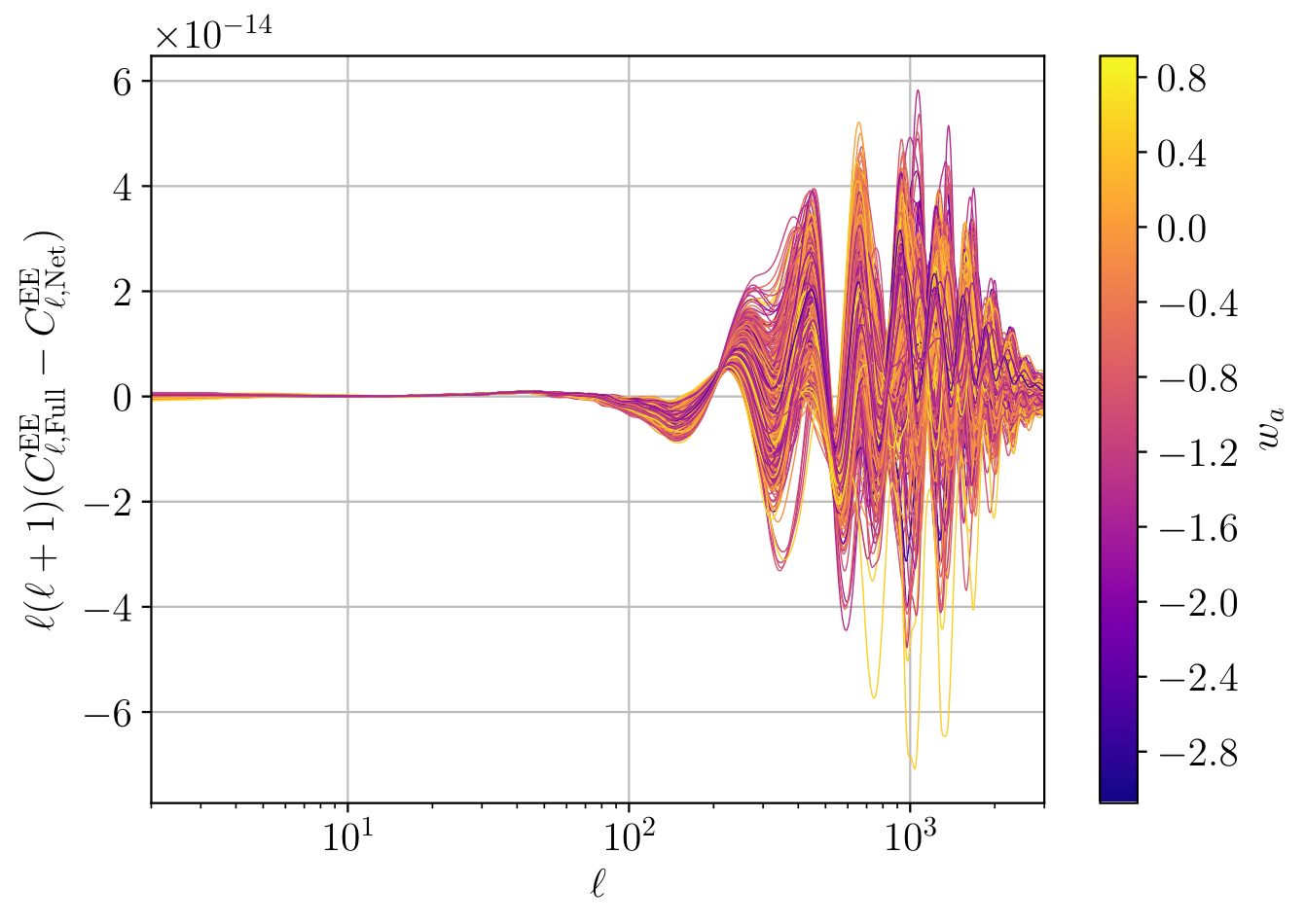}
	\end{subfigure}
	\caption{Same as \cref{fig:pardep_lcdm} but for the extended parameters (from top to bottom)  $\Omega_k$, $w_0$ and $w_a$\,.}
	\label{fig:pardep_cur_de}
\end{figure}

In \cref{fig:pardep_cur_de}, we repeat the exercise for the extended parameters  ($\Omega_k$, $w_0$, $w_a$). No clear correlation appears, apart from a small feature for the $C_l^{TT}$ residual as a function of $\Omega_k$, very similar to the feature already observed for $\Omega_\mathrm{m}$. However, we should recall that these residuals are plotted for random samples of our training domain, in which all parameters vary simultaneously (rather than all parameters but one being fixed). Since the ellipsoidal training domain is defined with a positive correlation between values of $\Omega_k$ and $\Omega_\mathrm{m}$, we are observing once again the correlation with $\Omega_\mathrm{m}$, caused by the limited accuracy of the ISW network [N1].

\begin{figure}[H]
	\centering
	\begin{subfigure}{0.45\linewidth}
		\includegraphics[width=\linewidth]{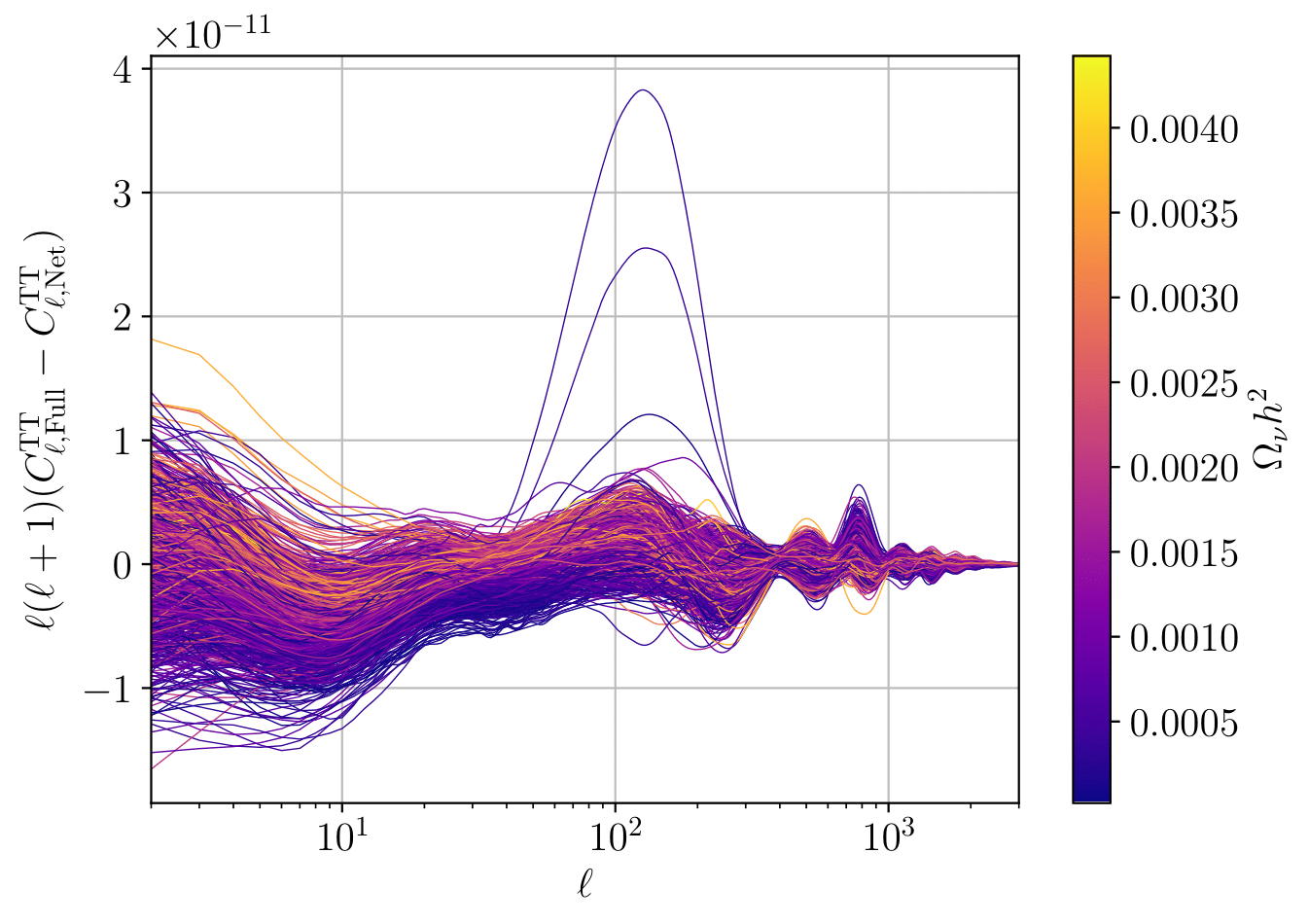}
	\end{subfigure}
	\begin{subfigure}{0.45\linewidth}
		\includegraphics[width=\linewidth]{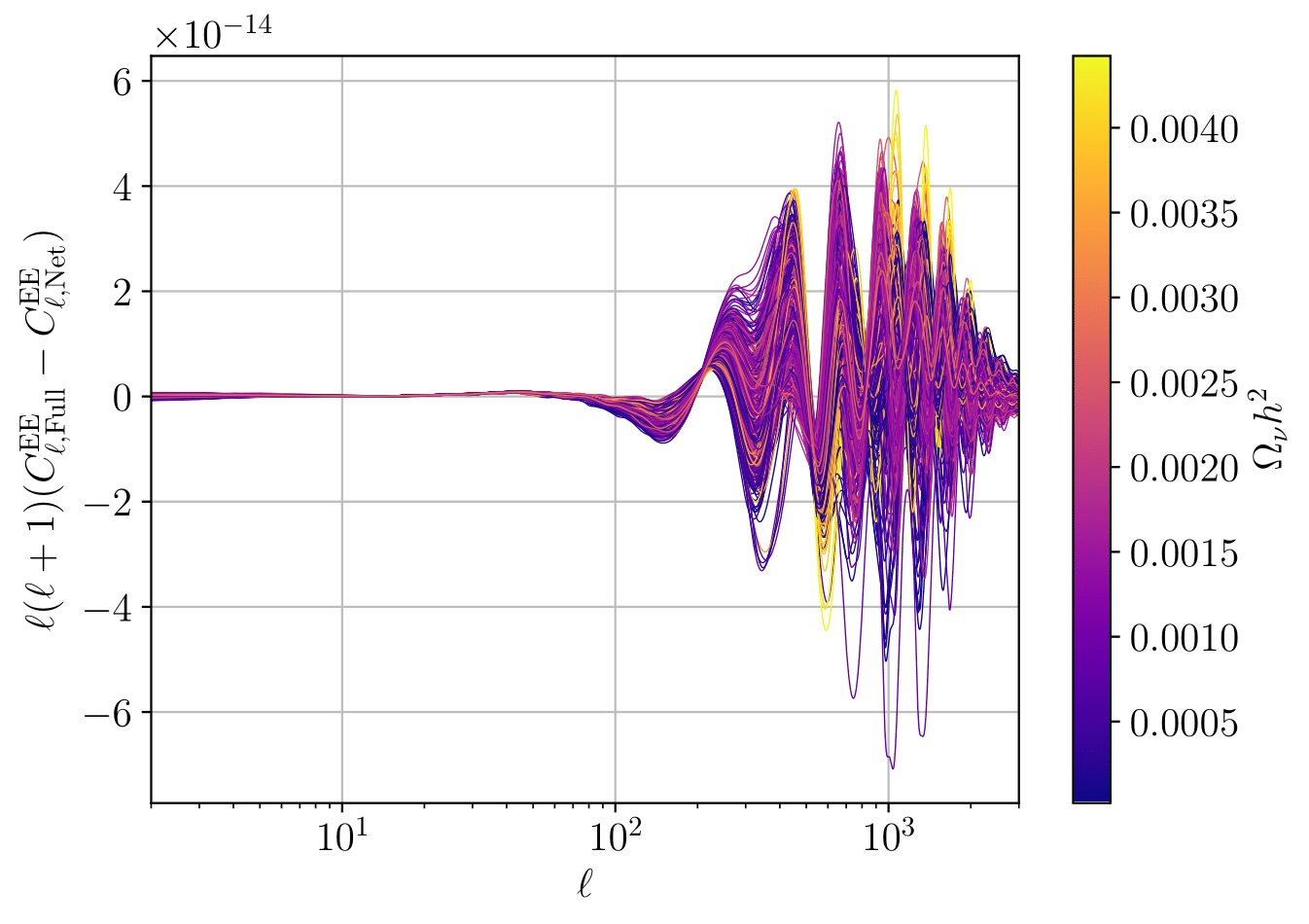}
	\end{subfigure}
	\begin{subfigure}{0.45\linewidth}
		\includegraphics[width=\linewidth]{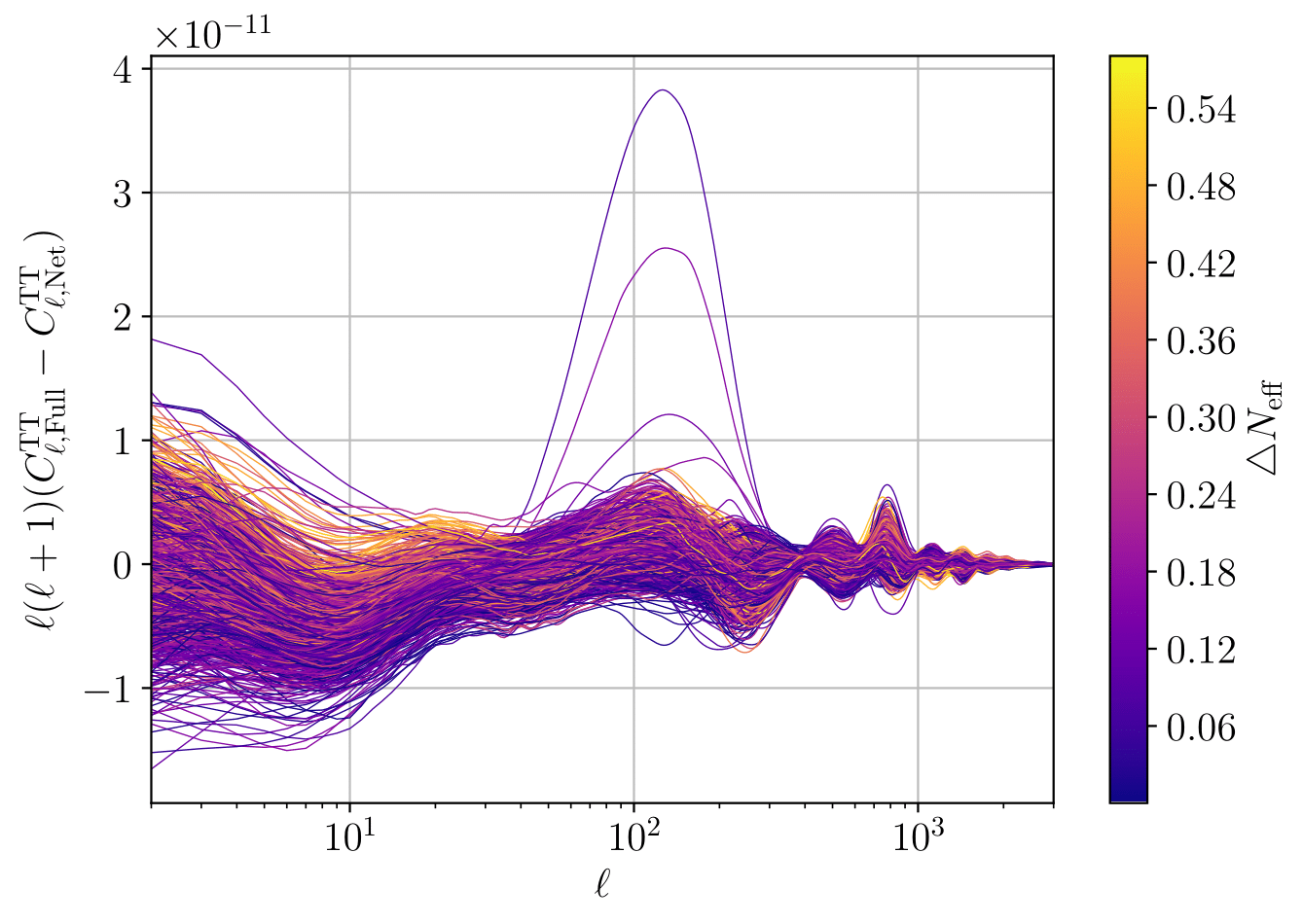}
	\end{subfigure}
	\begin{subfigure}{0.45\linewidth}
		\includegraphics[width=\linewidth]{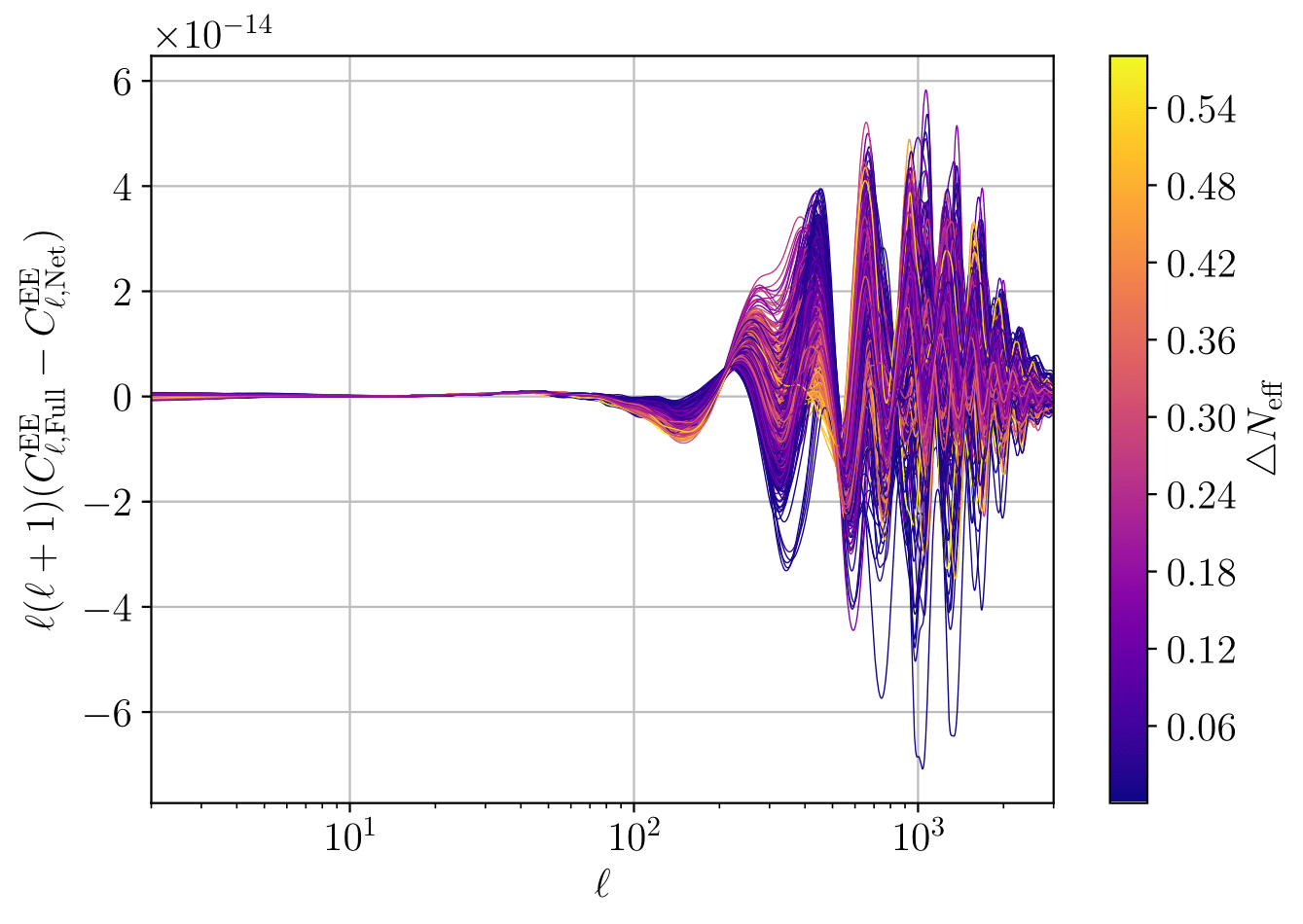}
	\end{subfigure}
		\caption{Same as \cref{fig:pardep_lcdm} but for the extended parameters $\Omega_\nu h^2$ (top) and $\Delta N_\mathrm{eff}$ (bottom).}
	\label{fig:pardep_neu}
\end{figure}

\noindent Finally, in \cref{fig:pardep_neu}, we color-code the residuals as a function of the parameters $\Omega_\nu$ and $\Delta N_\mathrm{eff}$. The color impression of these plots is very different, because we are now dealing with parameters with a lower bound in zero and a half-Gaussian distribution across the training region. It is thus expected that the majority of residuals have a blue-ish color (corresponding to values close to zero/the peak of the mode). Still, we can see small correlations. For $C_\ell^{TT}$, we observe features at low $\ell$ which are again reminiscent of the correlation between the residuals and $\Omega_\mathrm{m}$. Since the ellipsoidal training domain is defined with a positive correlation between values of $\Omega_\nu$ and $\Omega_\mathrm{m}$, and also between $\Delta N_\mathrm{eff}$ and $\Omega_\mathrm{m}$, we expect that we are observing again the same feature as before. The same applies to $C_\ell^{EE}$.\footnote{We observe however an additional effect: for the largest value of $\Omega_\nu h^2$, \classnet seems to systematically mis-predict the phase and amplitude of the acoustic peaks in $C_\ell^{EE}$. This is likely related to the fact that for large $\Omega_\nu h^2$ values, neutrinos are already non-relativistic at decoupling, and the gravitational interactions between photon perturbations and neutrino perturbations prior to recombination become significantly different. It is likely that our networks are not trained with enough high-neutrino-mass models for accurately modeling this peculiar effect. However, this occurs only for very large values $\Omega_\nu \geq  0.004$ (i.e., $\sum m_\nu\geq0.4$~eV) that are excluded by current data sets.}

In summary, these tests provide a plausible explanation for part of the bias observed in $\Omega_\mathrm{b} h^2$ and $\Omega_\mathrm{m} h^2$, likely due to the limited accuracy with which the network [N1] predicts the early ISW effect, and with which the network [N5] predicts the effects of the baryon-to-photon ratio and of the redshift of equality on the amplitude and phase of the acoustic oscillations relevant for the polarisation spectrum. The other biases on $n_s$\,, $H_0$\,, $\Omega_\nu h^2$\,, and $\Delta N_\mathrm{eff}$ seem to originate more indirectly, either due to a propagation of the former biases through parameter correlations, or to an average shape of the residuals that can be counter-acted by a shift in some parameters.
\pagebreak[20]

\subsection{Posterior slices}
\label{ssec:trouble}

The analysis of residuals in the last section did not allow us to:
\begin{itemize}
\item investigate the origin of parameter bias for parameters that are not varied in the training set, such as the primordial tilt $n_s$ (on which the source functions do not depend),
\item check which observable has a dominant contribution to the bias (e.g. temperature spectrum, polarisation spectrum),
\item prove explicitly which networks contribute the most to parameter bias and need to be improved in priority in future versions.
\end{itemize}
To address these points, one could in principle repeat multiple parameter estimations in which one source function is predicted with \classnet and all others with \classfull, and in which different likelihood subsets are used. However, going through this method after each new network training would require lot of computing power, while our whole point is precisely to save computing time.

A much more economic approach consists in comparing posterior slices, that is, evaluating the likelihood and the posterior with either \classnet or \classfull when one parameter is varied while the others are all fixed close to their best-fit value. Since we adopt top-hat priors on the cosmological parameters of our basis, the logarithm of the likelihood and of the posterior are identical up to a constant term, and the difference between the log-posteriors, $\log P^\mathrm{Net} - \log P^\mathrm{Full}$, is equal to the difference between the log-likelihoods, $\log {\cal L}^\mathrm{Net} - \log {\cal L}^\mathrm{Full}$.

Here we perform such an investigation for the two parameters whose mean value, confidence limit or upper bound are the most biased by \classnet, namely, $n_s$ and $\Delta N_\mathrm{eff}$\,. For each of them, we compute the full Planck likelihood \texttt{planck\_2018\_highl\_plick.TTTEEE} as well as the individual contribution from the low -$\ell$ likelihood, high-$\ell$ TT likelihood, high-$\ell$ TE likelihood and high-$\ell$ EE likelihood. We perform this exercise for 20 values of $n_s$ or 20 values of $\Delta N_\mathrm{eff}$\,, with all source functions but one calculated by \classfull. This amounts in 2 (cases) $\times$ 2 (parameters) $\times$ 20 (values) $\times$ 9 (sources) $=720$ \class calls and likelihood evaluations in Cobaya, which is nothing compared to an MCMC run.\\

\begin{figure}[t]
	\centering
	\begin{subfigure}{0.49\linewidth}
		\includegraphics[width=\linewidth]{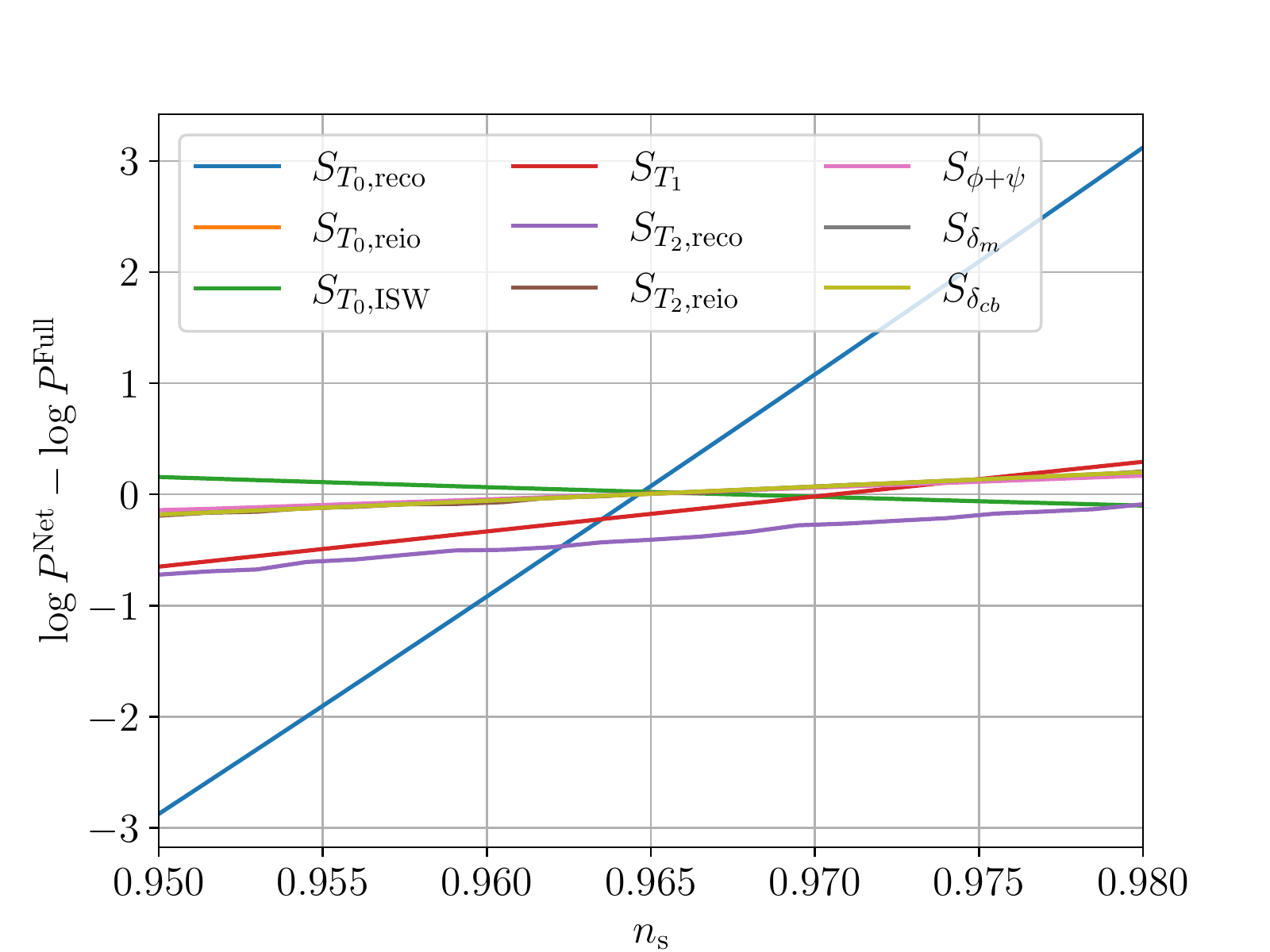}
	\end{subfigure}
	\begin{subfigure}{0.49\linewidth}
		\includegraphics[width=\linewidth]{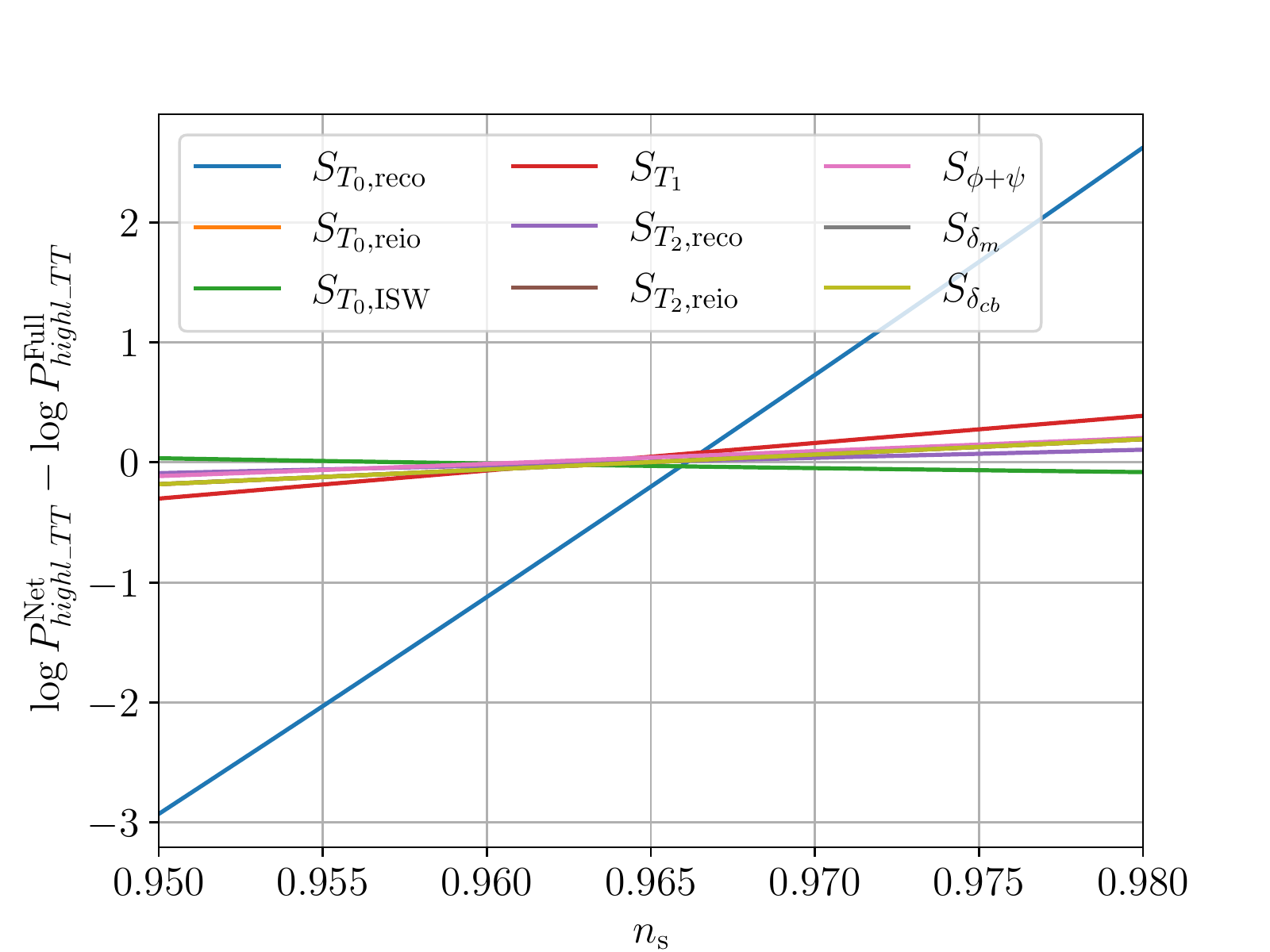}
	\end{subfigure}
	\caption{\textbf{Left:} Difference between log-posteriors derived from the full Planck likelihood as a function of $n_s$ (all other parameters being fixed close to their best-fit value) with one source function predicted by \classnet while all others are calculated by \classfull. \textbf{Right:} Same for the the  Planck high-$\ell$ temperature-only likelihood.}
	\label{fig:lcdm_ns_pc}
\end{figure}

{\bf Scalar tilt $\mathbf{n_s}$\,:} The left plot in \cref{fig:lcdm_ns_pc} shows the difference $\log P^\mathrm{Net} - \log P^\mathrm{Full}$ as a function of $n_s$, with just one source function predicted by the network at a time. Under the approximation of a Gaussian posterior, a linear term in $n_s$ induces a shift between the means ($\bar{n}_s^\mathrm{Net}$, $\bar{n}_s^\mathrm{Full}$) while a quadratic term additionally introduces a bias in the errors ($\sigma_{n_s}^\mathrm{Net}$, $\sigma_{n_s}^\mathrm{Full}$). Any constant offset is not observable in Bayesian parameter inference.\footnote{In fact quadratic, linear and constant terms can result in shifts in means and errors. Consider a Gaussian posterior: 
\begin{equation}
	\log P^\mathrm{Net} - \log P^\mathrm{Full} = \left(\frac{x-\mu_\mathrm{Net}}{\sigma_\mathrm{Net}}\right)^2 -  \left(\frac{x-\mu_\mathrm{Full}}{\sigma_\mathrm{Full}}\right)^2 = x^2 \underbrace{\left(\frac{1}{\sigma_\mathrm{Net}^2} - \frac{1}{\sigma_\mathrm{Full}^2}\right)}_{a}-2x\underbrace{\left(\frac{\mu_\mathrm{Net}}{\sigma_\mathrm{Net}^2} - \frac{\mu_\mathrm{Full}}{\sigma_\mathrm{Full}^2}\right)}_b+\underbrace{\left(\frac{\mu_\mathrm{Net}^2}{\sigma_\mathrm{Net}^2} - \frac{\mu_\mathrm{Full}^2}{\sigma_\mathrm{Full}^2}\right)}_c
\end{equation}
Without a quadratic contribution ($a=0$) the errors coincide and the linear term describes a shift in means. The constant term describes the same shift but at second order and can thus be neglected. Without a linear term ($b=0$) the quadratic contribution introduces a bias in the errors. From the definition of $b$ one infers a bias in the means which is modulated by $a$ since $\mu_\mathrm{Net}/\mu_\mathrm{Full} = 1/(1+a \sigma_\mathrm{Full}^2)$. A mixture of quadratic and linear contributions lead to biases in both means and errors.
}

We do observe a linear dependence on $n_s$, as expected from the fact that the mean $\bar{n}_s^\mathrm{Net}$ was biased by 0.10$\sigma$ to 0.17$\sigma$ in the Cobaya runs of \cref{sec:parameter_estimation}. It is immediately obvious from the figure that the network [N2] contributing to $S_{T_0,\mathrm{reco}}$ is the dominant source of bias on $n_s$ (note that we usually omit the \texttt{no\_isw} suffix for shortness). The dependence of $\log P^\mathrm{Net} - \log P^\mathrm{Full}$ on $n_s$ is almost linear, because  the bias on the error $\sigma_{n_s}$ is insignificant compared to the bias on the mean $\bar{n}_s$.

\begin{figure}[t]
	\centering
	\begin{subfigure}{0.49\linewidth}
		\includegraphics[width=\linewidth]{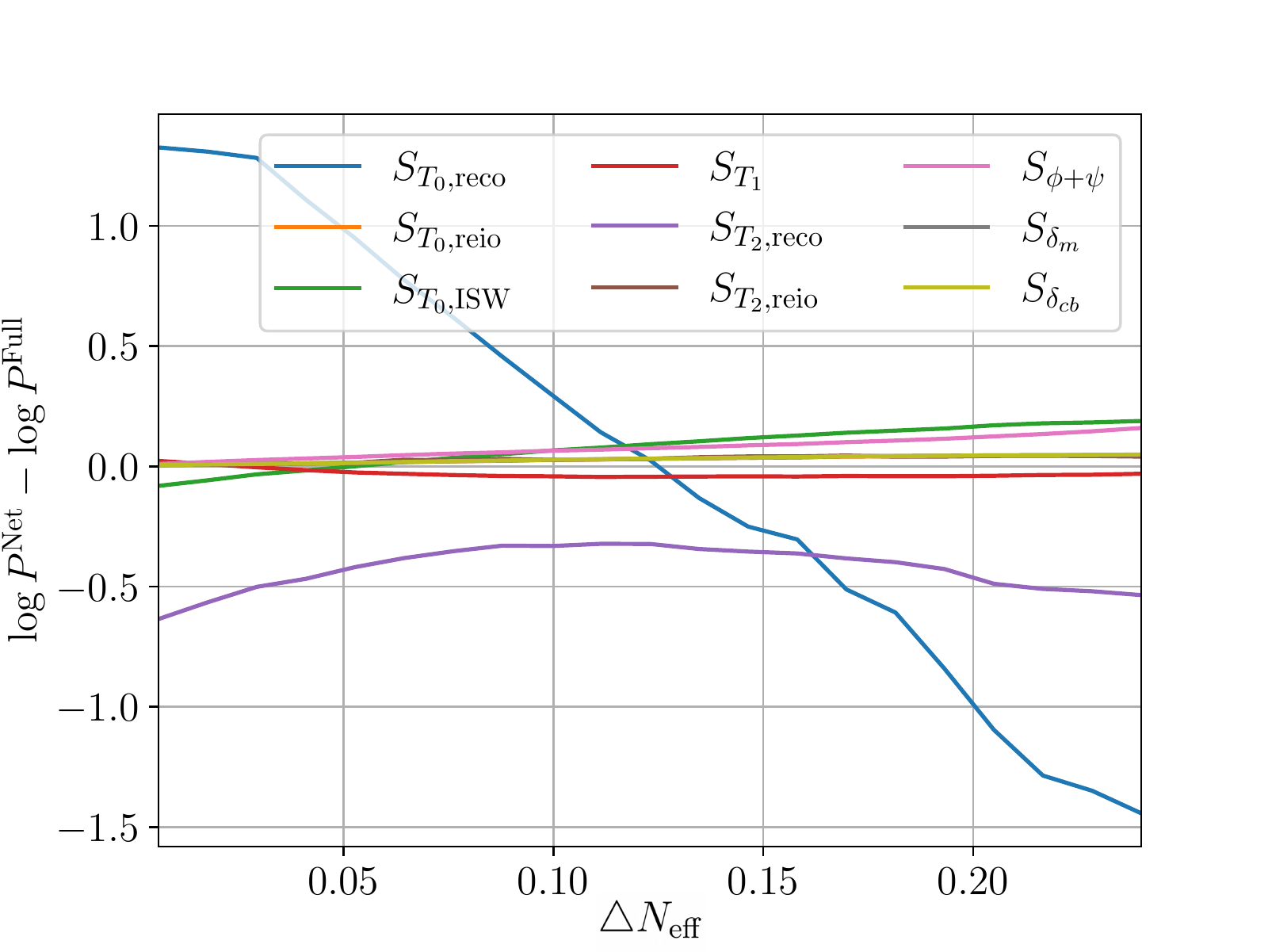}
	\end{subfigure}
	\caption{Difference between log-posteriors derived from the full Planck likelihood as a function of $\Delta N_\mathrm{eff}$ (all other parameters being fixed close to their best-fit value) with one source function predicted by \classnet while all others are calculated by \classfull.}
	\label{fig:11p_Nur_pc}
\end{figure}

Since $S_{T_0,\mathrm{reco}}$ is only relevant for temperature, the bias must be related to the predicted shape of the spectra $C_\ell^{TT}$ and $C_\ell^{TE}$. By splitting the likelihood in several contributions, we checked that this bias appears mainly through the \texttt{planck\_2018\_highl\_plick.TT} likelihood, i.e.\ in the temperature power spectrum for $\ell>30$, and also through the TE likelihood. As a matter of fact, the right plot in \cref{fig:lcdm_ns_pc}, which show the difference $\log P^\mathrm{Net} - \log P^\mathrm{Full}$ for just the TT likelihood, looks very similar to the left plot. We conclude that the network [N2] tends to systematically tilt the shape of $S_{T_0,\mathrm{reco}}$ as a function of $k$, with a small deficit of power in the large-$k$ region. This tilting propagates to $C_\ell^\mathrm{TT}$ (and $C_\ell^\mathrm{TE}$) at large $\ell$, and gets compensated during parameter evaluation by a slightly higher value of $n_s$. By eye, this is not totally obvious in the residual plot for $S_{T_0,\mathrm{reco}}$ (\cref{fig:source_functions_t0}), $C_\ell^\mathrm{TT}$ (\cref{fig:power_spectra_tt_lines}) and $C_\ell^\mathrm{TE}$ (\cref{fig:power_spectra_te_lines}), but \cref{fig:lcdm_ns_pc} proves it unambiguously.
An expansion and/or improvement of the [N2]~network could be considered in the future if higher accuracy becomes necessary.
\\[1\baselineskip]
{\bf Number of effective relativistic degrees of freedom $\mathbf{\Delta N_\mathrm{eff}}$\,:} \Cref{fig:11p_Nur_pc} shows the contribution of each network to the difference $\log P^\mathrm{Net} - \log P^\mathrm{Full}$. Like for $n_s$\,, the network [N2] (that predicts $S_{T_0,\mathrm{reco}}$) contributes the most to the shifting of the upper bound on $\Delta N_\mathrm{eff}$, followed by the network [N5] (that predicts $S_{T_2,\mathrm{reco}}$ and thus $S_P = \sqrt{6} S_{T_2}$ as well). By splitting the likelihood in different parts, we checked that the [N2]-error propagates to the posterior through the high-$\ell$ TT and TE likelihoods, while the [N5]-error propagates through the high-$\ell$ EE and TE likelihoods. Our interpretation is that during parameter estimation the parameters $n_s$ and $\Delta N_\mathrm{eff}$ are both at play for fitting the small systematic error induced by the [N2] an [N5] networks in the high-$k$ source functions and thus in the high-$\ell$ temperature and polarisation spectra. The fiducial data is better fitted with slightly too large values~of~$n_s$~and~$\Delta N_{\rm eff}$\,. A larger value of $n_s$  raises power at all wavenumbers larger than the pivot wavenumber (and thus at all large $\ell$'s), while a larger $\Delta N_{\rm eff}$ can be used to keep a sharp exponential drop of power at large $\ell$ through the Silk damping effect. We are talking here of very small effects, corresponding to baises  at the level of at most 0.08$\sigma$ to $0.17\sigma$ in both $n_s$ and $\Delta N_\mathrm{eff}$, which are difficult to anticipate visually when just looking at the shape of the residuals e.g. in \cref{fig:power_spectra_tt_lines,fig:power_spectra_te_lines}.

\vspace*{1\baselineskip}
{\bf Other Parameters:}
We followed the same method for other parameters. Like for $n_s$ and $\Delta N_\mathrm{eff}$, we found that the network [N2] ($S_{T_0,\mathrm{reco}}$) has the largest impact -- through the TT and TE spectra -- on differences between the \classnet and \classfull posteriors when cutting along a given parameter direction. Like for $\Delta N_\mathrm{eff}$, we also found that [N5] ($S_{T_2,\mathrm{reco}}$) has the second largest contribution -- through the TE and EE spectra.\\
 
To summarize, in \cref{sec:parameter_estimation}, we have proven that \classnet performs very well in parameter estimation for various models. In this appendix, we have identified the most important networks to improve in case more accurate parameter estimation is needed. In \cref{sec:resi_cor}, the tests based on correlations in the residuals suggested that [N1] and [N5] are worth improving, but this test only brings some hints and no definite conclusion. In \cref{ssec:trouble}, the posterior slices unambiguously pin-pointed which networks contribute the most to parameter bias. There, we found that the impact on the likelihoods of the parameter dependence induced by [N1] is negligible. If more accuracy becomes necessary, the network that should be improved in priority is [N2] (due to its induced biasing of $n_s, \Delta N_\mathrm{eff}$), and then, to a lesser extent, [N5].
\section{Implementation of the networks in the released CLASS}

\label{sec:implement}

In this appendix, we explain the detailed implementation of the \classnet specificities into \class, mainly to the attention of readers who want to retrain or modify \classnet. Access to \classnet can currently be found in the git branch {\tt classnet} of the public github repository \url{https://github.com/lesgourg/class_public}.

\subsection{File Structure and Implementation}

All the data relevant for \classnet is stored in a new directory {\tt classnet\_workspace/}. This includes the array of wavenumbers at which the source functions are predicted by the NNs, the fixed precision settings used for the generation of the training/validation/testing data, the normalisation of all NN input, the weights of the NNs, and the domain on which the NNs are trained and assumed to be valid. Furthermore, per default, all new data that is generated by the use of \classnet is stored in this folder. This can include data sets to train new neural networks, benchmark scripts, data to test the \classnet performance, as well as corresponding plots.

The pieces of code specific to \classnet -- including all the scripts used to generate the plots included in this paper -- are stored in the \texttt{python/nn/} folder (relative to the main {\tt class/} directory)  and sorted into subfolders. The \texttt{models} and \texttt{example} subfolders are particularly relevant for tests and modifications. The former defines the architecture and input of each individual NN, while the latter hosts some documented evaluation, testing and training scripts.
The \class code is entirely written in \texttt{C} and embedded in {\tt python} through the \texttt{classy} wrapper located in \texttt{python/classy.pyx}. Additional pieces of code specific to \classnet are entirely contained in (or interfaced with) this wrapper, with the exception of a few conditional statements in the {\tt perturbations.c} module of the \texttt{C} code, allowing to skip the full ODE integration when the NNs are used instead. As a result, \classnet is currently (and for the foreseeable future)\footnote{While there exist a few \texttt{C} or \texttt{C++} APIs for neural networks, calling them through \texttt{python} is at this point far more convenient and equally fast.} only accessible through {\tt python}. 
In the \texttt{classy} wrapper, each module of \class is called sequentially within the \texttt{.compute()} function. Before the {\tt perturbation.c} module  is executed, the wrapper checks whether the user supplied any parameter related to the NNs. If not, the code proceeds with the usual integration of the coupled ODEs (\classfull mode). If such parameters are found, the code instead loads the domain stored in the workspace folder in order to check whether the requested model parameters lie within the validation region -- and are thus suitable for NN acceleration.
If this is the case, the \texttt{C} flag \texttt{perform\_NN\_skip} is set to {\tt \_TRUE\_}, allowing the \texttt{C} code to only compute the time sampling vector and the wavenumber sampling vector relevant for the given cosmology, while skipping the integration of the ODEs. In a next step, the \classnet environment is set up by creating an instance of the \texttt{predictor}-class. As detailed in the \texttt{class/python/nn/predictors.py} file, the role of this class is to load the NNs, their $k$ array, and the cosmological parameters. Then, the neural networks are evaluated for each of the time samples of the given cosmology, predicting the source function for all the $k$ values of the network. However, these do not a priori coincide with the $k$ values of the given cosmology. Thus, the source function are further interpolated and finally stored at the correct time and $k$ values. At this point, the susequent \class modules can be run in the same way as in \classfull.

\subsection{Single \classnet evaluation}

The \class input parameters specific to \classnet are defined and set in the {\tt classy} wrapper. The first one is the workspace directory path and name \texttt{workspace\_path}, set by default to
\begin{center}
\texttt{<path-to-class>/classnet\_workspace}
\end{center}
\enlargethispage*{1\baselineskip}
The second is the flag \texttt{use\_nn} set to by default to \texttt{False}. Note that the precision parameters should not differ from the ones used during the training process that are stored in within \texttt{<path-to-workspace>/data/manifest.json}. 

At this point, we did no implement any automatic check that these parameters coincide, and it is the responability of the user to control the NN precision parameters in situations where accuracy is critical.

An explanatory python script with a single \classnet evaluation can be found within the \class folder \texttt{<path-to-class>/python/nn/examples/example\_evaluate.py}.

When the user wants to test and compare different versions of the same network, it is possible to switch between them using a \enquote{generation} parameter. For example, passing the parameter \texttt{Net\_ST0\_ISW=12} would attempt to load network \texttt{Net\_ST0\_ISW\_12.pt} in the directory \texttt{<path-to-workspace>/models/}. By default, \classnet will load the network \texttt{Net\_ST0\_ISW.pt} (without a \enquote{generation} index). 

Besides, there is an input parameter \texttt{nn\_verbose} -- set to \texttt{1} by default -- that controls the verbosity of the output. Finally, one can use the flag  \texttt{nn\_force} to force \texttt{classy} to use \classnet also outside of the validation region, to be used only for testing or debugging.

\subsection{Multiple \classnet evaluations within parameter inference code}



\noindent {\bf General usage.} The usage of \classnet in the framework of parameter inference packages such as Cobaya~\cite{cobaya,cobaya_code}, MontePython~\cite{montepython_code,Brinckmann:2018cvx} or others is straightforward. We provide a Cobaya example script at
\begin{flushleft}
\texttt{<path-to-class>/python/nn/examples/example\_mcmc\_cobaya.py}
\end{flushleft}
and some MontePython input files at:
\begin{flushleft}
\texttt{<path-to-class>/python/nn/examples/example\_nn.param} \\
\texttt{<path-to-class>/python/nn/examples/base2018TTTEEE\_lensing\_bao\_nn.param}
\end{flushleft}
that show how to perform a parameter scan of the $\Lambda$CDM model using the Planck likelihoods. 

In fact, parameter inference packages can be used in combination with \classnet in the exact same way as \classfull, excepted that a few additional \class parameters need to be specified in input:
\begin{flushleft}
{\tt
\textquotesingle use\_nn\textquotesingle:\textquotesingle yes\textquotesingle,\\
\textquotesingle workspace\_path\textquotesingle:\textquotesingle <path-to-class>/classnet\_workspace\textquotesingle 
}
\end{flushleft}
The setting  {\tt \textquotesingle use\_nn\textquotesingle:\textquotesingle yes\textquotesingle } means that \classnet will try to use the network each time that a new model in the chain falls within the region of trusted accuracy, and will default to the \classfull mode otherwise. No further considerations about the switching between \classnet and \classfull have to be specified explicitly by the user.\\

\noindent {\bf Checking the actual use of the networks.} When doing preliminary tests, the user may wish to control whether the code is using more often the \classnet or \classfull mode. This information can be either obtained in the standard output with the verbose parameter {\tt \textquotesingle nn\_verbose\textquotesingle } set to 1 or more, or by asking the inference code to store the derived quantity {\tt \textquotesingle nn\_chi2\textquotesingle }. This value of the $\triangle \chi^2$ (see \cref{eq:chi2_parameter_region}) is computed at the beginning of a \classnet call to decide which mode is going to be used. It will then be stored as an extra column in the chains.\footnote{As explained in \cref{ssec:domain_cn}, in the current released version, \classnet is used each time that this $\triangle \chi^2$ is below 46.12.} Note that it is expected that during the burn-in phase a significant fraction of the calls require \classfull. Nonetheless, we find that compared to the total number of sampled points, this happens only a small number of times and the total acceleration of the MCMC is marginally affected. \\

\noindent {\bf Single load of the networks during parameter inference runs.} The parameter inference software will normally create one \texttt{Class} instance per chain at the beginning of the run. During the run, before calculating a new model, the input parameter and the output quantitites of the previous run get cleared, but the \texttt{Class} instance itself is not deleted. It should be noted that the networks get loaded when the \classnet acceleration is performed, but remain loaded as long as the instance is not deleted. This saves a non-negligible fraction of runtime for subsequent computations. \\

\noindent {\bf Choice of cosmological parameter basis.}
The network was trained with a given set of model parameters.
However, there is no obligation to stick to the same parameter basis during parameter inference. Indeed, once the user has defined a list of varied and fixed parameters in the input file of the parameter inference package, these parameters are passed to \classnet and interpreted by the {\tt input} module like in any \class call. Later, these parameter are converted to the basis used by the networks and passed to them. This means, for instance, that the user is still free to 
use indifferently big or small omega's for each species (with $\omega_x \equiv \Omega_x h^2$), or to
specify {\tt m\_nu} instead of {\tt omega\_ncdm} or {\tt Omega\_ncdm}. In all cases, the networks gets in input the total $\Omega_\nu h^2$, summed internally over all non-cold dark matter species, and the total $N_\mathrm{eff}$, summed over all the species that are non-relativistic at the initial time used by \class.

However, \classnet does an automatic check that a few model parameters have been set correctly in order to match some key assumptions used during training. An error is returned whenever the user omitts to fix the following three parameters:
\begin{flushleft}
{\tt
\textquotesingle Omega\_Lambda\textquotesingle :0,\\
\textquotesingle N\_ncdm\textquotesingle :1,\\
\textquotesingle deg\_ncdm\textquotesingle :3}
\end{flushleft}
As a matter of fact, the plain cosmological constant must be switched off with $\Omega_\Lambda=0$ in order to activate the dark radiation fluid labelled as {\tt fld}, whose density {\tt Omega\_fld} gets computed automatically in the {\tt input} module using the budget equation, and whose equation-of-state parameters {\tt w0\_fld} and {\tt wa\_fld} can be either fluctuated or set to fixed values. When {\tt w0\_fld=-1} and {\tt wa\_fld=0}, the dark energy fluid is {\it de facto} equivalent to a plain cosmological constant. The above settings for the parameters {\tt N\_ncdm} and {\tt deg\_ncdm} ensure that the user assumes three massive neutrinos degenerate in mass. Since this assumption was performed during training, it is required in order to guarantee an accurate use of the networks.\footnote{Thus, if the user wants a single massive neutrino, the networks need to be retrained. Note however that, in order to get some CMB and matter power spectra close to the predictions of realistic neutrino mass schemes (normal or inverted hierarchy), it is always much better to assume three degenerate massive neutrinos as we are doing here, see e.g. \cite{Lesgourgues:2006nd}.} With such settings, the \class parameter {\tt N\_ur} (the effective number of ultra-relativistic species) accounts for additional relativistic degrees of freedom, up to a shift of 0.0044 that can be seen as a ``fudge factor'' (accounting for non-thermal distortions in the active neutrino phase-space distribution inferred from the latest studies of neutrino decoupling). This means that  a model with three active neutrinos plus $\triangle N_\mathrm{eff}$ non-standard relativistic relics corresponds to {\tt N\_ur}~$= 0.0044 +  \triangle N_\mathrm{eff}$. After summing over the contribution of massless and massive neutrinos, \class finds internally $N_\mathrm{eff} = 3.0440 +  \triangle N_\mathrm{eff}$, consistently with \cite{Froustey:2020mcq,Bennett:2020zkv}. 

\classnet also does an automatic check that some requirements are met at the level of precision parameters. By default, the parameter {\tt \textquotesingle compute damping scale\textquotesingle } is set to {\tt \textquotesingle no\textquotesingle } by \class. However, when using \classnet, it must be activated in order to pass an analytic approximation to the photon damping scale as an input to the networks. Thus the input file should include
\begin{flushleft}
{\tt \textquotesingle compute damping scale\textquotesingle :\textquotesingle yes\textquotesingle }
\end{flushleft}
Finally, in its current released version, \classnet checks that the parameters {\tt P\_k\_max\_1/Mpc} or {\tt P\_k\_max\_h/Mpc} have been set in such a way that the matter power spectra are not requested beyond $k=100 h\,$Mpc$^{-1}$, since the networks have been trained with \texttt{\textquotesingle P\_k\_max\_1/Mpc\textquotesingle:100}. \\

\noindent {\bf Default parameter values.} When some parameters are omitted, \classnet defaults to the same values as the standard \class code, excepted for the neutrino mass, whose default value is set to $10^{-5}$eV in \classnet instead of 0 in \class. (In practise, this makes no difference, since a neutrino of mass $10^{-5}$eV is still ultra-relativistic today and thus indistinguishable from a massless one). The $\Lambda$CDM parameters have default values close to the Planck 2018 best fit. The default value of {\tt N\_ur} is 3.046 and should never be used, since \classnet always assumes already three massive neutrinos. Thus, the user not interested in extra relativistic degrees of freedom should set manually {\tt N\_ur}  to 0.0044. The default value of spatial curvature is zero, and that of dark energy parameters is ({\tt w0\_fld=-1}, {\tt wa\_fld=0}), corresponding to a plain cosmological constant.

\subsection{Training}

Training can be split into three parts:
\begin{enumerate}
\item generating a parameter sample within a domain for the training/validation/test sets (domain sampling), 
\item computing the \classfull results at each sample (data generation), 
\item training the weights of the neural networks (training).
\end{enumerate}
We outline these steps in the following subsections. Documented step-by-step scripts showing how to train the networks for an arbitrary set of models and parameters can be found in the python files \texttt{example\_train\_full.py} and \texttt{example\_retrain.py} in the\\ \texttt{<path-to-class>/python/nn/examples} directory. The former script has all its settings written explicitly in the file, and needs to be edited when used for different purposes. Instead, the latter can be called with additional command-line arguments. 

All the tasks related to a given version of the networks -- including their training, storage and evaluation -- are handled by a single instance of a python class {\tt Workspace}, defined in {\tt python/nn/workspace.py}. This class can be seen as a wrapper of a given workspace directory -- like the default  directory {\tt classnet\_workspace/} provided by default, or another one with a path and name \texttt{workspace\_path} defined by the user. Thus, all the training steps are coded as some functions of one instance of the {\tt Workspace} class.

In the next paragraphs, we provide the necessary functions to retrain the networks using the methods outlined in this paper. All python scripts are stored in a python package \texttt{classynet} that is installed when compiling \class from the {\tt classnet} branch. To create a workspace environment, one calls:
\begin{lstlisting}
	import classynet
	my_workspace = classynet.workspace.Workspace(
		workspace_path)
\end{lstlisting}
This includes the \texttt{workspace\_path} (to either an existing or new workspace directory). There is a more advanced function taking an additional argument {\tt generation}, in case one wants to train only one or a few specific networks (out of the list of networks defined in \cref{sec:architecture} ) with a specific version number:
\begin{lstlisting}
	import classynet
	my_workspace = classynet.workspace.GenerationalWorkspace(
		workspace_path, generations)
\end{lstlisting}
The optional argument \texttt{generations} is a dictionary with keys:
\begin{lstlisting}
	Net_ST0_Reco, Net_ST0_Reio, Net_ST0_ISW, Net_ST1, 
	Net_ST2_Reco, Net_ST2_Reio, Net_phi_plus_psi
\end{lstlisting}
For each key one can pass an integer value referring to the network generation to be trained or used, e.g. {\tt \textquotesingle Net\_ST0\_Reco\textquotesingle :2} if we want to train or use {\tt Net\_ST0\_Reco\_2.pt}.

\subsubsection{Domain sampling}

The generation of a \texttt{Domain} instance can be called from the \texttt{Workspace} instance, either by creating a domain based on a bestfit point and a covariance matrix that might have been obtained from an MCMC run, or by loading a domain which has been created in advance and has its properties stored in a workspace. 

The former can be done by calling the \texttt{save} function of the \texttt{Domain} class:
\begin{lstlisting}
	domain = my_workspace.domain_from_path(bestfit_path, 
		covmat_path, pnames, sigma_train=6, 
		sigma_validation=5, sigma_test=5)	
	domain.save() 
\end{lstlisting}
with appropriate paths to the bestfit and covariance matrix, as well as a list of parameters \texttt{pnames} to be sampled. The parameters \texttt{sigma\_train}, \texttt{sigma\_validate}  and \texttt{sigma\_test} specify the size of the ellipsoid on which we want to train, validate or test the networks.

Instead of being created from scratch, the domain can also be loaded from an existing workspace as 
\begin{lstlisting}
	domain = my_workspace.domain()
	domain.save() 
\end{lstlisting}
Next, the process of populating the domain with a given number of samples drawn from adapted Latin Hypercube Sampling (which was outlined in \cref{sec:training_cn}) is achieved by the function \texttt{.sample\_save()}:
\begin{lstlisting}
	domain.sample_save(training_count, 
		validation_count, test_count)
\end{lstlisting}
The arguments are the targeted number of samples. The samples are stored within the workspace within individual data directories
\begin{lstlisting}
	<path-to-workspace>/training_data/parameter_sample.h5
	<path-to-workspace>/validation_data/parameter_sample.h5
	<path-to-workspace>/test_data/parameter_sample.h5 
\end{lstlisting}

\subsubsection{Data generation}
The process of data generation is achieved by the nested \texttt{Generator} class, which can be called via a function of the \texttt{Workspace} class. It reads the samples that were created in the previous step, runs \classfull for each cosmology and saves the quantities used to train \classnet. These include the source functions and all the input parameters/functions described in \cref{sec:architecture} (saved as \texttt{sources\_xxx.h5}). To perform these steps, one first loads the sampled parameters with:
\begin{lstlisting}
	training, validation, test = 
		my_workspace.loader().cosmological_parameters()
\end{lstlisting}
These parameter samples are used to calculate the corresponding source source functions with:
\begin{lstlisting}
	my_workspace.generator().generate_source_data(
		fixed, training, validation, test, 
		fixed_training_only, processes)
\end{lstlisting}
Next to the lists of cosmological input parameters in each sample (\texttt{training, validation, test}), we additionally provide the function with dictionaries of fixed \class parameters \texttt{fixed} that specify the output and the precision of the generator.\textsuperscript{\ref{footnote:settings}} Additionally we provide extra precision parameters \texttt{fixed\_training\_only} used only for the training data set, to ensure the training quality.\footnote{While generating the training data, we reduce the numerical noise in the calculation of the source functions by setting the following precision parameters \texttt{\textquotesingle k\_min\_tau0\textquotesingle :1e-4}, \texttt{\textquotesingle tol\_background\_integration\textquotesingle :1e-12}, \texttt{\textquotesingle tol\_perturbation\_integration\textquotesingle :1e-6}, \texttt{\textquotesingle reionization\_optical\_depth\_tol\textquotesingle :1e-5}.} 
The argument {\tt processes} is the number of threads to be run in parallel.
Note that storing the source functions and the related cosmological quantities requires a large amount of disc space (300GB for 11\,000 source functions). 

It is interesting to comment on the last two steps performed within the python function \mbox{\color{deepblue} \tt generate\_source\_data()}.
Once all source functions and cosmological data have been generated, the $k$-array of the networks that are about to be trained is determined by combing the $k$-arrays of all generated data sets\footnote{In order to do so, the $k$-array with the most entries among the data set is extended by evenly sampling $k$ values in logspace until reaching the lowest $k_\mathrm{min}$ over the data set.} with:
\begin{lstlisting}
	workspace.generator().generate_k_array()
\end{lstlisting}
This array is stored both in the network weights and in the \texttt{manifest.json} dictionary which is generated by calling:
\begin{lstlisting}
	workspace.generator().write_manifest(fixed, pnames)
\end{lstlisting}
The last step also stores the minimal and maximal value of each source function and input parameter/function across all samples. These values are used later to automatically rescale the network input/output -- usually to values varying between -1 and 1. Note that overwriting the $k$-array or the normalization information within a given workspace would corrupt an older training.

\subsubsection{Training}

Once the previous files have been generated, the actual training of the network can start.
All steps are embedded in a nested \texttt{Trainer} class. As a preliminary step, the source functions are interpolated at the $k$-array of the networks and all NN input/output quantities are rescaled.\footnote{Quantities which are denoted as \enquote{cosmo inputs} in \cref{sec:architecture} are normalized by min-max to the range $[-1,1]$. The quantities $\tau,\tau/\tau_\mathrm{reio},\tau/\tau_\mathrm{reco}$ and the source functions $\delta_{\mathrm{m}}, \delta_\mathrm{cb}$ are traded against their logarithm in base 10. The visibility functions $g_\mathrm{reio},g'_\mathrm{reio}$ and all other source functions are simply rescaled by their largest absolute value, such that they do not exceed the range $[-1,1]$.} Then, the training of the NN weights proceeds as described in \cref{sec:training_cn}. For instance, the training of all networks is achieved with:
\begin{lstlisting}
	my_workspace.trainer().train_all_models(workers)
\end{lstlisting}
where {\tt workers} is the number of threads to be used. Intead, a specific training of e.g.
two networks \texttt{Net\_ST1, Net\_phi\_plus\_psi} can be achieved with:
\begin{lstlisting}
	from classynet.models import Net_ST1, Net_phi_plus_psi
	list = {'Net_ST1':Net_ST1, 
		'Net_phi_plus_psi': Net_phi_plus_psi}
	my_workspace.trainer().train_models(list, workers)
\end{lstlisting}
The number of epochs and the learning rate as a function of epoch can be specified for each individual network within the corresponding network specification file in the
\begin{center}
\texttt{<path-to-class>/python/nn/models/}
\end{center}
folder. After each epoch, a checkpoint is saved in the 
\begin{center}
\texttt{<path-to-workspace>/models/}
\end{center}
directory as a pytorch \texttt{.pt} compressed file. We additionally store the final NN weights as a separate \texttt{.pt} file for easier access. With the network architecture adopted in this work, we expect that it might be necessary to re-train the most critical networks [N2] and [N5] multiple times, since the result of a single training is not necessarily optimal (see \cref{sec:training_cn}). Posterior cuts through the Planck likelihoods can be further used to select the version yielding the smallest parameter biases.

\clearpage

\bibliography{Biblio}{}

\providecommand{\href}[2]{#2}\begingroup\raggedright\begin{thebibliography}{10}

\bibitem{camb_code}
A.~Lewis and A.~Challinor, ``Camb: Code for anisotropies in the microwave
  background.'' Astrophysics Source Code Library.

\bibitem{class_overview}
J.~Lesgourgues, \emph{The cosmic linear anisotropy solving system (class) i:
  Overview},  2011.

\bibitem{class_approximations}
D.~Blas, J.~Lesgourgues and T.~Tram, \emph{The cosmic linear anisotropy solving
  system (class). part ii: Approximation schemes},
  \href{http://dx.doi.org/10.1088/1475-7516/2011/07/034}{\emph{Journal of
  Cosmology and Astroparticle Physics} {\bf 2011} (2011) 034–034}.

\bibitem{Campagne:2017xps}
J.~E. Campagne, J.~Neveu and S.~Plaszczynski, \emph{{Angpow: a software for the
  fast computation of accurate tomographic power spectra}},
  \href{http://dx.doi.org/10.1051/0004-6361/201730399}{\emph{Astron.
  Astrophys.} {\bf 602} (2017) A72},
  [\href{https://arxiv.org/abs/1701.03592}{{\tt 1701.03592}}].

\bibitem{Schoneberg:2018fis}
N.~Sch\"oneberg, M.~Simonovi\'c, J.~Lesgourgues and M.~Zaldarriaga,
  \emph{{Beyond the traditional Line-of-Sight approach of cosmological angular
  statistics}},
  \href{http://dx.doi.org/10.1088/1475-7516/2018/10/047}{\emph{JCAP} {\bf 10}
  (2018) 047}, [\href{https://arxiv.org/abs/1807.09540}{{\tt 1807.09540}}].

\bibitem{Heitmann:2008eq}
K.~Heitmann, M.~White, C.~Wagner, S.~Habib and D.~Higdon, \emph{{The Coyote
  Universe I: Precision Determination of the Nonlinear Matter Power Spectrum}},
  \href{http://dx.doi.org/10.1088/0004-637X/715/1/104}{\emph{Astrophys. J.}
  {\bf 715} (2010) 104--121}, [\href{https://arxiv.org/abs/0812.1052}{{\tt
  0812.1052}}].

\bibitem{Heitmann:2009cu}
K.~Heitmann, D.~Higdon, M.~White, S.~Habib, B.~J. Williams and C.~Wagner,
  \emph{{The Coyote Universe II: Cosmological Models and Precision Emulation of
  the Nonlinear Matter Power Spectrum}},
  \href{http://dx.doi.org/10.1088/0004-637X/705/1/156}{\emph{Astrophys. J.}
  {\bf 705} (2009) 156--174}, [\href{https://arxiv.org/abs/0902.0429}{{\tt
  0902.0429}}].

\bibitem{Lawrence:2009uk}
E.~Lawrence, K.~Heitmann, M.~White, D.~Higdon, C.~Wagner, S.~Habib et~al.,
  \emph{{The Coyote Universe III: Simulation Suite and Precision Emulator for
  the Nonlinear Matter Power Spectrum}},
  \href{http://dx.doi.org/10.1088/0004-637X/713/2/1322}{\emph{Astrophys. J.}
  {\bf 713} (2010) 1322--1331}, [\href{https://arxiv.org/abs/0912.4490}{{\tt
  0912.4490}}].

\bibitem{Agarwal:2012ew}
S.~Agarwal, F.~B. Abdalla, H.~A. Feldman, O.~Lahav and S.~A. Thomas,
  \emph{{PkANN - I. Non-linear matter power spectrum interpolation through
  artificial neural networks}},
  \href{http://dx.doi.org/10.1111/j.1365-2966.2012.21326.x}{\emph{Mon. Not.
  Roy. Astron. Soc.} {\bf 424} (2012) 1409--1418},
  [\href{https://arxiv.org/abs/1203.1695}{{\tt 1203.1695}}].

\bibitem{Agarwal:2013aea}
S.~Agarwal, F.~B. Abdalla, H.~A. Feldman, O.~Lahav and S.~A. Thomas,
  \emph{{pkann \textendash{} II. A non-linear matter power spectrum
  interpolator developed using artificial neural networks}},
  \href{http://dx.doi.org/10.1093/mnras/stu090}{\emph{Mon. Not. Roy. Astron.
  Soc.} {\bf 439} (2014) 2102--2121},
  [\href{https://arxiv.org/abs/1312.2101}{{\tt 1312.2101}}].

\bibitem{Heitmann:2013bra}
K.~Heitmann, E.~Lawrence, J.~Kwan, S.~Habib and D.~Higdon, \emph{{The Coyote
  Universe Extended: Precision Emulation of the Matter Power Spectrum}},
  \href{http://dx.doi.org/10.1088/0004-637X/780/1/111}{\emph{Astrophys. J.}
  {\bf 780} (2014) 111}, [\href{https://arxiv.org/abs/1304.7849}{{\tt
  1304.7849}}].

\bibitem{Lawrence:2017ost}
E.~Lawrence, K.~Heitmann, J.~Kwan, A.~Upadhye, D.~Bingham, S.~Habib et~al.,
  \emph{{The Mira-Titan Universe II: Matter Power Spectrum Emulation}},
  \href{http://dx.doi.org/10.3847/1538-4357/aa86a9}{\emph{Astrophys. J.} {\bf
  847} (2017) 50}, [\href{https://arxiv.org/abs/1705.03388}{{\tt 1705.03388}}].

\bibitem{Euclid:2018mlb}
{\scshape Euclid} collaboration, M.~Knabenhans et~al., \emph{{Euclid
  preparation: II. The EuclidEmulator -- A tool to compute the cosmology
  dependence of the nonlinear matter power spectrum}},
  \href{http://dx.doi.org/10.1093/mnras/stz197}{\emph{Mon. Not. Roy. Astron.
  Soc.} {\bf 484} (2019) 5509--5529},
  [\href{https://arxiv.org/abs/1809.04695}{{\tt 1809.04695}}].

\bibitem{Ho:2021tem}
M.-F. Ho, S.~Bird and C.~R. Shelton, \emph{{Multifidelity emulation for the
  matter power spectrum using Gaussian processes}},
  \href{http://dx.doi.org/10.1093/mnras/stab3114}{\emph{Mon. Not. Roy. Astron.
  Soc.} {\bf 509} (2021) 2551--2565},
  [\href{https://arxiv.org/abs/2105.01081}{{\tt 2105.01081}}].

\bibitem{Euclid:2020rfv}
{\scshape Euclid} collaboration, M.~Knabenhans et~al., \emph{{Euclid
  preparation: IX. EuclidEmulator2 \textendash{} power spectrum emulation with
  massive neutrinos and self-consistent dark energy perturbations}},
  \href{http://dx.doi.org/10.1093/mnras/stab1366}{\emph{Mon. Not. Roy. Astron.
  Soc.} {\bf 505} (2021) 2840--2869},
  [\href{https://arxiv.org/abs/2010.11288}{{\tt 2010.11288}}].

\bibitem{Arico:2021izc}
G.~Aric\`o, R.~E. Angulo and M.~Zennaro, \emph{{Accelerating
  Large-Scale-Structure data analyses by emulating Boltzmann solvers and
  Lagrangian Perturbation Theory}},
  \href{https://arxiv.org/abs/2104.14568}{{\tt 2104.14568}}.

\bibitem{DeRose:2018xdj}
J.~DeRose, R.~H. Wechsler, J.~L. Tinker, M.~R. Becker, Y.-Y. Mao, T.~McClintock
  et~al., \emph{{The Aemulus Project I: Numerical Simulations for Precision
  Cosmology}},
  \href{http://dx.doi.org/10.3847/1538-4357/ab1085}{\emph{Astrophys. J.} {\bf
  875} (2019) 69}, [\href{https://arxiv.org/abs/1804.05865}{{\tt 1804.05865}}].

\bibitem{McClintock:2018uyf}
T.~McClintock, E.~Rozo, M.~R. Becker, J.~DeRose, Y.-Y. Mao, S.~McLaughlin
  et~al., \emph{{The Aemulus Project II: Emulating the Halo Mass Function}},
  \href{http://dx.doi.org/10.3847/1538-4357/aaf568}{\emph{Astrophys. J.} {\bf
  872} (2019) 53}, [\href{https://arxiv.org/abs/1804.05866}{{\tt 1804.05866}}].

\bibitem{Zhai:2018plk}
Z.~Zhai, J.~L. Tinker, M.~R. Becker, J.~DeRose, Y.-Y. Mao, T.~McClintock
  et~al., \emph{{The Aemulus Project III: Emulation of the Galaxy Correlation
  Function}},
  \href{http://dx.doi.org/10.3847/1538-4357/ab0d7b}{\emph{Astrophys. J.} {\bf
  874} (2019) 95}, [\href{https://arxiv.org/abs/1804.05867}{{\tt 1804.05867}}].

\bibitem{Zennaro:2021bwy}
M.~Zennaro, R.~E. Angulo, M.~Pellejero-Ib\'a\~nez, J.~St\"ucker, S.~Contreras
  and G.~Aric\`o, \emph{{The BACCO simulation project: biased tracers in real
  space}},  \href{https://arxiv.org/abs/2101.12187}{{\tt 2101.12187}}.

\bibitem{Kobayashi:2020zsw}
Y.~Kobayashi, T.~Nishimichi, M.~Takada, R.~Takahashi and K.~Osato,
  \emph{{Accurate emulator for the redshift-space power spectrum of dark matter
  halos and its application to galaxy power spectrum}},
  \href{http://dx.doi.org/10.1103/PhysRevD.102.063504}{\emph{Phys. Rev. D} {\bf
  102} (2020) 063504}, [\href{https://arxiv.org/abs/2005.06122}{{\tt
  2005.06122}}].

\bibitem{dsh_code}
M.~Kaplinghat, L.~Knox and C.~Skordis, \emph{{Rapid calculation of theoretical
  cmb angular power spectra}},
  \href{http://dx.doi.org/10.1086/342656}{\emph{Astrophys. J.} {\bf 578} (2002)
  665}, [\href{https://arxiv.org/abs/astro-ph/0203413}{{\tt
  astro-ph/0203413}}].

\bibitem{cmbwarp_code}
R.~Jimenez, L.~Verde, H.~Peiris and A.~Kosowsky, \emph{{Fast cosmological
  parameter estimation from microwave background temperature and polarization
  power spectra}},
  \href{http://dx.doi.org/10.1103/PhysRevD.70.023005}{\emph{Phys. Rev. D} {\bf
  70} (2004) 023005}, [\href{https://arxiv.org/abs/astro-ph/0404237}{{\tt
  astro-ph/0404237}}].

\bibitem{pico_code}
B.~Fendt, Chad:~Wandelt, ``Pico: Parameters for the impatient cosmologist.''
  Astrophysics Source Code Library.

\bibitem{WMAP:2012nax}
{\scshape WMAP} collaboration, G.~Hinshaw et~al., \emph{{Nine-Year Wilkinson
  Microwave Anisotropy Probe (WMAP) Observations: Cosmological Parameter
  Results}},
  \href{http://dx.doi.org/10.1088/0067-0049/208/2/19}{\emph{Astrophys. J.
  Suppl.} {\bf 208} (2013) 19}, [\href{https://arxiv.org/abs/1212.5226}{{\tt
  1212.5226}}].

\bibitem{Planck:2013pxb}
{\scshape Planck} collaboration, P.~A.~R. Ade et~al., \emph{{Planck 2013
  results. XVI. Cosmological parameters}},
  \href{http://dx.doi.org/10.1051/0004-6361/201321591}{\emph{Astron.
  Astrophys.} {\bf 571} (2014) A16},
  [\href{https://arxiv.org/abs/1303.5076}{{\tt 1303.5076}}].

\bibitem{planck2015}
P.~A.~R. Ade, N.~Aghanim, M.~Arnaud, M.~Ashdown, J.~Aumont, C.~Baccigalupi
  et~al., \emph{Planck2015 results},
  \href{http://dx.doi.org/10.1051/0004-6361/201525830}{\emph{Astronomy and
  Astrophysics} {\bf 594} (2016) A13}.

\bibitem{planck2018}
N.~Aghanim, Y.~Akrami, M.~Ashdown, J.~Aumont, C.~Baccigalupi, M.~Ballardini
  et~al., \emph{Planck 2018 results},
  \href{http://dx.doi.org/10.1051/0004-6361/201833910}{\emph{Astronomy and
  Astrophysics} {\bf 641} (2020) A6}.

\bibitem{Auld:2006pm}
T.~Auld, M.~Bridges, M.~P. Hobson and S.~F. Gull, \emph{{Fast cosmological
  parameter estimation using neural networks}},
  \href{http://dx.doi.org/10.1111/j.1745-3933.2006.00276.x}{\emph{Mon. Not.
  Roy. Astron. Soc.} {\bf 376} (2007) L11--L15},
  [\href{https://arxiv.org/abs/astro-ph/0608174}{{\tt astro-ph/0608174}}].

\bibitem{Auld:2007qz}
T.~Auld, M.~Bridges and M.~P. Hobson, \emph{{CosmoNet: Fast cosmological
  parameter estimation in non-flat models using neural networks}},
  \href{http://dx.doi.org/10.1111/j.1365-2966.2008.13279.x}{\emph{Mon. Not.
  Roy. Astron. Soc.} {\bf 387} (2008) 1575},
  [\href{https://arxiv.org/abs/astro-ph/0703445}{{\tt astro-ph/0703445}}].

\bibitem{Albers:2019rzt}
J.~Albers, C.~Fidler, J.~Lesgourgues, N.~Sch\"oneberg and J.~Torrado,
  \emph{{CosmicNet. Part I. Physics-driven implementation of neural networks
  within Einstein-Boltzmann Solvers}},
  \href{http://dx.doi.org/10.1088/1475-7516/2019/09/028}{\emph{JCAP} {\bf 09}
  (2019) 028}, [\href{https://arxiv.org/abs/1907.05764}{{\tt 1907.05764}}].

\bibitem{Manrique-Yus:2019hqc}
A.~Manrique-Yus and E.~Sellentin, \emph{{Euclid-era cosmology for everyone:
  neural net assisted MCMC sampling for the joint 3 \texttimes{} 2
  likelihood}}, \href{http://dx.doi.org/10.1093/mnras/stz3059}{\emph{Mon. Not.
  Roy. Astron. Soc.} {\bf 491} (2020) 2655--2663},
  [\href{https://arxiv.org/abs/1907.05881}{{\tt 1907.05881}}].

\bibitem{SpurioMancini:2021ppk}
A.~Spurio~Mancini, D.~Piras, J.~Alsing, B.~Joachimi and M.~P. Hobson,
  \emph{{CosmoPower: emulating cosmological power spectra for accelerated
  Bayesian inference from next-generation surveys}},
  \href{http://dx.doi.org/10.1093/mnras/stac064}{\emph{Mon. Not. Roy. Astron.
  Soc.} {\bf 511} (2022) 1771--1788},
  [\href{https://arxiv.org/abs/2106.03846}{{\tt 2106.03846}}].

\bibitem{Nygaard:2022wri}
A.~Nygaard, E.~B. Holm, S.~Hannestad and T.~Tram, \emph{{CONNECT: A neural
  network based framework for emulating cosmological observables and
  cosmological parameter inference}},
  \href{https://arxiv.org/abs/2205.15726}{{\tt 2205.15726}}.

\bibitem{Bevins:2021eah}
H.~T.~J. Bevins, W.~J. Handley, A.~Fialkov, E.~d.~L. Acedo and K.~Javid,
  \emph{{globalemu: a novel and robust approach for emulating the sky-averaged
  21-cm signal from the cosmic dawn and epoch of reionization}},
  \href{http://dx.doi.org/10.1093/mnras/stab2737}{\emph{Mon. Not. Roy. Astron.
  Soc.} {\bf 508} (2021) 2923--2936},
  [\href{https://arxiv.org/abs/2104.04336}{{\tt 2104.04336}}].

\bibitem{Mootoovaloo:2021rot}
A.~Mootoovaloo, A.~H. Jaffe, A.~F. Heavens and F.~Leclercq, \emph{{Kernel-based
  emulator for the 3D matter power spectrum from CLASS}},
  \href{http://dx.doi.org/10.1016/j.ascom.2021.100508}{\emph{Astron. Comput.}
  {\bf 38} (2022) 100508}, [\href{https://arxiv.org/abs/2105.02256}{{\tt
  2105.02256}}].

\bibitem{Lesgourgues:2013bra}
J.~Lesgourgues and T.~Tram, \emph{{Fast and accurate CMB computations in
  non-flat FLRW universes}},
  \href{http://dx.doi.org/10.1088/1475-7516/2014/09/032}{\emph{JCAP} {\bf 09}
  (2014) 032}, [\href{https://arxiv.org/abs/1312.2697}{{\tt 1312.2697}}].

\bibitem{gauges}
C.-P. Ma and E.~Bertschinger, \emph{Cosmological perturbation theory in the
  synchronous and conformal newtonian gauges},
  \href{http://dx.doi.org/10.1086/176550}{\emph{The Astrophysical Journal} {\bf
  455} (1995) 7}.

\bibitem{Tram:2013ima}
T.~Tram and J.~Lesgourgues, \emph{{Optimal polarisation equations in FLRW
  universes}},
  \href{http://dx.doi.org/10.1088/1475-7516/2013/10/002}{\emph{JCAP} {\bf 10}
  (2013) 002}, [\href{https://arxiv.org/abs/1305.3261}{{\tt 1305.3261}}].

\bibitem{Pitrou:2020lhu}
C.~Pitrou, T.~S. Pereira and J.~Lesgourgues, \emph{{Optimal Boltzmann
  hierarchies with nonvanishing spatial curvature}},
  \href{http://dx.doi.org/10.1103/PhysRevD.102.023511}{\emph{Phys. Rev. D} {\bf
  102} (2020) 023511}, [\href{https://arxiv.org/abs/2005.12119}{{\tt
  2005.12119}}].

\bibitem{DiDio:2013bqa}
E.~Di~Dio, F.~Montanari, J.~Lesgourgues and R.~Durrer, \emph{{The CLASSgal code
  for Relativistic Cosmological Large Scale Structure}},
  \href{http://dx.doi.org/10.1088/1475-7516/2013/11/044}{\emph{JCAP} {\bf 11}
  (2013) 044}, [\href{https://arxiv.org/abs/1307.1459}{{\tt 1307.1459}}].

\bibitem{Chevallier:2000qy}
M.~Chevallier and D.~Polarski, \emph{{Accelerating universes with scaling dark
  matter}}, \href{http://dx.doi.org/10.1142/S0218271801000822}{\emph{Int. J.
  Mod. Phys. D} {\bf 10} (2001) 213--224},
  [\href{https://arxiv.org/abs/gr-qc/0009008}{{\tt gr-qc/0009008}}].

\bibitem{Linder:2002et}
E.~V. Linder, \emph{{Exploring the expansion history of the universe}},
  \href{http://dx.doi.org/10.1103/PhysRevLett.90.091301}{\emph{Phys. Rev.
  Lett.} {\bf 90} (2003) 091301},
  [\href{https://arxiv.org/abs/astro-ph/0208512}{{\tt astro-ph/0208512}}].

\bibitem{Hu:1995em}
W.~T. Hu, \emph{{Wandering in the Background: A CMB Explorer}},  other thesis,
  8, 1995.

\bibitem{hu_eisenstein}
D.~J. Eisenstein and W.~Hu, \emph{Baryonic features in the matter transfer
  function}, \href{http://dx.doi.org/10.1086/305424}{\emph{The Astrophysical
  Journal} {\bf 496} (1998) 605–614}.

\bibitem{planck2018_1}
N.~Aghanim, Y.~Akrami, M.~Ashdown, J.~Aumont, C.~Baccigalupi, M.~Ballardini
  et~al., \emph{Planck 2018 results. v. cmb power spectra and likelihoods},
  \href{http://dx.doi.org/10.1051/0004-6361/201936386}{\emph{Astronomy and
  Astrophysics} {\bf 641} (2020) A5}.

\bibitem{planck2018_2}
N.~Aghanim, Y.~Akrami, M.~Ashdown, J.~Aumont, C.~Baccigalupi, M.~Ballardini
  et~al., \emph{Planck2018 results. viii. gravitational lensing},
  \href{http://dx.doi.org/10.1051/0004-6361/201833886}{\emph{Astronomy and
  Astrophysics} {\bf 641} (2020) A8}.

\bibitem{bao_sixdf}
F.~Beutler, C.~Blake, M.~Colless, D.~H. Jones, L.~Staveley-Smith, L.~Campbell
  et~al., \emph{The 6df galaxy survey: baryon acoustic oscillations and the
  local hubble constant},
  \href{http://dx.doi.org/10.1111/j.1365-2966.2011.19250.x}{\emph{Monthly
  Notices of the Royal Astronomical Society} {\bf 416} (2011) 3017–3032}.

\bibitem{bao_dr7}
A.~J. Ross, L.~Samushia, C.~Howlett, W.~J. Percival, A.~Burden and M.~Manera,
  \emph{The clustering of the sdss dr7 main galaxy sample – i. a 4 per cent
  distance measure at z = 0.15},
  \href{http://dx.doi.org/10.1093/mnras/stv154}{\emph{Monthly Notices of the
  Royal Astronomical Society} {\bf 449} (2015) 835–847}.

\bibitem{bao_dr12}
S.~Alam, M.~Ata, S.~Bailey, F.~Beutler, D.~Bizyaev, J.~A. Blazek et~al.,
  \emph{The clustering of galaxies in the completed sdss-iii baryon oscillation
  spectroscopic survey: cosmological analysis of the dr12 galaxy sample},
  \href{http://dx.doi.org/10.1093/mnras/stx721}{\emph{Monthly Notices of the
  Royal Astronomical Society} {\bf 470} (2017) 2617–2652}.

\bibitem{pantheon}
D.~M. Scolnic, D.~O. Jones, A.~Rest, Y.~C. Pan, R.~Chornock, R.~J. Foley
  et~al., \emph{The complete light-curve sample of spectroscopically confirmed
  sne ia from pan-starrs1 and cosmological constraints from the combined
  pantheon sample}, \href{http://dx.doi.org/10.3847/1538-4357/aab9bb}{\emph{The
  Astrophysical Journal} {\bf 859} (2018) 101}.

\bibitem{cosmopower}
A.~S. Mancini, D.~Piras, J.~Alsing, B.~Joachimi and M.~P. Hobson,
  \emph{Cosmopower: emulating cosmological power spectra for accelerated
  bayesian inference from next-generation surveys},  2021.

\bibitem{adam}
D.~P. Kingma and J.~Ba, \emph{Adam: A method for stochastic optimization},
  2017.

\bibitem{Brinckmann:2018cvx}
T.~Brinckmann and J.~Lesgourgues, \emph{{MontePython 3: boosted MCMC sampler
  and other features}},
  \href{http://dx.doi.org/10.1016/j.dark.2018.100260}{\emph{Phys. Dark Univ.}
  {\bf 24} (2019) 100260}, [\href{https://arxiv.org/abs/1804.07261}{{\tt
  1804.07261}}].

\bibitem{cobaya}
J.~Torrado and A.~Lewis, \emph{Cobaya: code for bayesian analysis of
  hierarchical physical models},
  \href{http://dx.doi.org/10.1088/1475-7516/2021/05/057}{\emph{Journal of
  Cosmology and Astroparticle Physics} {\bf 2021} (2021) 057}.

\bibitem{cobaya_code}
J.~Torrado and A.~Lewis, ``Cobaya: Bayesian analysis in cosmology.''
  Astrophysics Source Code Library.

\bibitem{Lewis:2002ah}
A.~Lewis and S.~Bridle, \emph{{Cosmological parameters from CMB and other data:
  A Monte Carlo approach}},
  \href{http://dx.doi.org/10.1103/PhysRevD.66.103511}{\emph{Phys. Rev.} {\bf
  D66} (2002) 103511}, [\href{https://arxiv.org/abs/astro-ph/0205436}{{\tt
  astro-ph/0205436}}].

\bibitem{Lewis:2013hha}
A.~Lewis, \emph{{Efficient sampling of fast and slow cosmological parameters}},
  \href{http://dx.doi.org/10.1103/PhysRevD.87.103529}{\emph{Phys. Rev.} {\bf
  D87} (2013) 103529}, [\href{https://arxiv.org/abs/1304.4473}{{\tt
  1304.4473}}].

\bibitem{Eisenstein:1997jh}
D.~J. Eisenstein and W.~Hu, \emph{{Power spectra for cold dark matter and its
  variants}}, \href{http://dx.doi.org/10.1086/306640}{\emph{Astrophys. J.} {\bf
  511} (1997) 5}, [\href{https://arxiv.org/abs/astro-ph/9710252}{{\tt
  astro-ph/9710252}}].

\bibitem{Rossi:2014nea}
G.~Rossi, C.~Y\`eche, N.~Palanque-Delabrouille and J.~Lesgourgues,
  \emph{{Constraints on dark radiation from cosmological probes}},
  \href{http://dx.doi.org/10.1103/PhysRevD.92.063505}{\emph{Phys. Rev. D} {\bf
  92} (2015) 063505}, [\href{https://arxiv.org/abs/1412.6763}{{\tt
  1412.6763}}].

\bibitem{Grandon:2022gdr}
D.~Grand\'on and E.~Sellentin, \emph{{Bayesian error propagation for neural-net
  based parameter inference}},  \href{https://arxiv.org/abs/2205.11587}{{\tt
  2205.11587}}.

\bibitem{montepython_code}
J.~B. K. P.~S. Audren, Benjamin;~Lesgourgues, ``Monte python: Monte carlo code
  for class in python.'' Astrophysics Source Code Library.

\bibitem{Lesgourgues:2006nd}
J.~Lesgourgues and S.~Pastor, \emph{{Massive neutrinos and cosmology}},
  \href{http://dx.doi.org/10.1016/j.physrep.2006.04.001}{\emph{Phys. Rept.}
  {\bf 429} (2006) 307--379},
  [\href{https://arxiv.org/abs/astro-ph/0603494}{{\tt astro-ph/0603494}}].

\bibitem{Froustey:2020mcq}
J.~Froustey, C.~Pitrou and M.~C. Volpe, \emph{{Neutrino decoupling including
  flavour oscillations and primordial nucleosynthesis}},
  \href{http://dx.doi.org/10.1088/1475-7516/2020/12/015}{\emph{JCAP} {\bf 12}
  (2020) 015}, [\href{https://arxiv.org/abs/2008.01074}{{\tt 2008.01074}}].

\bibitem{Bennett:2020zkv}
J.~J. Bennett, G.~Buldgen, P.~F. De~Salas, M.~Drewes, S.~Gariazzo, S.~Pastor
  et~al., \emph{{Towards a precision calculation of $N_{\rm eff}$ in the
  Standard Model II: Neutrino decoupling in the presence of flavour
  oscillations and finite-temperature QED}},
  \href{http://dx.doi.org/10.1088/1475-7516/2021/04/073}{\emph{JCAP} {\bf 04}
  (2021) 073}, [\href{https://arxiv.org/abs/2012.02726}{{\tt 2012.02726}}].

\end{thebibliography}\endgroup
\bibliographystyle{JHEP}

\end{document}